\documentclass{article}

\usepackage[english]{babel}
\usepackage[a4paper,top=2cm,bottom=2cm,left=3cm,right=3cm,marginparwidth=1.75cm]{geometry}

\usepackage{amsmath}
\usepackage{xcolor,graphicx}
\usepackage{hyperref}
\hypersetup{%
  colorlinks=true,
  breaklinks=true,
  linkcolor=blue,
  anchorcolor=black,
  citecolor=brown,
  filecolor=blue,
  menucolor=red,
  urlcolor=blue,
}
\usepackage[authoryear]{natbib}
\usepackage{subcaption}

\providecommand{\keywords}[1]{\textbf{\textit{Keywords---}} #1}

\title{Is Stephen Curry really a guard? A new perspective on player typologies using functional data analysis}
\author{%
Steven Golovkine\thanks{Département de mathématiques et statistique, Université Laval, Canada \href{mailto:steven.golovkine@mat.ulaval.ca}{steven.golovkine@mat.ulaval.ca}}
\and
Edward Gunning\thanks{Department of Biostatistics and Epidemiology, University of Pennsylvania, USA \href{mailto:edward.gunning@pennmedicine.upenn.edu}{edward.gunning@pennmedicine.upenn.edu}}
}
\date{\today}

\begin{document}
\maketitle

\begin{abstract}
We present a novel representation of NBA players' shooting patterns based on Functional Data Analysis (FDA). Each player's charts of made and missed shots are treated as smooth functional data defined over a two-dimensional domain corresponding to the offensive half-court. This continuous representation enables a parsimonious multivariate functional principal components analysis (MFPCA) decomposition, producing a set of common principal component functions that capture the primary modes of variability in shooting patterns, along with player-specific scores that quantify individual deviations from the average behavior. We first interpret the principal component functions to characterize the main sources of variation in shooting tendencies. We then apply $k$-medoids clustering to the principal component scores to construct a data-driven taxonomy of players. Comparing our empirical clusters to conventional NBA position labels reveals low agreement, suggesting that our shooting-pattern representation might capture aspects of playing style not fully reflected in official designations. The application of FDA to this area introduces a flexible, interpretable, and continuous framework for analyzing player tendencies, with potential applications in coaching, scouting, and historical player or match comparisons.

\end{abstract}


\section{Introduction} 
\label{sec:introduction}

With the increasing availability of large, detailed, and complex sports performance datasets, sports analytics has become a rapidly expanding field of both research and application, drawing significant interest from the statistics and machine learning communities. In team sports, traditional approaches have often focused on predicting game outcomes based on team characteristics \citep{dixonModellingAssociationFootball1997,babootaPredictiveAnalysisModelling2019}; for recent reviews, see \cite{horvatUseMachineLearning2020,bunkerApplicationMachineLearning2022}.
More recently, the availability of tracking data has enabled researchers to delve into more specific aspects of various sports, such as evaluating individual player performance in soccer \citep{pappalardoPlayeRankDatadrivenPerformance2019,decroosActionsSpeakLouder2019} or analyzing defensive performance in American football passing plays \citep{schmidSimulatingDefensiveTrajectories2021}. See \cite{kovalchikPlayerTrackingData2023} for a recent review on machine learning approaches to analyze tracking data. Basketball, in particular, presents opportunities for analysis, supported by extensive data provided by the National Basketball Association (NBA).

Basketball is a team sport in which two teams of five players compete to score points by shooting a ball through the opponent's hoop while defending their own. The game has been extensively studied through various analytical perspectives. For example, researchers have examined shooting mechanics and movements \citep{millerRelationshipBasketballShooting1996,liuChangesBasketballShooting1999,okazakiReviewBasketballJump2015}, estimated pre-game and in-game winning probabilities \citep{maddoxBayesianEstimationIngame2022}, and modeled outcomes of possessions \citep{cervoneMultiresolutionStochasticProcess2016}. Statistical models have also been applied to analyze defensive strategies \citep{csataljayEffectsDefensivePressure2013} and spatial performance on the court \citep{zuccolottoSpatialPerformanceAnalysis2023}. For a comprehensive review of performance analysis in basketball, see \cite{ternerModelingPlayerTeam2021}.

In basketball, game outcomes are heavily influenced by shooting accuracy, as a player's effectiveness is often defined by their ability to consistently score from various court locations. 
Analyzing shooting patterns thus serves as a powerful tool, not only for assessing individual performance but also for refining team strategies. An important tool for examining shooting patterns is shooting charts, which are visual representations that display all the shots a player has attempted and made over a season or a single game. Shooting charts are extensively studied in basketball research, with much of the work focusing on modeling them as spatial point processes. This approach typically involves creating a count matrix of shots taken by each player on a discretized court, then estimating an intensity surface over this grid by fitting a log-Gaussian Cox process. A low-rank approximation of these intensity surfaces is achieved using non-negative matrix factorization, as described in \citet{millerFactorizedPointProcess2014,franksCharacterizingSpatialStructure2015}. Non-parametric estimation of the intensity surface is also considered in \cite{yinBayesianNonparametricLearning2022}. These low-rank intensity surfaces help characterize NBA players’ shooting tendencies and evaluate their scoring efficiency. Some models do not require the intensity surface fitting approach. For example, \cite{reichSpatialAnalysisBasketball2006} employed conditionally autoregressive models to capture spatial correlations, while \cite{huZeroinflatedPoissonModel2023} fitted a Bayesian zero-inflated Poisson model to represent shooting attempts for players with distinct shooting styles. Although these methods effectively summarize player data, they are less suitable for comparing different players.

To compare players' shooting behaviors, specialized methods have been developed. \citet{fuHoopInSightAnalyzingComparing2024} employ (interactive) data visualization techniques to analyze and contrast shooting performance. In a different approach, \citet{huBayesianGroupLearning2021} use a log-Gaussian Cox process to model the shooting intensity with a specific similarity measure for comparing players. \citet{jiaoBayesianMarkedSpatial2021} propose a Bayesian marked spatial point process model to capture shot intensity, investigating whether players’ success rates are higher in areas where they attempt more shots and clustering players based on this pattern. In another study, \citet{wong-toiJointAnalysisField2022} analyze shooting patterns across players by dividing the court into a grid and using a Bayesian nonparametric mixture model to identify latent clusters among players. Similarly, \citet{yinAnalysisProfessionalBasketball2023} apply a Bayesian nonparametric matrix clustering technique to blue estimate intensity maps, offering insights into shot selection characteristics. All these methods require a discretized representation of the court.

To overcome the requirement of court discretization, we propose to estimate the shot density continuously across the offensive half-court and model these densities as functional data \citep{ramsayFunctionalDataAnalysis2005,kokoszkaIntroductionFunctionalData2017}.
This idea first appeared in our paper \cite{golovkineUseGramMatrix2024}. Both missed and made shot charts are treated as multivariate functional data over a two-dimensional domain (i.e., the offensive half-court). Functional Data Analysis (FDA) addresses the analysis of realizations of a continuous stochastic process, observed at discrete sampling points and defined on compact domains. 
Densities are computed separately for each shot type (made and missed) and each player and are therefore normalized for each individuals total number of each shots.
We prefer this over an un-normalized representation that contains information about overall shot volume, so that we can make make fair comparisons of shooting
patterns that are not affected by, e.g., number of games or minutes played.
By treating shot density as a stochastic process defined over the offensive half-court, we can consider each player’s shot density as a realization of this process. FDA tools allow us to summarize and capture variability within this process. To achieve this, we perform a basis decomposition using multivariate functional principal components analysis (MFPCA), and then describe players’ shooting styles based on this decomposition. Finally, representing player data onto the basis decomposition allows us to use clustering techniques to identify players with similar shooting behaviors. This approach to analyzing shooting patterns provides nuanced insights that can be valuable for coaches, analysts, and players alike.

The remainder of the article is organized as follows. In Section~\ref{sec:dataset}, we introduce the dataset covering the 2018-2019 through 2023-2024 regular NBA seasons. Section~\ref{sec:method} details our methodology, while Section~\ref{sec:results} presents the application of this methodology to NBA player data. Finally, we conclude with a discussion in Section~\ref{sec:discussion}.

\section{Dataset} 
\label{sec:dataset}

Our motivating dataset is derived from the NBA. To access comprehensive game data, we use the Python package \texttt{nba\_api}\footnote{\url{https://github.com/swar/nba_api/}}, which interfaces with the APIs provided by \url{nba.com}. The dataset includes both made and missed field goal attempt locations from the offensive half-court, covering all NBA games from the $2018-2019$ to the $2023-2024$ regular seasons. The offensive half-court is a $50$ ft (sideline to sideline) by $47$ ft (baseline to mid court line) rectangle. Filtering the dataset, we focus on players who attempted more than $1000$ field goals during this six-season period, resulting in a cohort of $173$ players out of the $573$ players who have stepped onto the court during these six seasons. Collectively, these players attempted $717381$ shots, of which $340466$ ($47\%$) were successful. To ensure data quality, we exclude attempts classified as impossible (e.g., out-of-bounds), leaving a dataset of $716114$ shots, with $340430$ ($48\%$) successful attempts.

In the NBA, players are typically categorized by their roles on the court into three primary positions: `guard', `forward', and `center'. Guards are responsible for ball handling, playmaking, and perimeter shooting. Centers, often referred to as ``big men'', primarily operate near the basket to score in the paint. Forwards are versatile players who contribute in multiple areas of the court. Additionally, the NBA recognizes hybrid roles for players with overlapping skill sets: `guard-forward', `forward-guard', `center-forward', and `forward-center'. To ensure a sufficient number of players in each group, we aggregate the `guard-forward' and `forward-guard' roles into a single `forward-guard' category, and the `center-forward' and `forward-center' roles into a single `forward-center' category. This results in five distinct positional groups (`guard', `forward-guard', `forward', `forward-center' and `center'), corresponding to the number of players on the court. The number of players across these categories is shown in Table~\ref{tab:player_position}.
\begin{table}
	\centering
	\caption{Number of players for each position according to the NBA.}
	\label{tab:player_position}
	\begin{tabular}{ c|ccccc }
		Position      & Guard & Forward-Guard & Forward & Forward-Center & Center \\
		\hline
		\# of players & 68    & 24            & 43      & 25             & 13     \\
	\end{tabular}
\end{table}

\begin{figure}
	\centering
	\includegraphics[scale=0.5]{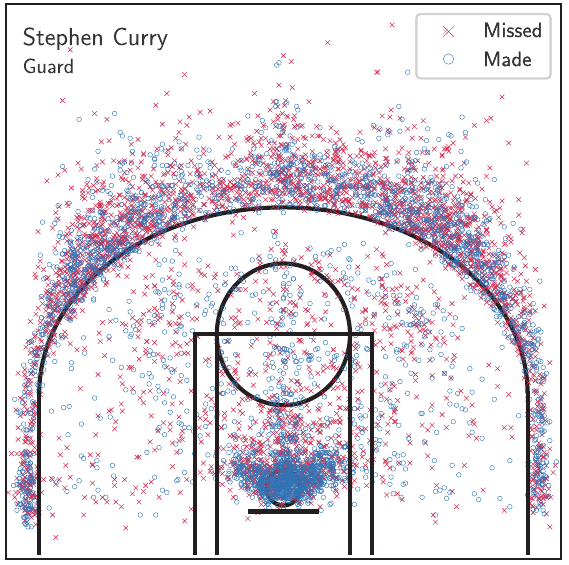}
	\includegraphics[scale=0.5]{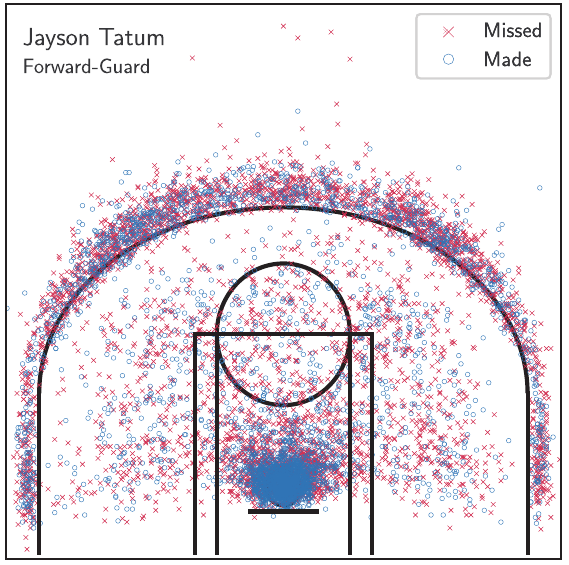}
	\includegraphics[scale=0.5]{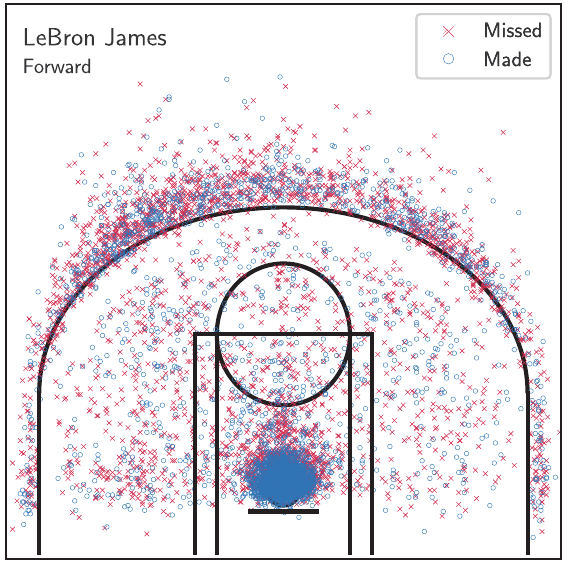}
	\includegraphics[scale=0.5]{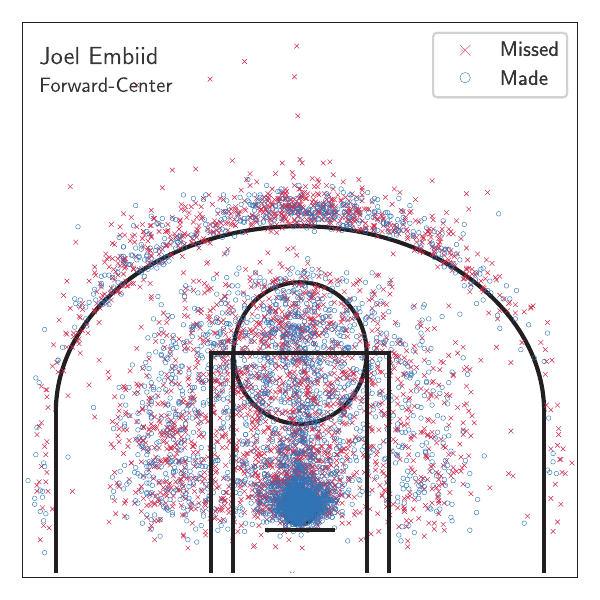}
	\includegraphics[scale=0.5]{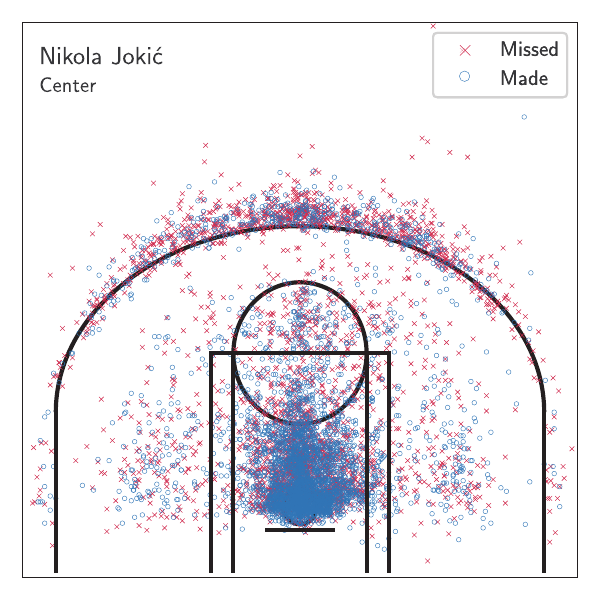}
	\caption{Shots charts for five selected NBA players.}
	\label{fig:examples_shooting_charts}
\end{figure}

We model both the locations of successful shots and unsuccessful shots for each player on the offensive half-court. Given that players can shoot from nearly any point within the well-defined boundaries of the half-court, the shot maps can be naturally represented as bivariate functional observations defined on two-dimensional rectangular grids. Figure~\ref{fig:examples_shooting_charts} displays raw shot charts for five players, each representing a different position: Stephen Curry (guard), Jayson Tatum (forward-guard), LeBron James (forward), Joel Embiid (forward-center) and Nikola Jokić (center). The code to reproduce this dataset and the analysis (including all the steps in the pre-processing pipeline) is available on \textsf{GitHub}\footnote{\url{https://github.com/StevenGolovkine/shooting_nba_fda/}}.


\section{Methodology} 
\label{sec:method}

\begin{figure}
	\centering
	\includegraphics[scale=0.35]{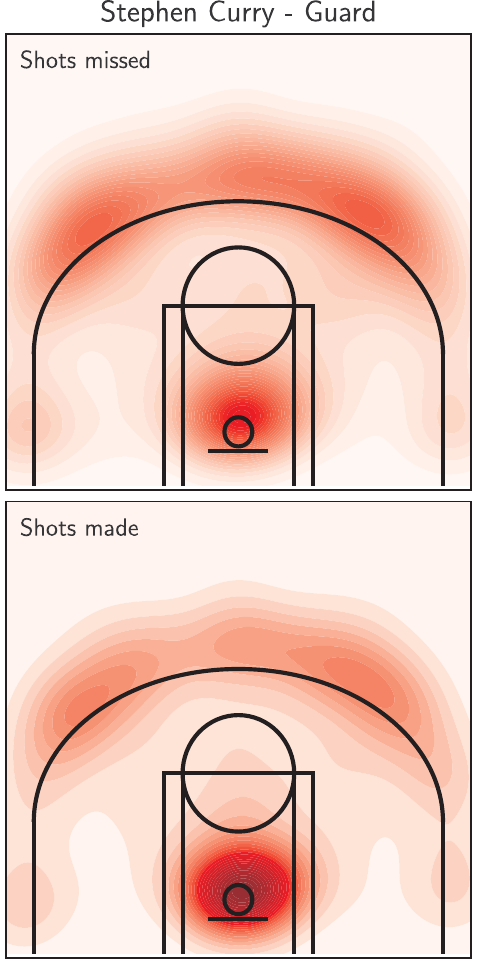}
	\includegraphics[scale=0.35]{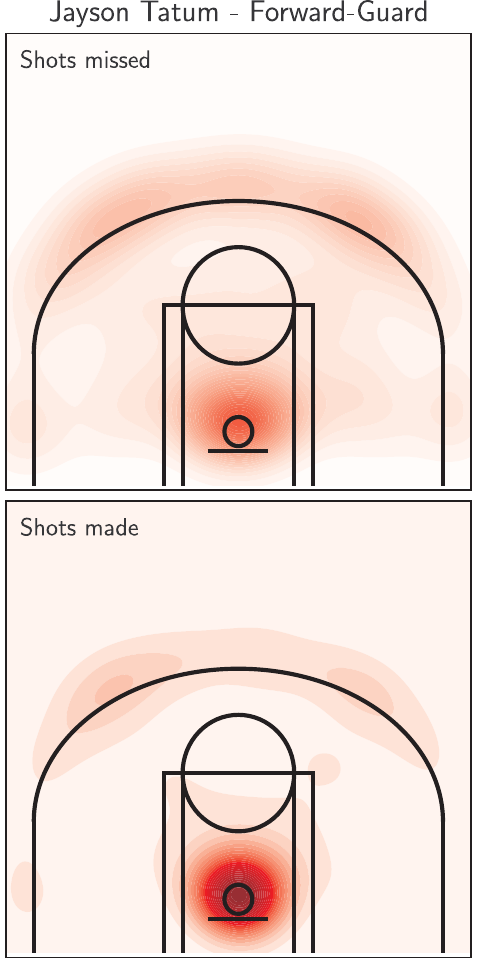}
	\includegraphics[scale=0.35]{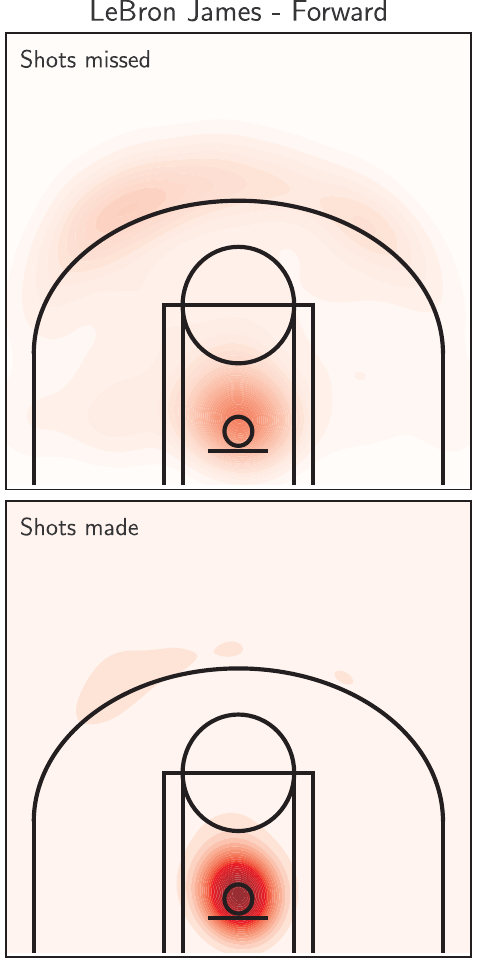}
	\includegraphics[scale=0.35]{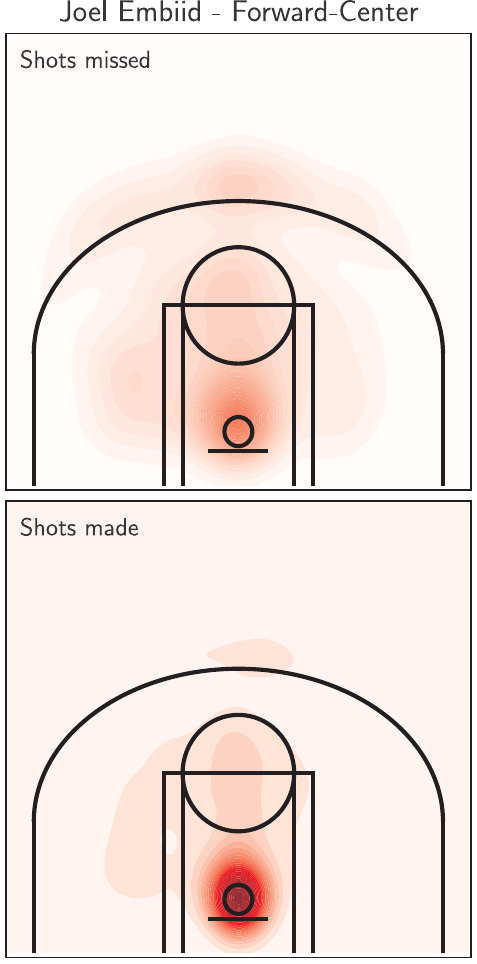}
	\includegraphics[scale=0.35]{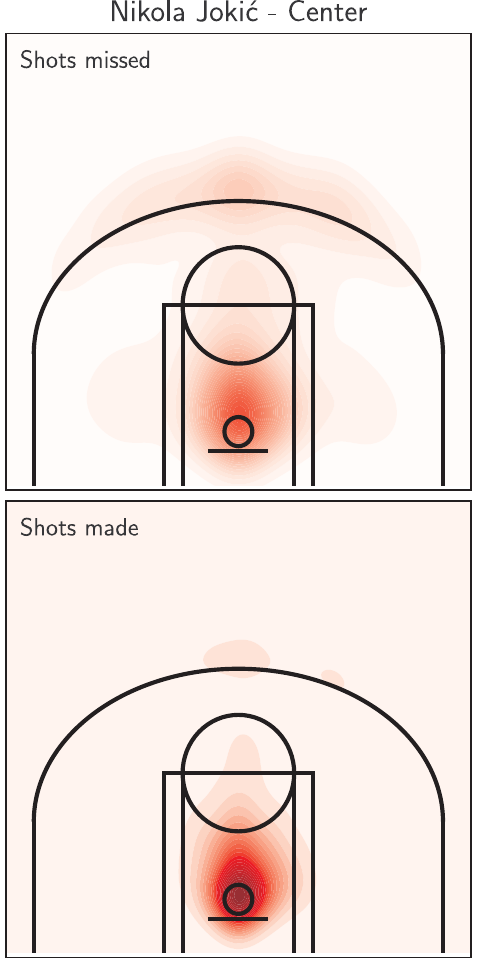}
	\\
	\includegraphics[scale=0.35]{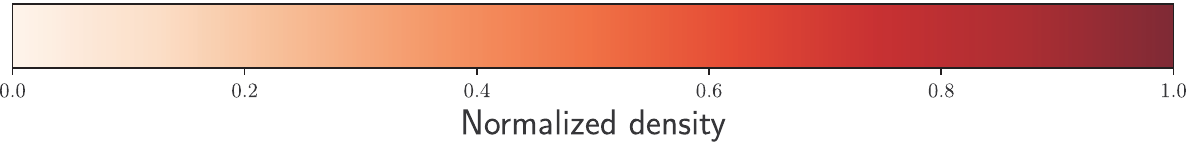}
  \caption{Smoothed shot chart densities for five selected NBA players. The densities have been normalized between $0$ and $1$. The normalization has been applied to the combined data across made and missed shots.}
	\label{fig:examples_shooting_density}
\end{figure}

The first step in our approach is to convert the raw shot charts into smooth functions suitable for analysis. 
To achieve this, we apply a two-dimensional kernel density estimator with a bivariate Gaussian kernel separately for each player and for each of missed and made shots.
After normalizing the court to the $[0, 1] \times [0, 1]$ range, density estimation is conducted on a regularly sampled grid of $201 \times 201$ points. 
Bandwidth selection is performed using Silverman's recommendation for $d$-dimensional data, also known as ``Silverman's multivariate factor'' \citep[pp.~86--87]{silvermanDensityEstimationStatistics1986}.
Silverman recommends a bandwidth factor of $A n^{-1 / (d + 4)}$ where $n$ is the number of observations and $$A = \left(\frac{4}{d+2}\right)^{1 / (d+4)}.$$
In our bivariate case, we have $d=2$ so that $A=1$, and our resulting bandwidth factor for every shot chart is $n^{-1 / 6}$, where $n$ is the number of shots the player has missed or made. The final bandwidth matrix is estimated by multiplying the bandwidth factor by the $2 \times 2$ sample covariance matrix of that player’s normalized shot coordinates.
We then represent the smoothed densities of missed and made shots for each player as the two features of a bivariate functional observation. Both features are defined on the same two-dimensional rectangular domain and observed over the $201 \times 201$ grid. Note that the choice of the grid resolution is arbitrary and can be adjusted as needed.
Sensitivity analyses to the choice of bandwidth and grid are contained in Appendices \ref{sec:bandwidth} and \ref{sec:grid}, respectively.
Figure~\ref{fig:examples_shooting_density} shows the density estimates of the same five players whose shot charts are presented in Figure~\ref{fig:examples_shooting_charts}.

After the density estimation, our data consist of independent trajectories of a vector-valued stochastic process $X = (X^{(1)}, X^{(2)})^{\top}$. Let $\mathcal{T} = [0, 1] \times [0, 1]$. The first coordinate $X^{(1)}$ corresponds to the density of missed shots and the second coordinate $X^{(2)}$ corresponds to the density of made shots. Both of them are assumed to belong to $\mathcal{L}^2(\mathcal{T})$, the Hilbert space of square-integrable real-valued functions defined on $\mathcal{T}$ having the usual inner product. Thus, $X$ is a stochastic process indexed by $\mathbf{t} = (t_1, t_2)$ belonging to the $2$-fold Cartesian product $\mathcal{T} \times \mathcal{T}$ and taking values in the $2$-fold Cartesian product space $\mathcal{H} = \mathcal{L}^2(\mathcal{T}) \times \mathcal{L}^2(\mathcal{T})$. In this space, we define the inner product as the sum of the inner products in $\mathcal{L}^2(\mathcal{T})$. Let $\mu: \mathcal{T} \times \mathcal{T} \rightarrow \mathcal{H}$ denote the mean function of the process $X$ and let $C$ denote the matrix-valued covariance function of the process $X$. We can then define the covariance operator of $X$, $\Gamma: \mathcal{H} \rightarrow \mathcal{H}$, as an integral operator with kernel $C$.
Assuming that the covariance operator $\Gamma$ is a compact positive operator on $\mathcal{H}$ and using the results in \cite{happMultivariateFunctionalPrincipal2018}, there exists a complete orthonormal basis, the eigenfunctions, $\{\phi_k\}_{k \geq 1} \subset \mathcal{H}$ associated to a set of real numbers, the eigenvalues, $\{\lambda_k\}_{k \geq 1}$ that satisfy
$$\Gamma \phi_k = \lambda_k \phi_k, \quad\text{and}\quad \lambda_k \rightarrow 0 \quad\text{as}\quad k \rightarrow \infty.$$
Using the multivariate Karhunen-Loève theorem \citep{happMultivariateFunctionalPrincipal2018}, we obtain the decomposition
\begin{equation}\label{eq:kl_decomp_inf}
	X(s, t) = \mu(s, t) + \sum_{k = 1}^{\infty} c_k\phi_k(s, t), \quad s, t \in [0, 1] \times [0, 1],
\end{equation}
where $c_k$ are the projections of the centered curves onto the eigenfunctions.

We apply multivariate functional principal components analysis  \citep{happMultivariateFunctionalPrincipal2018} to decompose each estimated density into a common set of basis functions and player-specific coefficients using~\eqref{eq:kl_decomp_inf}.
We aim to find basis functions $\{\phi_k\}_{1 \leq k \leq K}$ that provide the most parsimonious representation of $X$, for a given $K$, as
\begin{equation}\label{eq:kl_decomp}
	X(s, t) = \mu(s, t) + \sum_{k = 1}^K c_k\phi_k(s, t), \quad s, t \in [0, 1] \times [0, 1],
\end{equation}
where $\mu(s, t)$ is the mean (average) function, and $c_k$ are the player-specific basis coefficients, referred to as scores. Each eigenfunction in the decomposition~\eqref{eq:kl_decomp} characterizes a variation away from the mean function, with the player-specific scores indicating the magnitude and direction of the deviation for each individual player. We estimate four components ($K = 4$), based on the approach of \cite{golovkineUseGramMatrix2024}, explaining approximately $90\%$ of the total variance in the data. The scores $c_k$ are estimated as the projection of the observations $X$ onto the eigenfunctions $\phi_k$ using the inner product in $\mathcal{H}$. These scores provide a finite-dimensional representation of the functional data, capturing player-specific characteristics, and can be used as scalar variables in subsequent analyses.
To create a data-driven taxonomy of players based on MFPCA, we apply a $k$-medoids clustering algorithm \citep{kaufmanFindingGroupsData2009} to the normalized functional principal component scores $c_k$, using $k = 5$ clusters, as the number of players on the court.
	Moreover, we will compare the estimated clusters to the $k=5$ nominal positions provided by the NBA.
		We do not expect perfect agreement -- defence, play-making and movement all play a role in addition to shooting.
		However, because shooting behavior is a key component of a player's overall game, the positions comparison can help inform the labels we provide to our clusters.
The $k$-medoids algorithm minimizes the dissimilarity between each player's scores and the medoid of its cluster. We use the weighted Euclidean distance as the measure of dissimilarity. This method offers a practical advantage over the $k$-means clustering algorithm: the center of each cluster is an actual player from the dataset rather than an averaged representation of all players within the cluster, enhancing interpretability. In our analysis, we consider two weighting schemes to calculate the distance: one assigns equal weights to all components, and the other weights components given the proportion of variance they explain.
After examining the cluster solution with $k=5$ and comparing it to the NBA-defined positions, we also employ a fully data-driven procedure where $k$ is selected based on the data---choosing $k \in \{2, \ldots, 10\}$ that maximizes the average Silhouette coefficient.

\section{Results} 
\label{sec:results}

\subsection{MFPCA decomposition of the densities} 
\label{sub:decomposition_of_the_densities}

\begin{figure}
	\centering
	\includegraphics[scale=0.35]{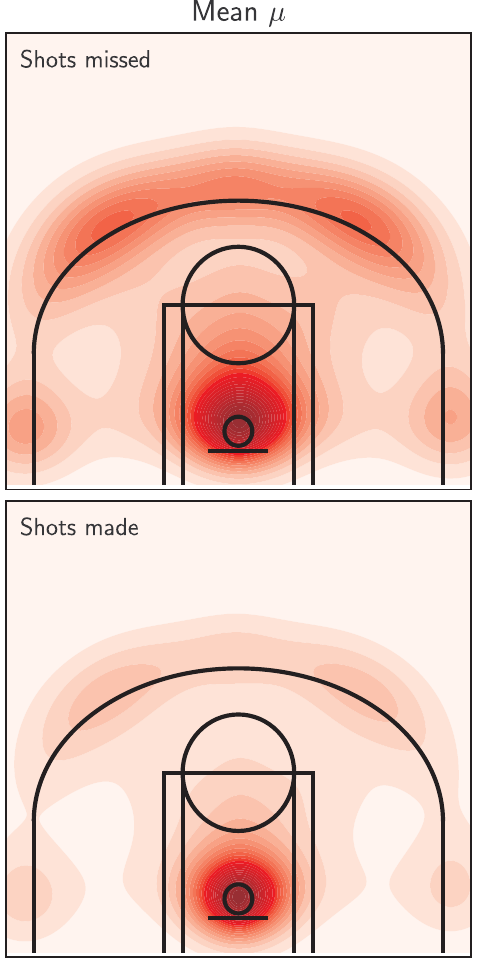}
	\includegraphics[scale=0.35]{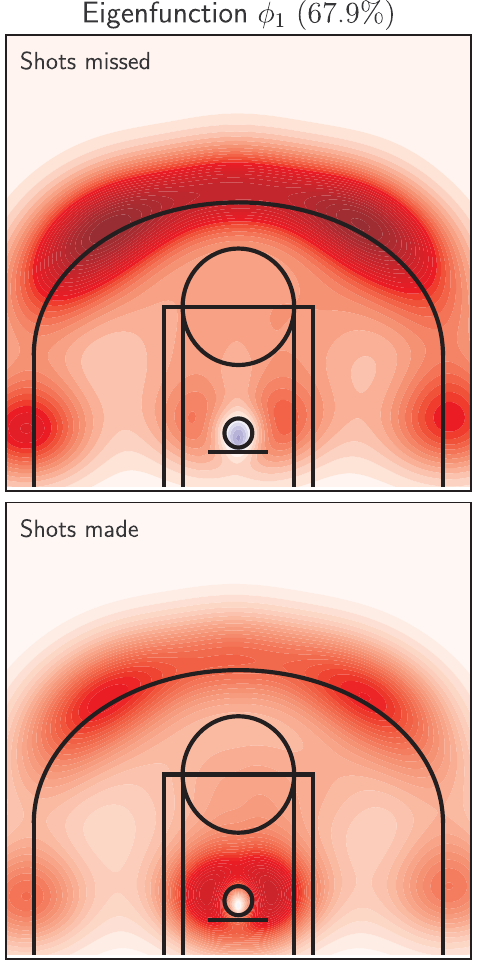}
	\includegraphics[scale=0.35]{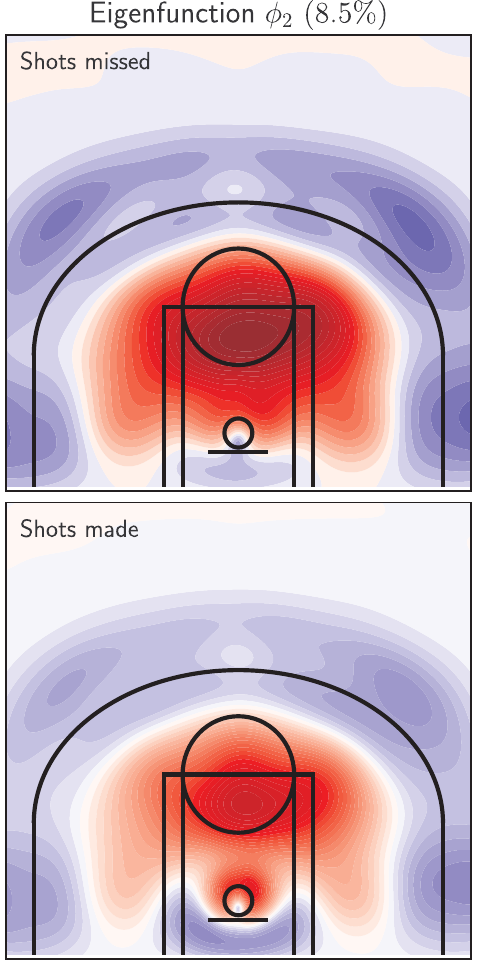}
	\includegraphics[scale=0.35]{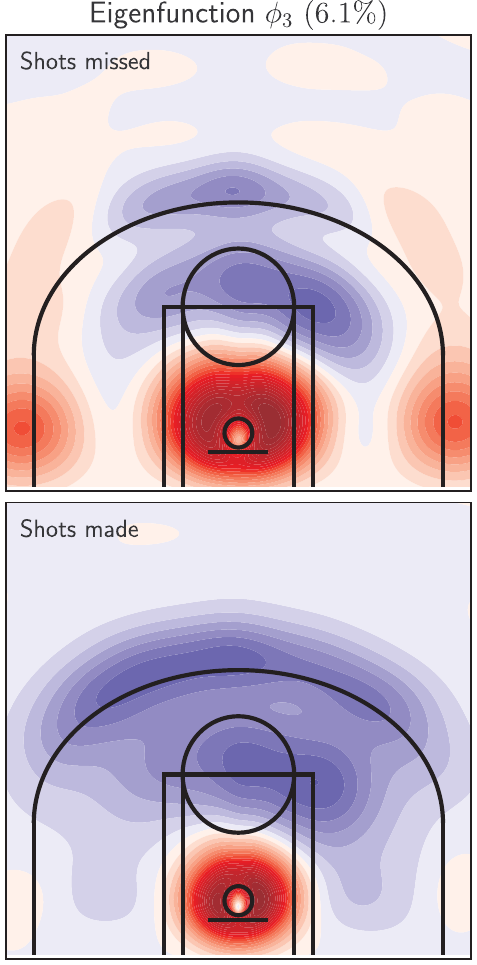}
	\includegraphics[scale=0.35]{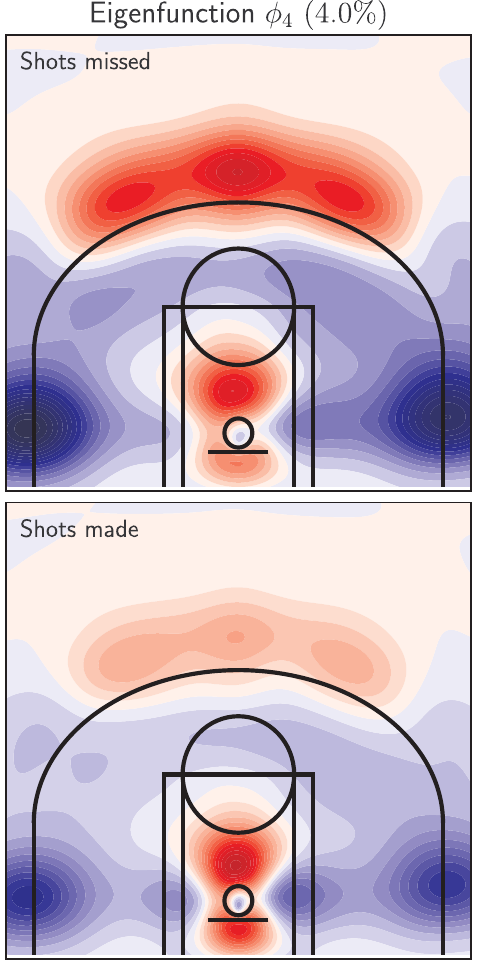}
	\\
	\includegraphics[scale=0.35]{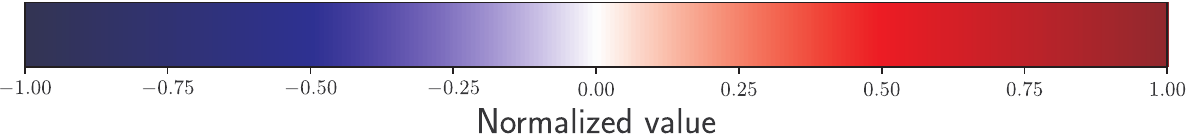}
  \caption{Estimates of the mean surface and the eigenfunctions of the shots density charts. The eigenfunctions represent deviations from the mean surface and are defined up to a sign, making their values at specific points not directly interpretable as densities. Note that the mean surface and eigenfunctions have been normalized to be between $-1$ and $1$ for presentation purposes. The normalization is applied to the combined functions across made and missed shots.}
	\label{fig:mean_eigenfunctions}
\end{figure}

We apply MFPCA to estimate the mean functions and eigenfunctions, providing a description of players' shooting behavior.
Figure~\ref{fig:mean_eigenfunctions} presents estimates of the mean functions alongside the estimated functional principal components, which capture deviations from the mean shooting density.
Unlike the shooting densities, the functional principal components may take on negative values as they represent deviations from the mean rather than densities themselves. For presentation purposes, the functional principal components have been normalized between $-1$ and $1$. Blue areas correspond to negative values of the functional components, and represent negative deviations from the mean density, while red areas correspond to positive values of the functional components, and thus positive deviations.
Because the individual densities have been normalized separately for each outcome (made and missed) and player, the patterns captured by the mean and functional principal components should be interpreted as reflecting where players tend to take made and missed shots, not where shots are more or less likely to be made (i.e., success probbability or efficiency). Table~\ref{tab:eigenfunction_contribution} presents the individual contribution of the univariate eigenfunctions to the variability explained by each eigenfunction.

	The mean chart represents the average shooting behavior of NBA players and is displayed in the first column of Figure \ref{fig:mean_eigenfunctions}.
		For shots made (second row, first column), the mean density displays very high values in the small region under the paint.
		This indicates that a large proportion of the shots that players make successfully are taken from this area.
		The next highest density region in the mean shots made chart is around the perimeter of the three-point line, but this density is lower than under the paint.
		For shots missed (first row, first column), there is also a high density region under the paint, indicating that a large number of the missed shots are also taken from this area.
		However, in comparison with the shots made density, the next highest density region -- around the perimeter of the three-point line -- is larger and closer in density values to the area under the paint.
		Therefore, we can deduce that although a large number of both made shots and missed shots are taken from under the paint and around the three point area, a comparatively larger proportion of \emph{missed} shots are taken from the three point line.
		This difference is intuitive, because three-point shots are inherently harder to make. Overall, the elbows and the short-corners are less important as shooting spots because these regions account for a smaller number of both missed and made shots.

The estimated functional components can be understood as characterizing different shooting behaviors. We first note that, except for $\phi_3$, the functional components for missed and made shots are similar.
	This suggests that the dominant patterns of player-to-player variation are driven by shot location, regardless of whether the shots are made or missed.
	The first component explains $67.9\%$ of the total variance and is similar in shape to the mean densities because it assigns higher positive density values to the areas under the paint and around the outside of the perimeter arc.
		Although the average player tends to have a large proportion of their made and missed shots from this area, players with a positive score $c_1$ tend to make and miss even more of their shots than average from these regions.
        In contrast, players with a negative score $c_1$ have a smaller-than-average proportion of their made and missed shots taken from these ``typical" shooting regions.
		In addition, for the missed shots, there is a small, compact blue area, appearing as a small dot, in the location directly under the basket.
		This means that players with a positive score $c_1$ have fewer of their missed shots from this location.
The second component, explaining $8.5\%$ of the total variance, highlights a contrast between shooting inside and outside the three-point line.
Players with a positive score $c_2$ for $\phi_2$ prefer to shoot closer to the basket, while those with negative scores prefer long-range shots.
	The third component (explaining $6.1\%$ of the total variance) contrasts players that shoot in the paint and the others, while also capturing some variation in missed shots from the corners.
		For made shots, positive scores $c_3$ are associated with frequent attempts and successes in the paint and fewer made shots from further out.
		For missed shots, positive scores $c_3$ also indicate frequent attempts and misses from around the paint, but interestingly also capture high proportions of missed shots from the corners, indicated by the two red regions on either side.
		This pattern indicates that these players who tend to take a large number of their shots (both made and missed) from inside the paint also tend to take shots from the corners but are generally unsuccessful (i.e., we see the high densities in the corners for the missed shots but not the made shots).
	Finally, the fourth component accounts for $4.0\%$ of the total variance and characterizes differences in shooting centrally (in the paint and along the central portion of the three-point arc) and shooting from the wings.
		Players with a positive (negative) score $c_4$ tend to make and miss more of their shots from central regions of the court, and fewer of their shots from the wings.
\begin{table}
	\centering
  \caption{Individual contribution (in \%) to the variability explained by each eigenfunction.}
	\label{tab:eigenfunction_contribution}
	\begin{tabular}{ c|cccc }
		Eigenfunction $k$ & $1$ & $2$ & $3$ & $4$\\
		\hline
    $\| \phi_{k, \text{missed}}\|^2$ & $70$ & $72$ & $55$ & $75$ \\
    $\| \phi_{k, \text{made}}\|^2$ & $30$ & $28$ & $45$ & $25$ \\
	\end{tabular}
\end{table}

\begin{figure}
	\centering
	\includegraphics[scale=0.295]{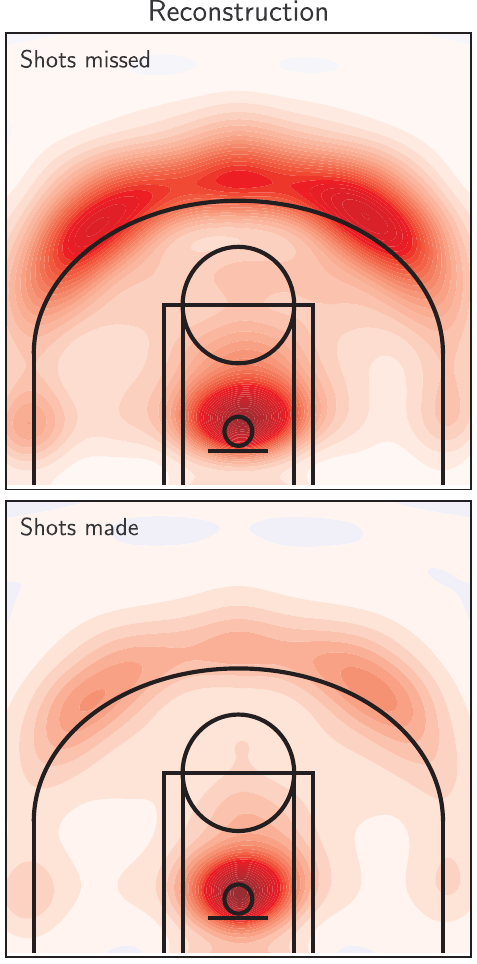}
	\includegraphics[scale=0.295]{figures/mean_normalized.pdf}
	\includegraphics[scale=0.295]{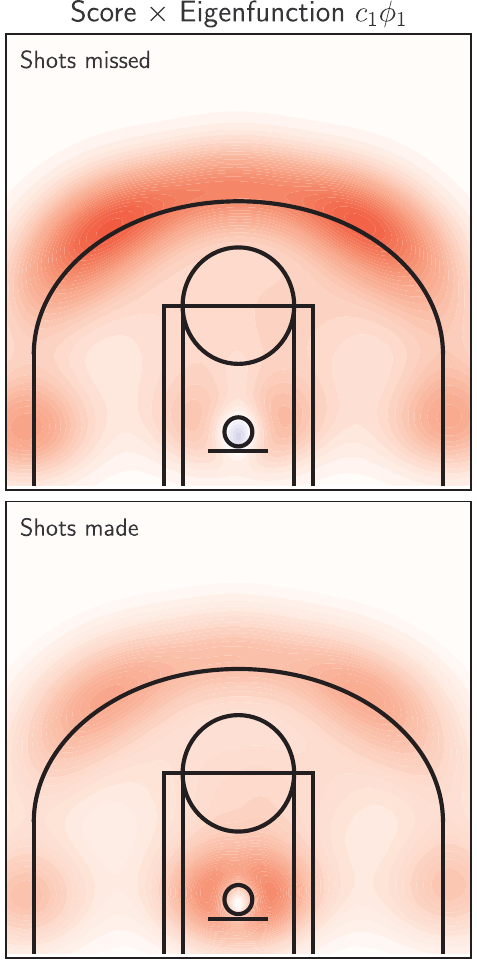}
	\includegraphics[scale=0.295]{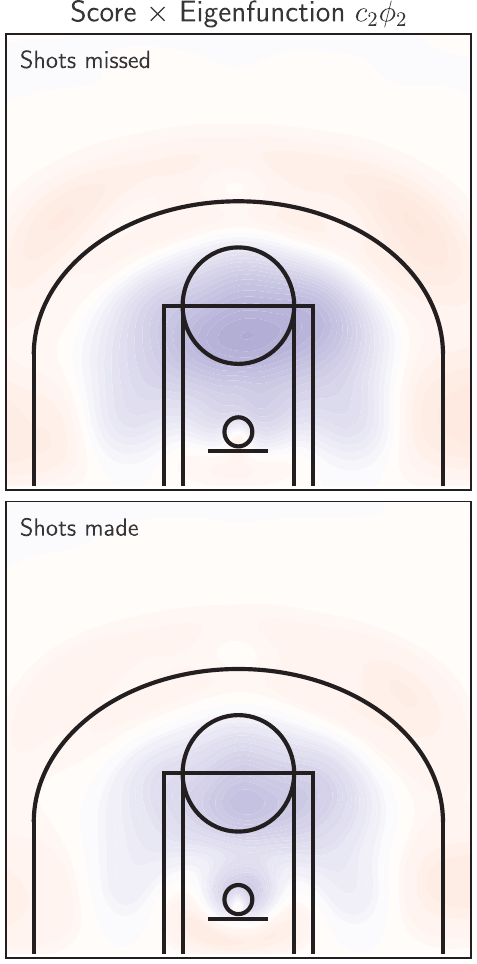}
	\includegraphics[scale=0.295]{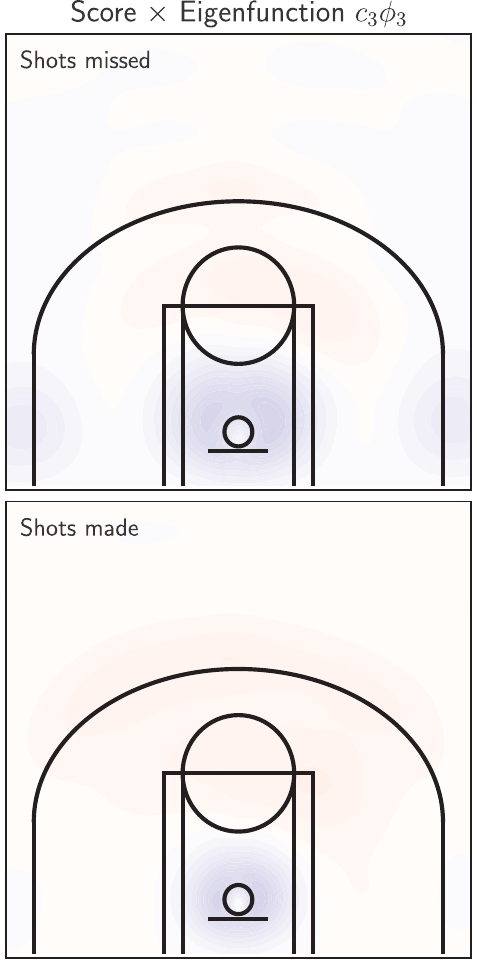}
	\includegraphics[scale=0.295]{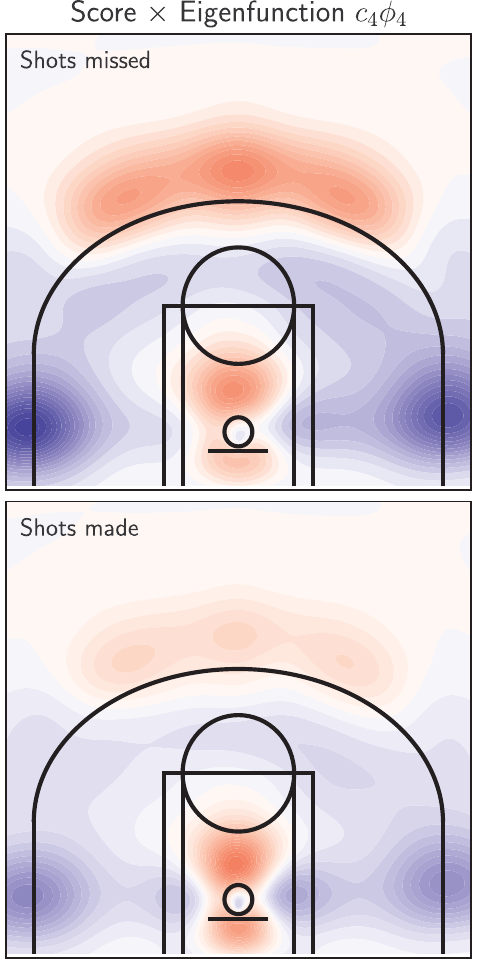}
	\\
	\includegraphics[scale=0.35]{figures/colorbar2.pdf}
  \caption{Decomposition of Stephen Curry's shot chart densities into the eigenfunctions basis. The normalization is applied to the combined functions across made and missed shots.}
	\label{fig:examples_players}
\end{figure}

As an example, Figure~\ref{fig:examples_players} illustrates the decomposition of Stephen Curry's density shot chart. Curry shoots more frequently than the average player, as indicated by his positive coefficient $c_1$, reflecting a higher number of both missed and made shots compared to the average player. His second coefficient $c_2$ is negative and close to zero, suggesting only a slight preference for shooting beyond the three-point line. The small negative value of his third coefficient $c_3$ indicates shooting behavior close to the average for this component. However, Curry prefers to shoot in the axis of the basket (fourth coefficient $c_4$ is positive) compared to the other players.

\begin{figure}
	\centering
	\includegraphics[scale=0.95]{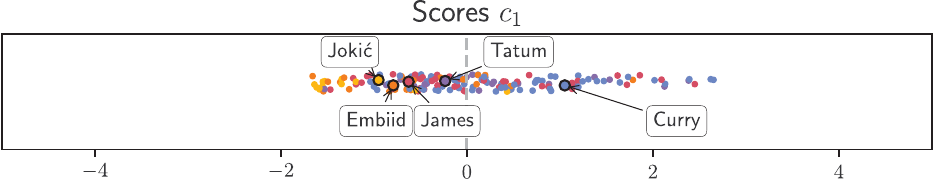}
	\includegraphics[scale=0.95]{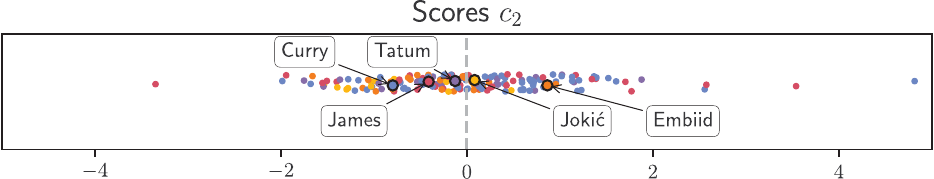}
	\includegraphics[scale=0.95]{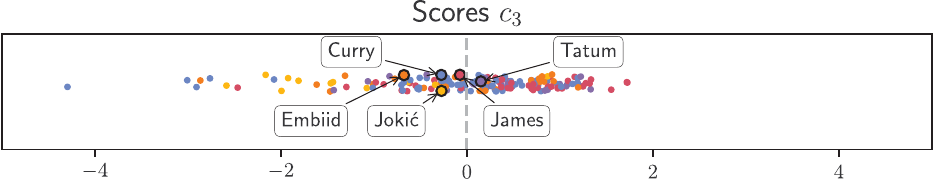}
	\includegraphics[scale=0.95]{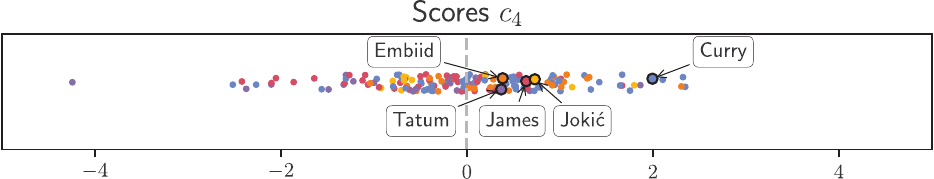}
	\includegraphics[scale=0.95]{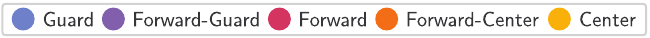}
	\caption{Scatter plots of the player-specific basis coefficients (scores) colored using the NBA defined typology of players. We highlighted one player per position. The scores have been standardized and jittered on the y-axis for visualization purposes. The first component discriminates players by their overall shooting volume, with positive scores indicating a higher-than-average number of shots. The second component contrasts between two-pointers and three-pointers, with positive scores associated with shots near the basket. The third component differentiates players that shots in the paint, combined with frequent missed shots from the corners and associated to positive scores, and the others. The fourth component contrasts between centrally located shooting and wing-area shooting, with positive scores indicating more central shooting.}
	\label{fig:clusters}
\end{figure}

Figure~\ref{fig:clusters} presents the estimated scores $c_k$ from~\eqref{eq:kl_decomp} colored by the NBA-defined playing positions. One representative player from each position is highlighted (same as Figures~\ref{fig:examples_shooting_charts} and~\ref{fig:examples_players}). Notably, the scores do not separate players by position. From Figure~\ref{fig:clusters}, it can be seen that some players exhibit scores (coefficients) that substantially differ from the general distribution. For example, for the second eigenfunction, Chris Paul (with a large coefficient) and Duncan Robinson (with a large negative coefficient) may be designated as outliers. Similarly, for the third eigenfunction, Chris Paul has a large negative coefficient. Although a positive $c_2$ typically indicates a preference for shooting near the basket and a negative $c_3$ suggests a preference for shooting beyond the three-point line, Chris Paul's scoring pattern appears contradictory. This inconsistency arises because Chris Paul's most common shooting position is the high post, located between the three-point line and the painted area. In this case, the eigenfunctions compensate for one another to capture his shooting style, which is relatively rare in the modern NBA. Finally, for the fourth eigenfunction, Reggie Bullock Jr. has a large negative coefficient, indicating a strong preference for shooting from the wings and corners.


\subsection{Clustering} 
\label{sub:clustering}

\subsubsection{Comparison with the five NBA positions}

The $k$-medoids algorithm was applied to the coefficients plotted in Figure~\ref{fig:clusters}, grouping players into five clusters using a Euclidean distance and different weighting schemes for the components. The choice of five clusters aligns with the number of players on the court for each team. Detailed clustering results, including player names and decomposition in eigenfunctions basis, are provided in Appendix~\ref{sec:clustering_results}. Figure~\ref{fig:average_cluster_no_weight} presents the medoids for each cluster when the components are weighted equally. The medoid, as the central observation in a cluster, represents the most representative player within that cluster. The decompositions of these representative players are given in Figure~\ref{fig:cluster_no_weight} in the Appendix. The clustering results can be interpreted based on shooting tendencies. Cluster $1$ groups high-volume perimeter shooters. Cluster $2$ includes players who primarily shoot from the center axis of the basket (represented by ), while players in Cluster $3$ predominantly score from the corners. Clusters $4$ and $5$ are similar, comprising players who primarily score in the paint but also attempt shots from beyond the three-point line, albeit with limited success.

\begin{figure}
	\centering
  \includegraphics[scale=0.35]{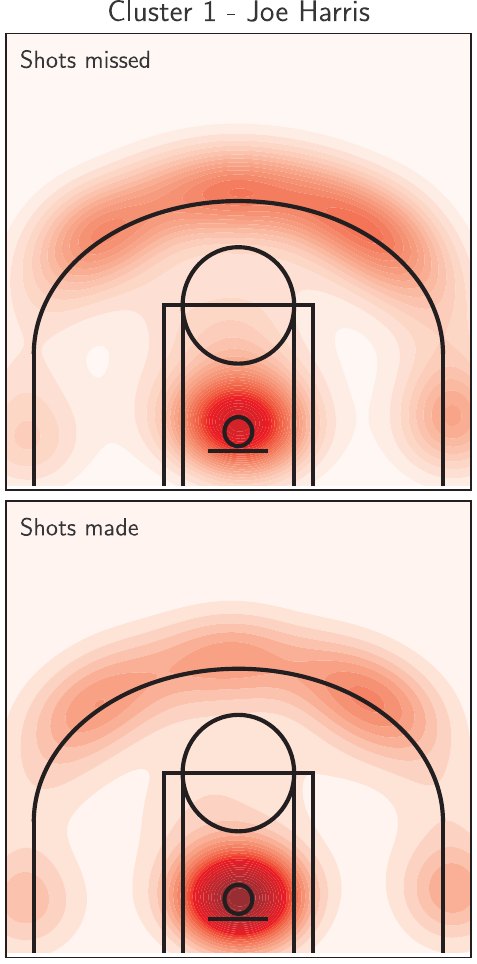}
	\includegraphics[scale=0.35]{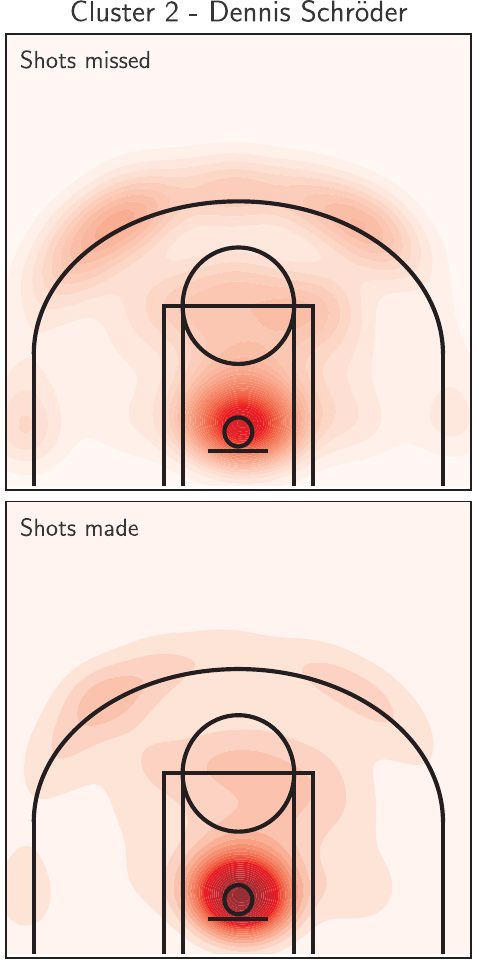}
	\includegraphics[scale=0.35]{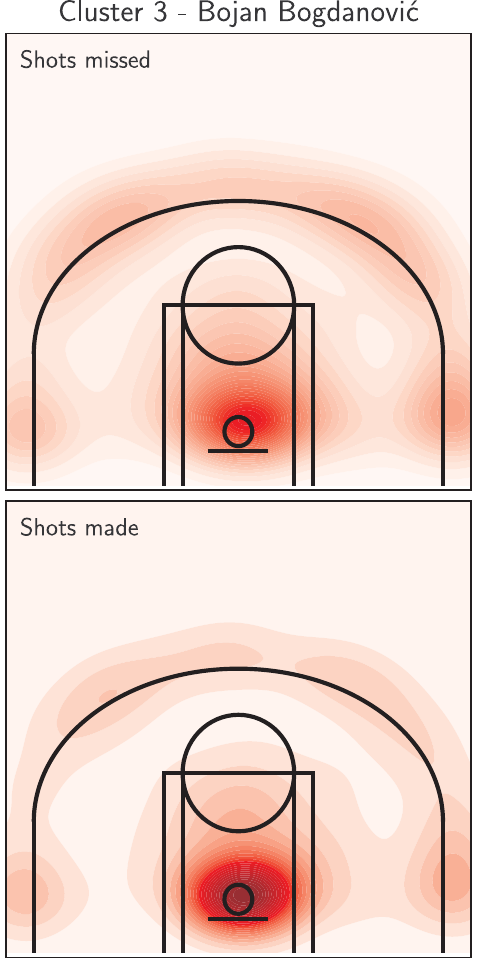}
	\includegraphics[scale=0.35]{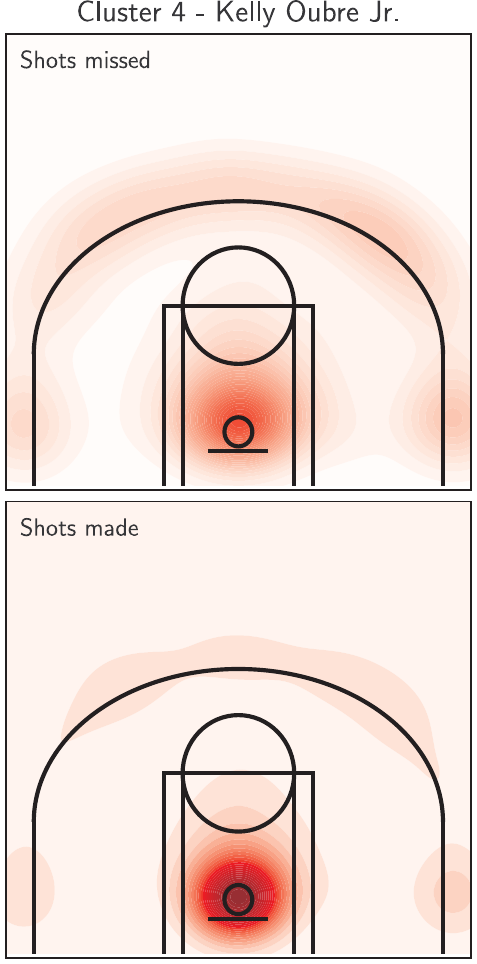}
	\includegraphics[scale=0.35]{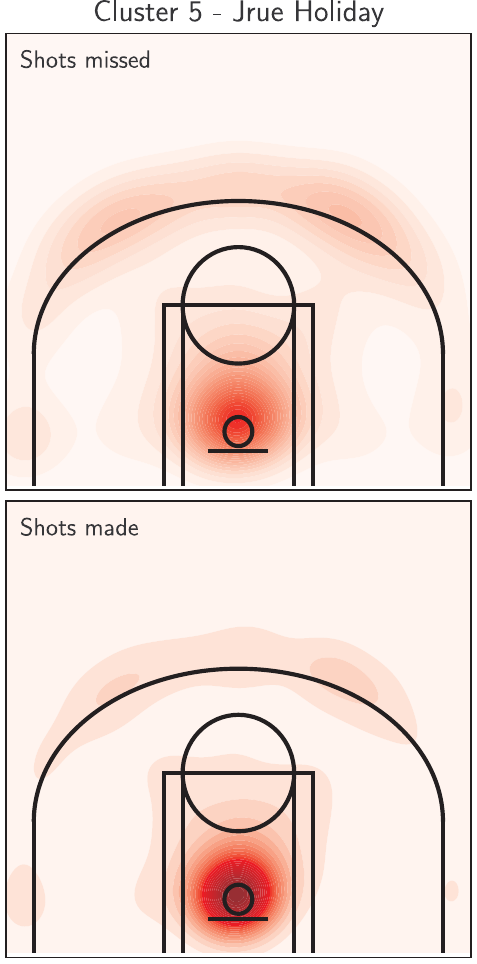}
	\caption{Most representative player of each cluster according to the $k$-medoids algorithm using a Euclidean distance with equal weights for each component. The normalization is applied to the combined functions across made and missed shots.}
	\label{fig:average_cluster_no_weight}
\end{figure}

\begin{figure}
	\centering
	\includegraphics[scale=0.35]{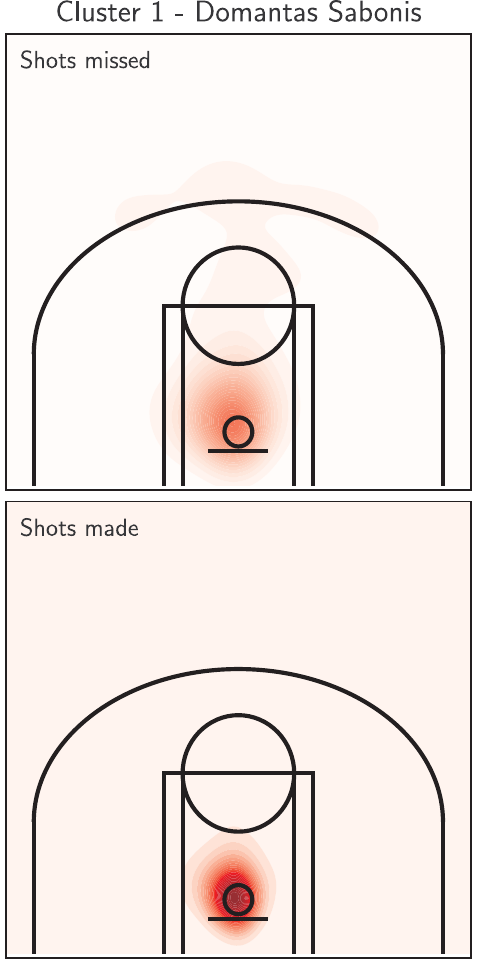}
	\includegraphics[scale=0.35]{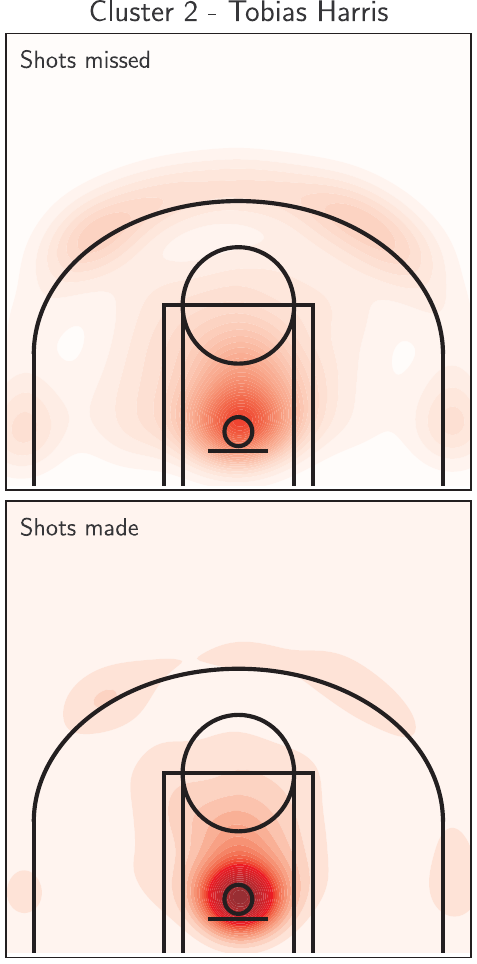}
	\includegraphics[scale=0.35]{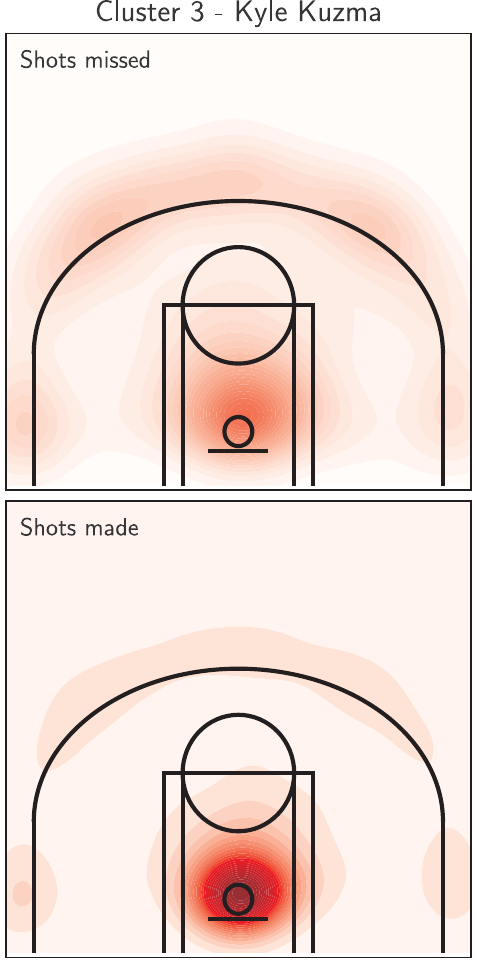}
	\includegraphics[scale=0.35]{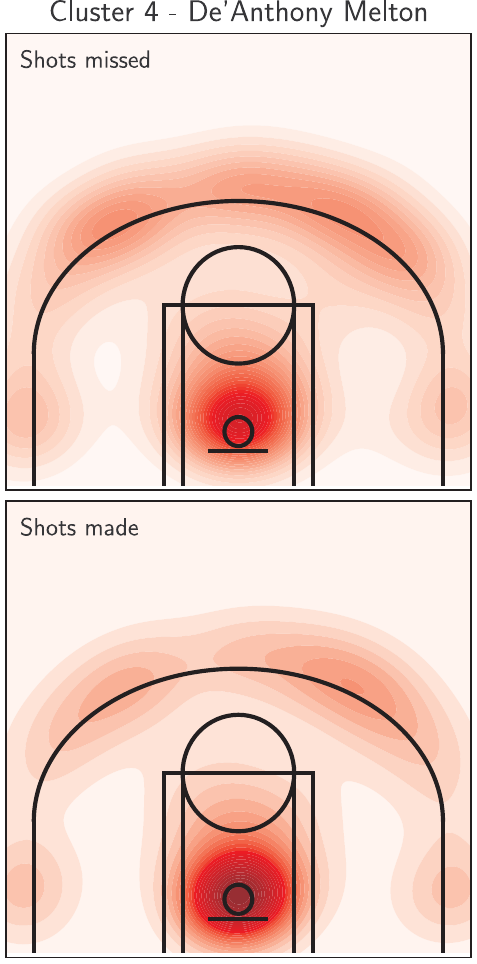}
	\includegraphics[scale=0.35]{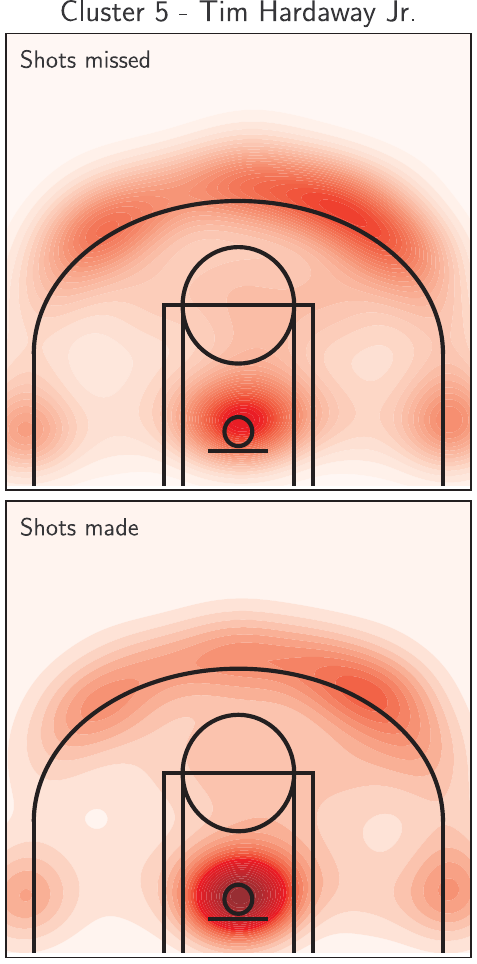}
	\caption{Most representative player from each cluster according to the $k$-medoids algorithm using a Euclidean distance with weights defined using the percentage of variance explained by each component. The normalization is applied to the combined functions across made and missed shots.}
	\label{fig:average_cluster_weight}
\end{figure}

Figure~\ref{fig:average_cluster_weight} presents the cluster medoids for the results obtained with components weighted using the percentage of variance they explain. Since the first component accounts for approximately $70\%$ of the total variance, the clusters are primarily driven by the first score $c_1$, reflecting the overall shooting frequency of the players. The clusters span a spectrum of shooting behaviors, from players who predominantly score in the paint (Cluster $3$) to those who are prolific three-point shooters (Cluster $4$). This effect is also evident in the decomposition shown in Figure~\ref{fig:cluster_weight} in the Appendix. The clusters can be ranked relative to the average player’s shooting tendencies. Compared to the average player, Cluster $1$ players shoot more frequently in the paint and attempt fewer three-pointers. Players in Cluster $2$ shoot slightly less than the average player, while those in Cluster $3$ align  with the average. Cluster $4$ comprises players who shoot slightly more than average, and Cluster $5$ players favor three-pointers significantly more.

\begin{figure}
	\centering
	\includegraphics[scale=0.475]{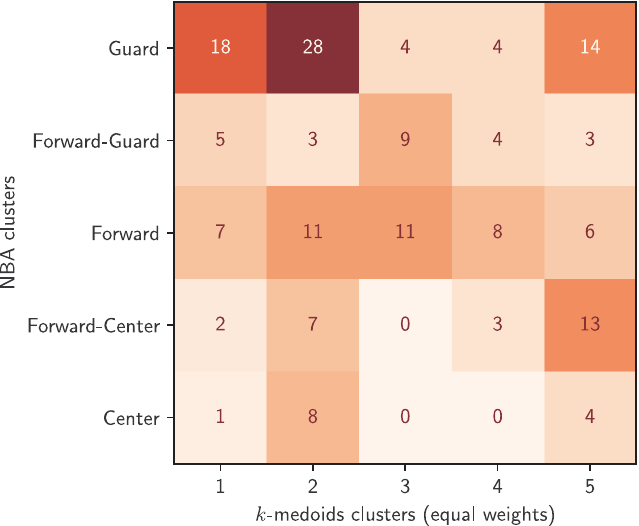}
	\hfill
	\includegraphics[scale=0.475]{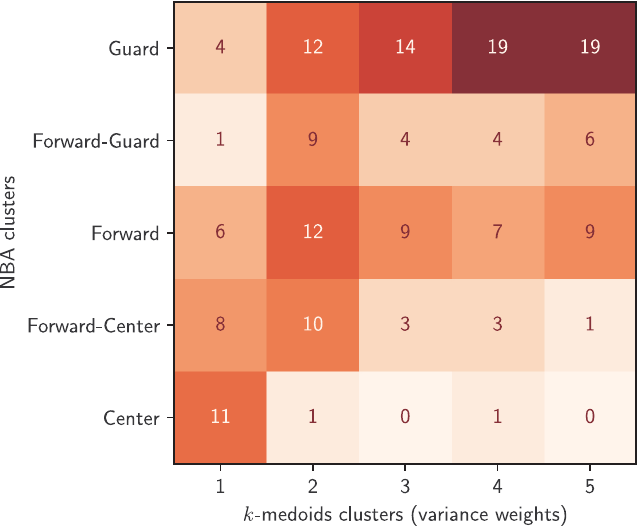}
	\hfill
	\includegraphics[scale=0.475]{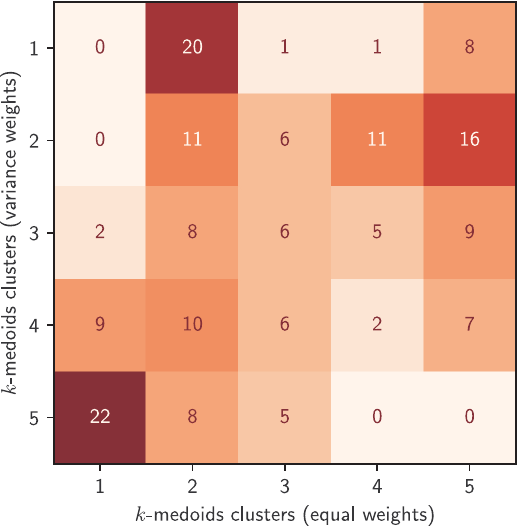}
	\caption{Agreement between the two clustering taxonomies (using equal weights and variance weights) and the conventional NBA position typology.}
	\label{fig:confusion_mat}
\end{figure}

Figure~\ref{fig:confusion_mat} presents the confusion matrices between the NBA-defined typology and the two clustering approaches we present. The NBA clusters are based on the usual player classification. Overall, there is minimal agreement between the NBA clusters and the $k$-medoids clusters, whether weights are equal or based on the percentage of variance explained. The Adjusted Rand Index (ARI, \cite{hubertComparingPartitions1985}) between the NBA clusters and the $k$-medoids clusters with equal weights is $0.05$ and between the NBA clusters and the $k$-medoids clusters with variance weights is $0.04$. This is expected as NBA clusters are declarative and not data-based. The only exception is the NBA-defined `center' category, which aligns with the cluster $1$ in the $k$-medoids results when weighted by the variance explained. Cluster $1$ of the weighted $k$-medoids contains players that shoot mostly under the basket, which is coherent with the usual scoring position of centers. Furthermore, there is little agreement between the two $k$-medoids clustering methods themselves. The ARI is equal to $0.1$. The Silhouette coefficient \citep{rousseeuwSilhouettesGraphicalAid1987} for the $k$-medoids with equal weights is equal to $0.04$, while the Silhouette coefficient for the $k$-medoids with variance weights is equal to $0.4$.
The Silhouette coefficient for the equal-weighted $k$-medoids partition indicates that the five groups are not clearly separated in the MFPCA score space. Most of the players lie close to the boundary between clusters, and their assignments should therefore be interpreted as descriptive summaries rather than as evidence for five naturally isolated shooting types. The one obtained with variance weights suggests a clearer partition, but this separation is driven mainly by the dominant components and therefore gives less emphasis to subtler spatial contrasts captured by the other components.
We also computed the Silhouette coefficient of the NBA clustering using the distance matrix used for the $k$-medoids with equal weights ($-0.05$) and using the distance matrix used for the $k$-medoids with variance weights ($-0.1$).
The negative Silhouette coefficients for the NBA position labels reinforce the same conclusion. The conventional positions do not form compact or well-separated groups when evaluated using shooting-pattern distances.

\subsubsection{Data-driven clustering}

\begin{figure}
	\centering
	\includegraphics[scale=0.5]{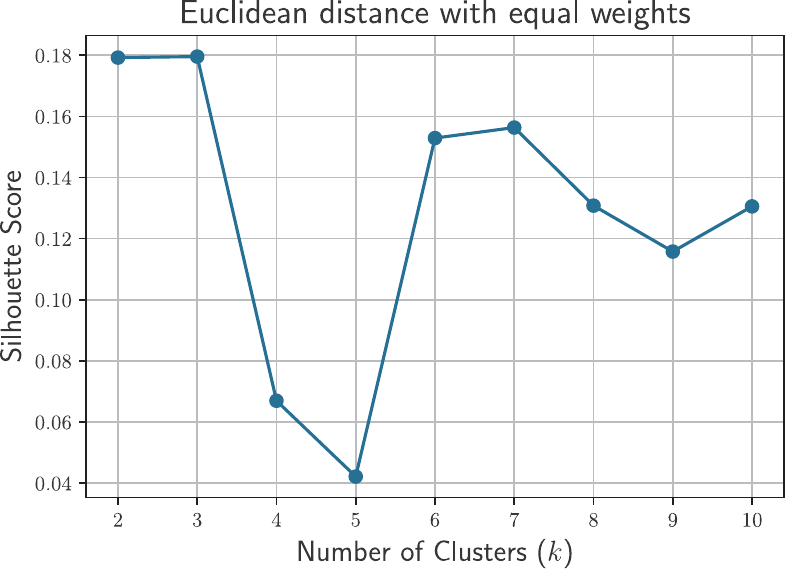}
	\hfill
	\includegraphics[scale=0.5]{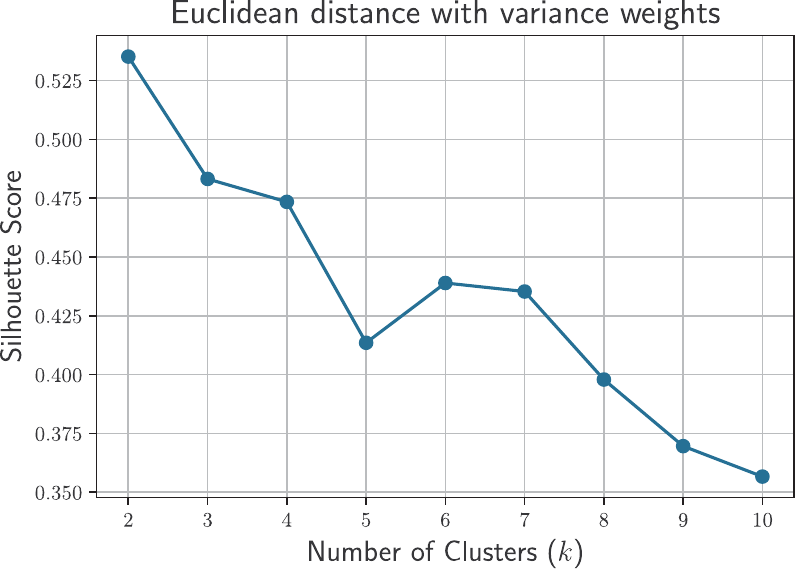}
	\hfill
	\caption{Silhouette coefficients for the $k$-medoids clustering with equal weights (left) and variance weights (right), computed for $k = 2, \ldots, 10$.}
	\label{fig:silhouette_plots}
\end{figure}

The previous clustering analysis fixed $k = 5$ in order to compare the empirical taxonomy with the five conventional NBA position groups. We also considered a fully data-driven choice of $k$ by computing the average Silhouette coefficient for $k = 2, \ldots, 10$, using the same $k$-medoids algorithm and the same distance matrices as above. This criterion selects the number of clusters that gives the clearest separation in the MFPCA score space, without imposing the basketball-motivated choice $k = 5$. The results are shown in Figure~\ref{fig:silhouette_plots}.

With equal weights for the four MFPCA scores, the maximum Silhouette coefficient is $0.18$ and is obtained for $k = 3$. The corresponding three clusters contain $41$, $57$ and $75$ players, with Norman Powell, Coby White and Dennis Schröder as medoids respectively. These groups reflect combinations of the secondary shooting-pattern contrasts represented by the later components, such as central versus wing shooting and paint versus perimeter tendencies.

\begin{figure}
  \centering
  \includegraphics[scale=0.35]{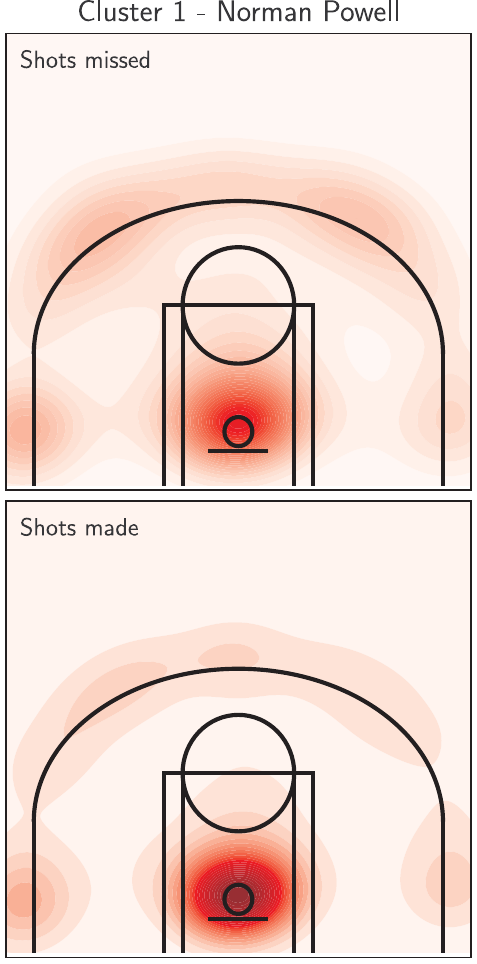}
  \includegraphics[scale=0.35]{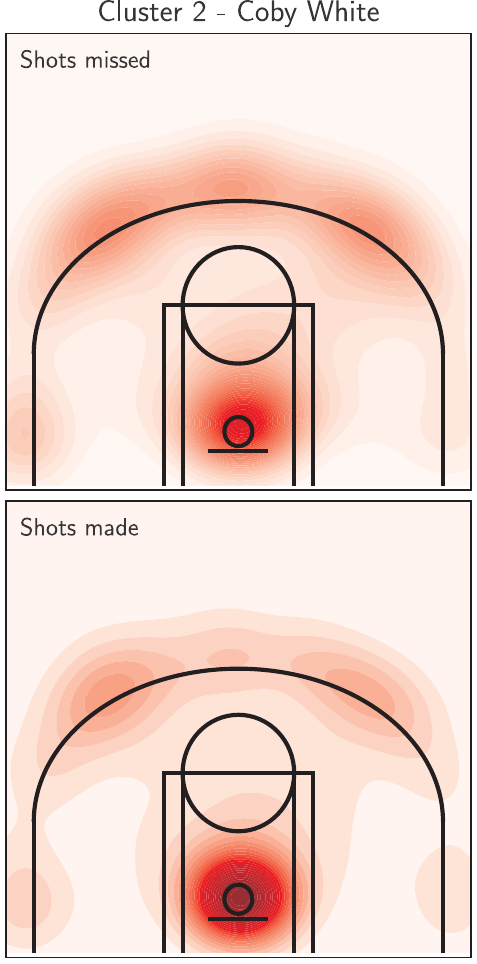}
  \includegraphics[scale=0.35]{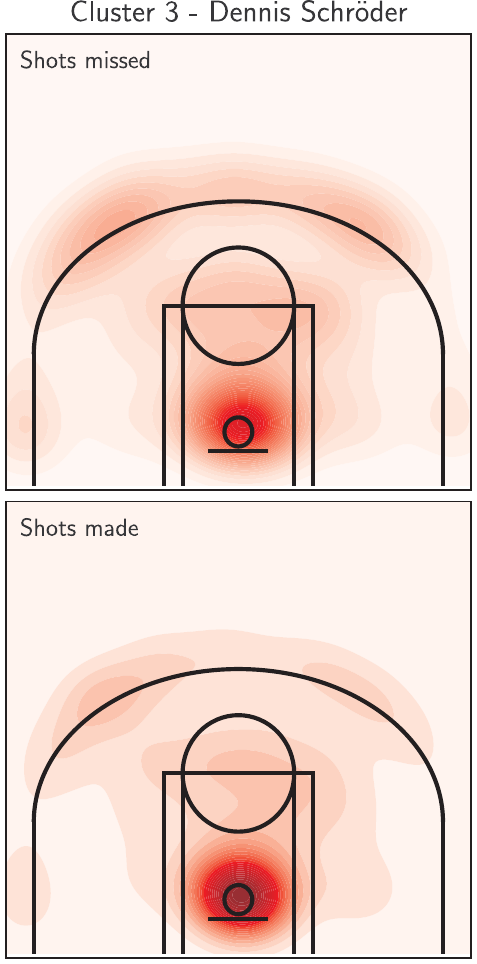}
  \caption{Most representative player from each cluster according to the $k$-medoids algorithm using a Euclidean distance with equal weights for each component and the number of clusters selected as the maximiser of the Silhouette coefficients. The normalization is applied to the combined functions across made and missed shots.}
  \label{fig:clustering_k3}
\end{figure}

In contrast, when the scores are weighted by the proportion of variance explained, the maximum Silhouette coefficient is $0.54$ and is obtained for $k = 2$. The two clusters contain $99$ and $74$ players, with Julius Randle and Jordan Poole as medoids respectively. Because the first component explains most of the variation, the variance-weighted data-driven clustering is mainly a split along the first score.

\begin{figure}
  \centering
  \includegraphics[scale=0.35]{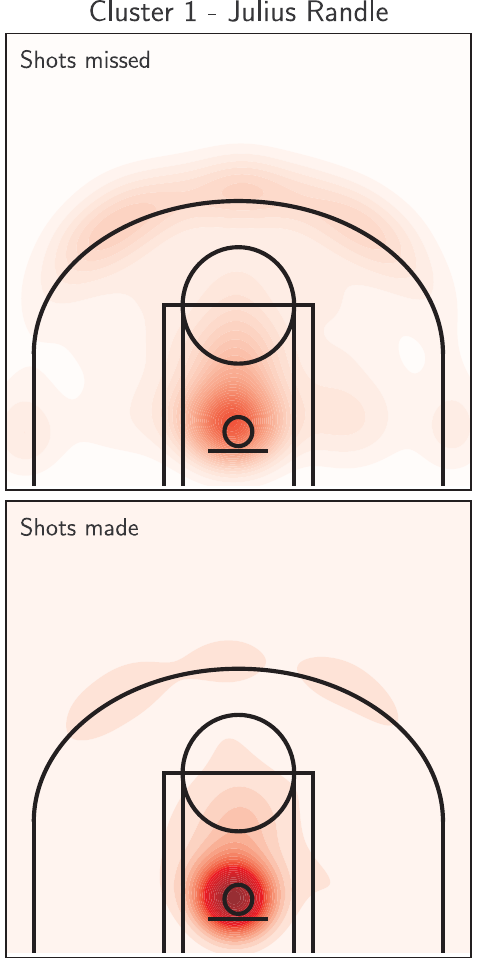}
  \includegraphics[scale=0.35]{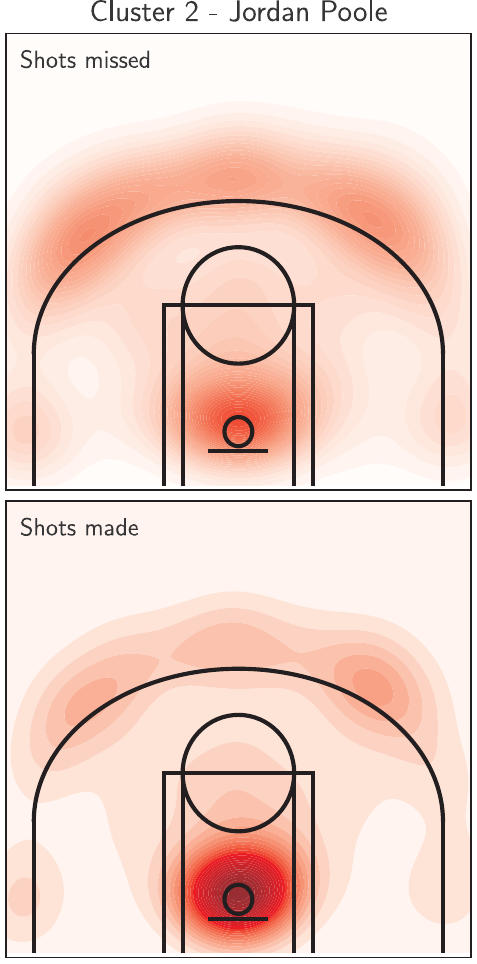}
  \caption{Most representative player from each cluster according to the $k$-medoids algorithm using a Euclidean distance with weights defined using the percentage of variance explained by each component and the number of clusters selected as the maximiser of the Silhouette coefficients. The normalization is applied to the combined functions across made and missed shots.}
  \label{fig:clustering_k3}
\end{figure}

Overall, the data-driven analysis suggests that the five-cluster solution is useful for comparison with NBA positions, but the strongest intrinsic grouping in the score space is coarser, especially when the leading component is given its variance-based weight.



\section{Discussion} 
\label{sec:discussion}

In this work, we presented a novel representation of NBA players’ shot density charts based on Functional Data Analysis (FDA). We defined the smoothed densities of each player’s made and missed shots as bivariate functional data over the entire offensive half-court. This functional representation enabled a principal component decomposition, in which a player’s variation from the mean shot chart is parsimoniously described by a common set of principal component functions, weighted by player-specific scores.

We first used these principal component functions to describe the major modes of variation in shooting behavior. The first component, which explained almost $70\%$ of the total variance, primarily captured shooting frequency, distinguishing between players who shoot with high and low volumes. The remaining components, which explain successively less variation, captured increasingly subtle but interpretable features of shot selection.

We then applied $k$-medoids clustering to the player-specific scores to construct a data-driven taxonomy of player shooting tendencies. The resulting clusters revealed diverse shooting profiles, which we interpreted through the typical functional principal component scores within each group. Although, we use the $k$-medoids clustering algorithm in our analysis, the MFPCA representation is not tied to a particular clustering algorithm. Actually, once each player's shot chart has been summarized by a finite vector of scores, all of the standard clustering algorithms could be applied. For exemple, $k$-means clustering could be used to identify groups around average score profiles, hierarchical clustering could show nested similarities among players, mixture models could provide probabilistic cluster memberships and density-based methods could be use in case of outliers. We use the $k$-medoids algorithm because the center of each cluster is an actual player from the NBA, rather than a potentially artificial mean profile. This provides a direct basketball interpretation of each cluster through its medoid.

Interestingly, the estimated clusters showed only weak agreement with the conventional NBA-defined player positions. This may be because shooting behavior represents only one aspect of a player’s role; play-making, defensive responsibilities, and off-the-ball movement also influence positional classification.
Given the weak agreement, future work could estimate fewer or more than five clusters.
Nevertheless, our analysis provides a novel data-driven perspective on how players differ in their shooting behaviors.

We envision a number of potential uses of both our functional shot-chart representation and the subsequent clustering of shooting patterns.
Firstly, the functional representation of a player's shot chart provides a comprehensive characterization of a player's shooting behavior, i.e., the shot chart is not reduced to one or more scalar summaries.
This representation could be used by coaches or front-office staff to monitor changes in a player's shooting patterns between or within seasons, to identify player-specific tendencies which inform specific training regimes, and also to identify players whose characteristics match the needs of the team during scouting and recruitment.
Because the charts can naturally be presented in a way that is interpretable to basketball fans who may not be familiar with the underlying mathematics and statistics, the representation can be used to tell stories to the general public about player tendencies, and enable historical comparisons between players, enriching the narrative around the sport.
While we used shots made and shots missed, other metrics defined over the offensive half-court (e.g., shooting efficiency or defensive shot charts, \cite{franksCharacterizingSpatialStructure2015}), could be represented in the same way for the aforementioned purposes.

There are also a number of possible methodological extensions to the modeling framework. 
We aggregated shooting behavior across the 2018-2019 through 2023-2024 regular season. However, this aggregation may mask meaningful season-to-season changes in shooting behavior, as a player's shot selection can evolve with time due to development, injuries, coaching strategy, etc. A natural extension of our framework would be to perform the analysis at the player-season level rather than at the aggregated player level. In this setting, the MFPCA scores could be used to represents the evolution of shooting-style over time. This could help to quantify the temporal evolution of a player's role, identify players whose shooting patterns change across seasons, and examine whether these changes correspond to team changes, coaching changes or something else. This could also strengthen the comparison between empirical shooting-pattern clusters and conventional NBA positions. We could assign a player-season to each cluster and study the stability of these assignments.
Another one of the most exciting avenues is to combine shot-chart representations with other high-throughput data sources that are now routinely collected within the NBA.
For example, temporal biomechanical data are now being measured by motion-tracking systems such as Hawk-Eye \citep{HawkeyeInnovation2025}, and the possibility of a multi-modal representation that reflects a player's tactical and physical tendencies -- and the interaction between them -- could provide more detailed and specific analyses.
Our modeling framework could also be applied to positional data in other sports, for example, to characterize a soccer player's ``natural position'' or a boxer's ``fighting style''.

Finally, some limitations of the proposed modelling framework are as follows.
	We employed standard FPCA which assumes that observations are realizations from an unconstrained space of functions.
		However, densities are constrained because they must be non-negative and integrate to $1$. Reconstructions from standard FPCA do not necessarily satisfy the non-negativity constraint so specialized methods have been developed \citep{petersenFunctionalDataAnalysis2016}. However, these are for the univariate case. Rather than develop or extend these methodological approaches to the bivariate case, we apply the more tractable, standard FPCA approach while acknowledging its limitations. If required, reconstructions that lie outside the space of densities can be projected back into the density space.
	Secondly, this work focused on characterising the spatial patterns of made or missed shots. The normalization in the density estimation step, performed separately for the made and missed shots, inherently removes information about total shot volume or shooting efficiency (e.g., percentage chance of success). Future work could include an overall measure of shot volume or shooting efficiency to improve the characterization of players' behaviors.

\newpage
\setcounter{equation}{0}
\renewcommand{\theequation}{A.\arabic{equation}}
\begin{appendix}

%

\section{Court description} 
\label{sec:description}

\begin{figure}[h]
	\centering
	\includegraphics[width=0.75\textwidth]{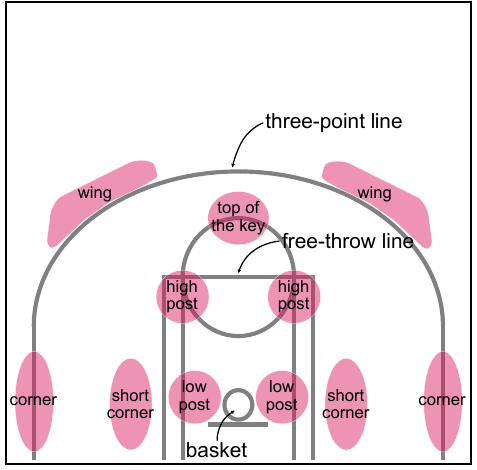}
	\caption{Description of the basketball court.}
	\label{fig:court_desc}
\end{figure}

Here, we present a description of the spatial regions of the basketball court referenced throughout the main text. These regions correspond to functionally important areas of play. Figure~\ref{fig:court_desc} presents the important locations on the offensive half-court. The basket is located at the center of the baseline, with the low-post regions surrounding it on both the left and right sides. Above these areas, the high-post zones are located near the free-throw line. The three-point line is the line that separates the two-point area from the three-point area, i.e, any shot made beyond this line counts as three point. The zone located between the free-throw line and the three-point line is commonly referred to as the top of the key. The corners of the court are divided into short-corner zones, that are closer to the basket, and corner three-point zones along the baseline. Finally, the wing areas are located on the side of the court between the sidelines and the extension of the free-throw line.


\section{Sensitivity analysis}

\subsection{Bandwidth selection} 
\label{sec:bandwidth}

We conducted a sensitivity check concerning the selection of  the smoothing bandwidth using cross-validation rather than fixing it through Silverman's rule. For each player and for each shot outcome (missed and made), the shot locations were first rescaled to $[0, 1] \times [0, 1]$, as in the main analysis. We then fitted isotropic product Gaussian kernel density estimators over a $20$-point logarithmic grid of candidate bandwidths from $10^{-3}$ to $1$, using the same scalar bandwidth in both spatial directions. Each candidate bandwidth was evaluated using five-fold cross-validation: within each fold, the kernel density estimator was estimated from the training shots and the held-out shots were scored by their log predictive density. The selected bandwidth was the value that maximized the average held-out log-likelihood. This criterion balances local detail and smoothness by favoring bandwidths that predict unseen shot locations well, rather than bandwidths that only reproduce the observed sample. After selecting the bandwidth separately for each player and shot outcome, we repeated the density estimation. The resulting densities show spatial patterns similar to those obtained with Silverman's rule, indicating that the main features of the analysis are not driven by the particular rule-of-thumb bandwidth used in the primary results.

\begin{figure}
	\centering
	\includegraphics[scale=0.35]{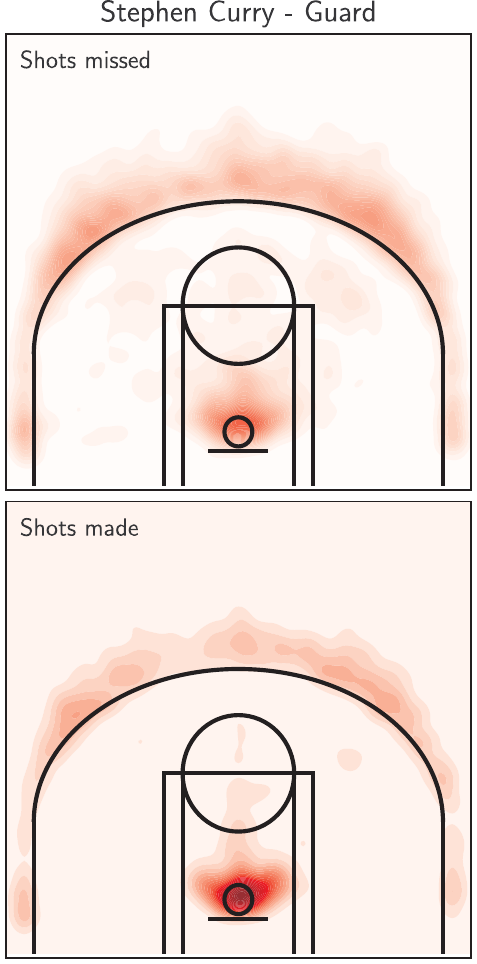}
	\includegraphics[scale=0.35]{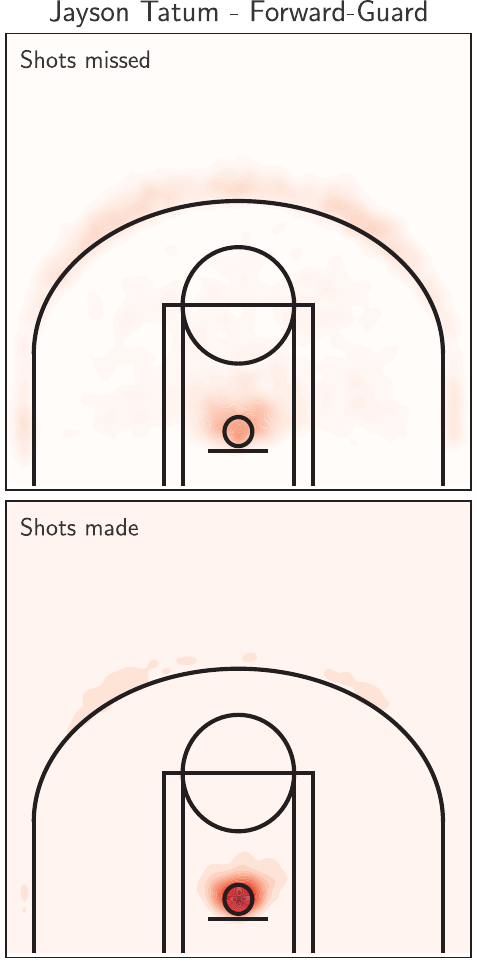}
	\includegraphics[scale=0.35]{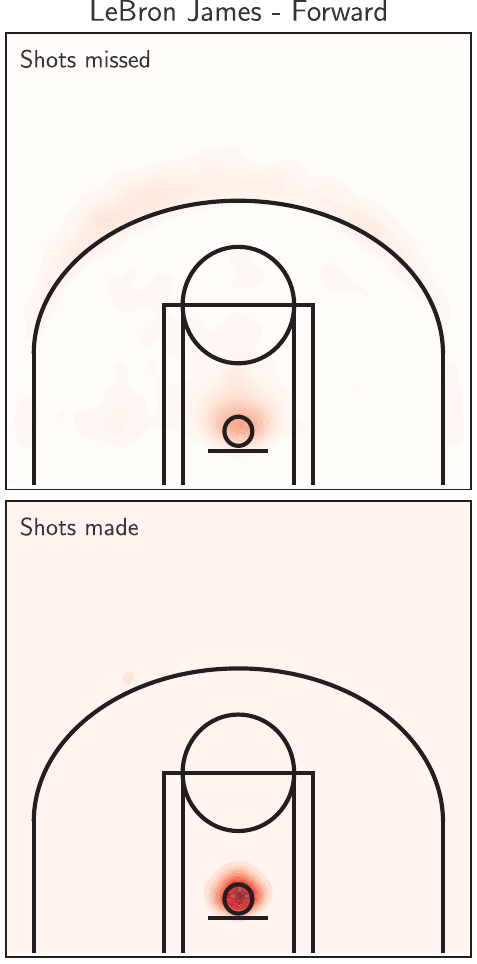}
	\includegraphics[scale=0.35]{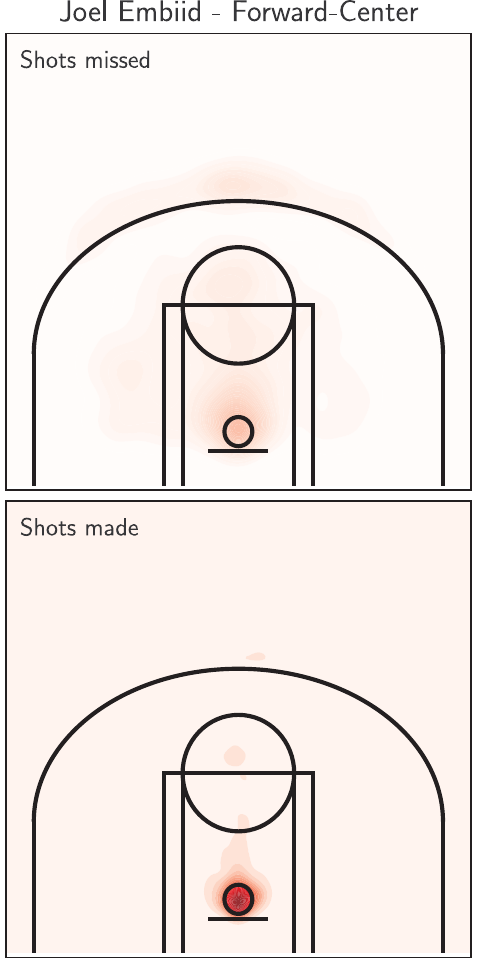}
	\includegraphics[scale=0.35]{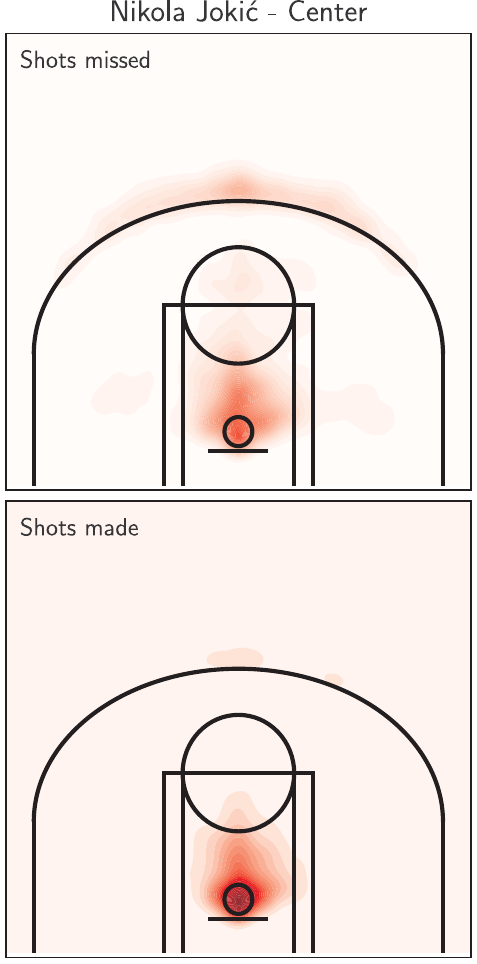}
	\\
	\includegraphics[scale=0.35]{figures/colorbar.pdf}
  \caption{Smoothed shot chart densities for five selected NBA players with a bandwitdh selected by cross-validation. The densities have been normalized between $0$ and $1$. The normalization has been applied to the combined data across made and missed shots.}
	\label{fig:examples_shooting_density_cv}
\end{figure}

\subsection{Grid} 
\label{sec:grid}

\begin{figure}
	\centering
	\includegraphics[scale=0.35]{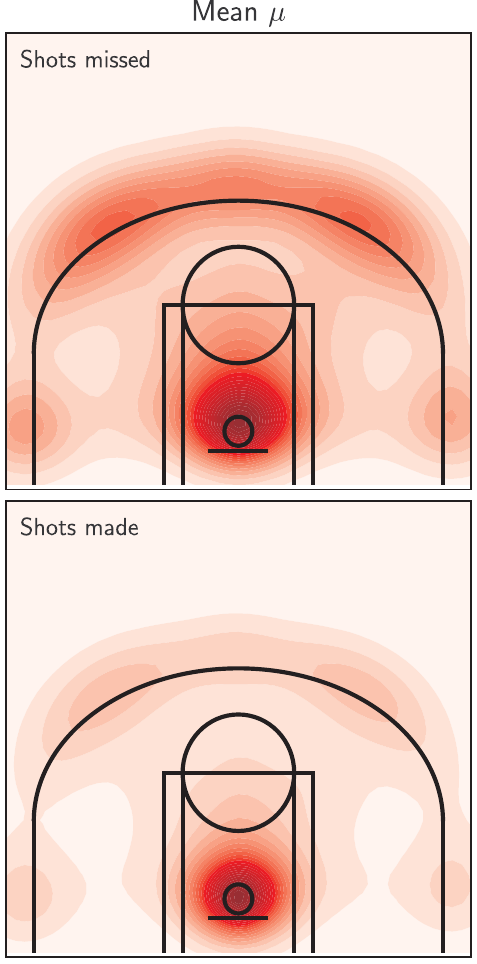}
	\includegraphics[scale=0.35]{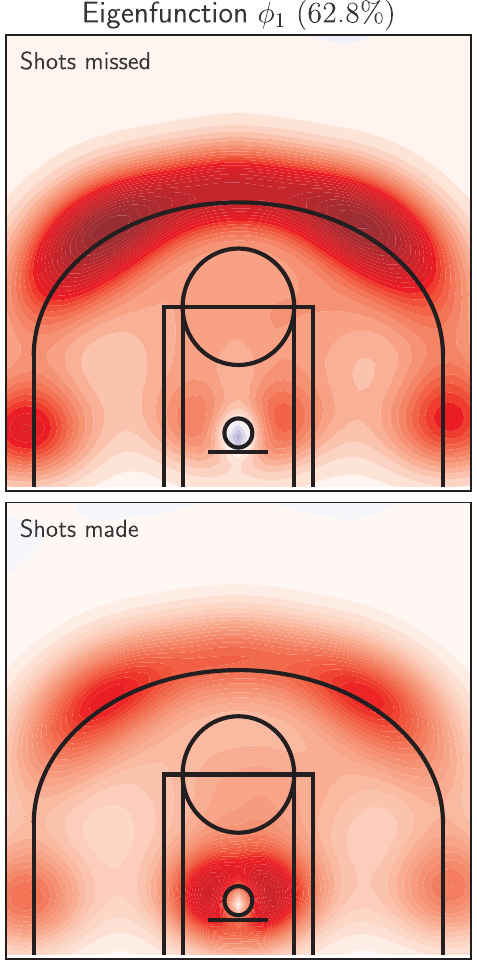}
	\includegraphics[scale=0.35]{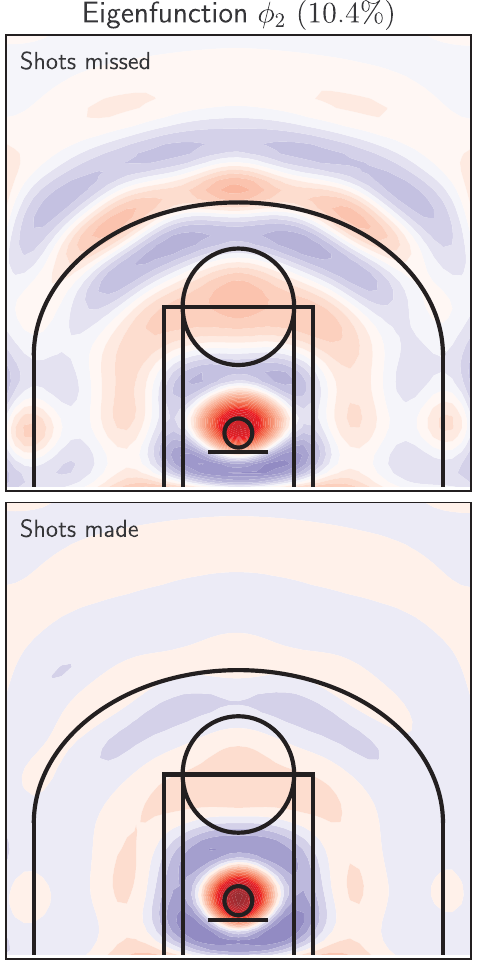}
	\includegraphics[scale=0.35]{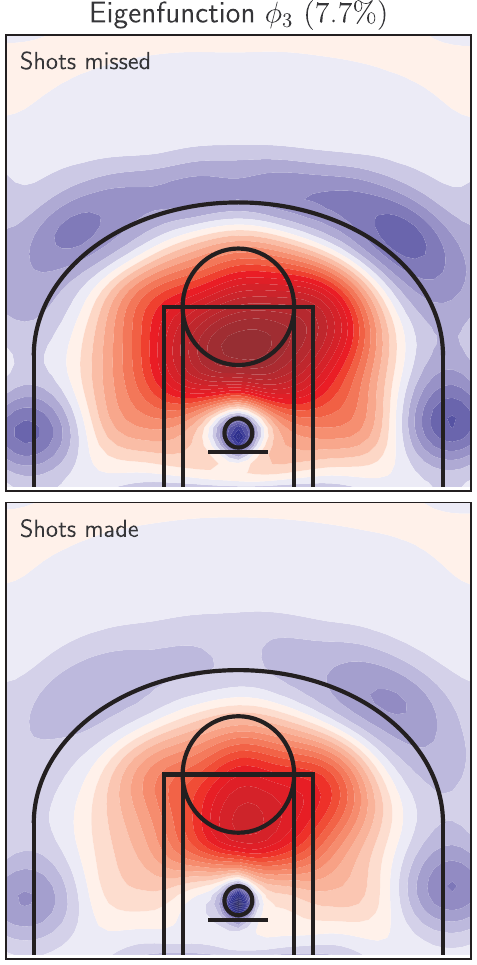}
	\includegraphics[scale=0.35]{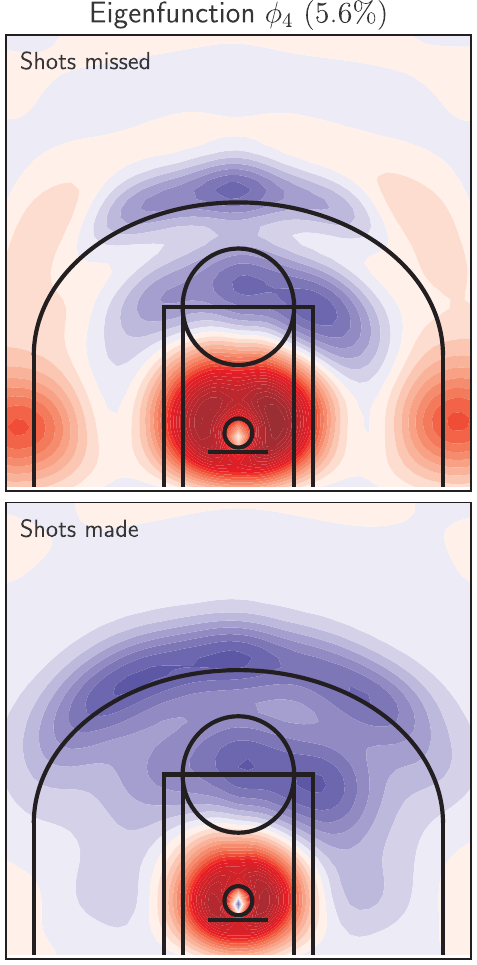}
	\\
	\includegraphics[scale=0.35]{figures/colorbar2.pdf}
	\caption{Estimation of the eigencomponents using a grid of $51 \times 51$.}
	\label{fig:grid-51}
\end{figure}

\begin{figure}
	\centering
	\includegraphics[scale=0.35]{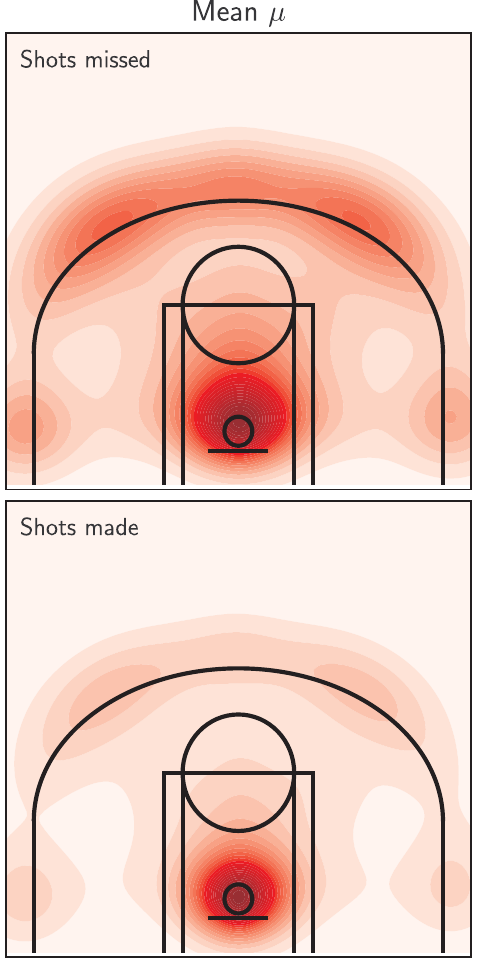}
	\includegraphics[scale=0.35]{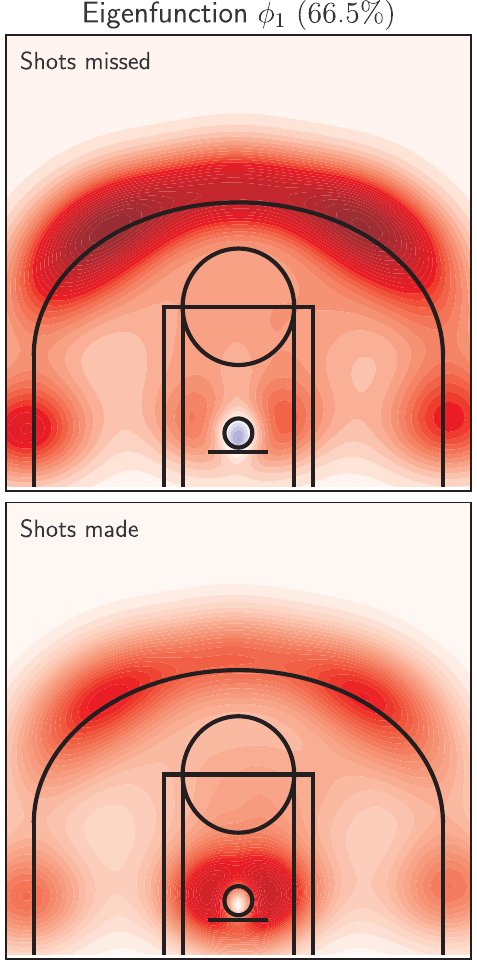}
	\includegraphics[scale=0.35]{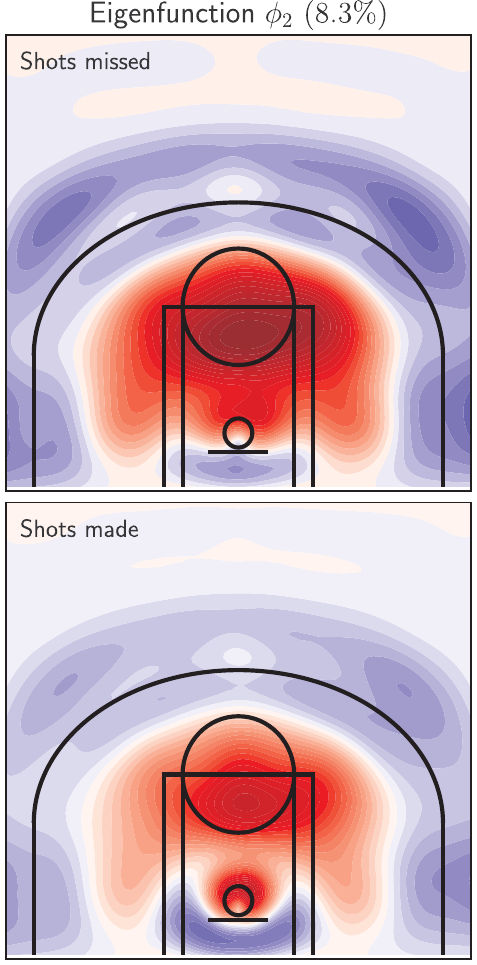}
	\includegraphics[scale=0.35]{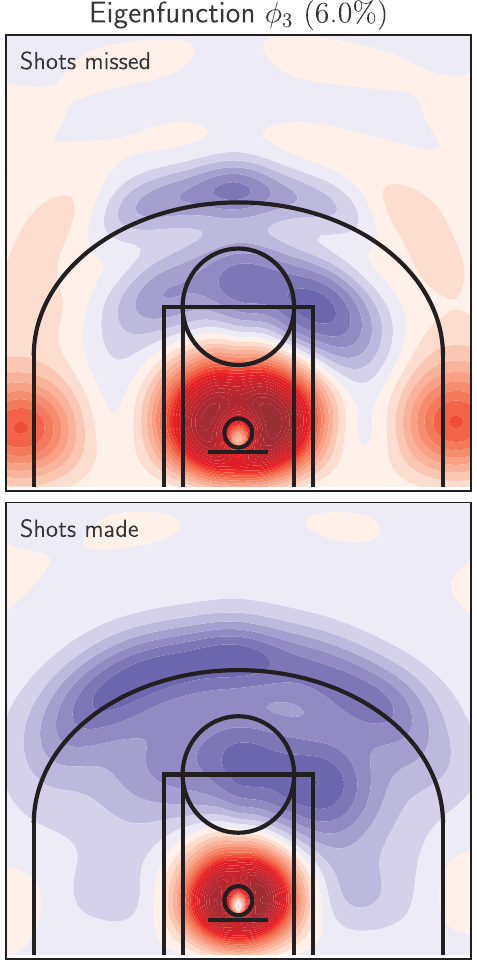}
	\includegraphics[scale=0.35]{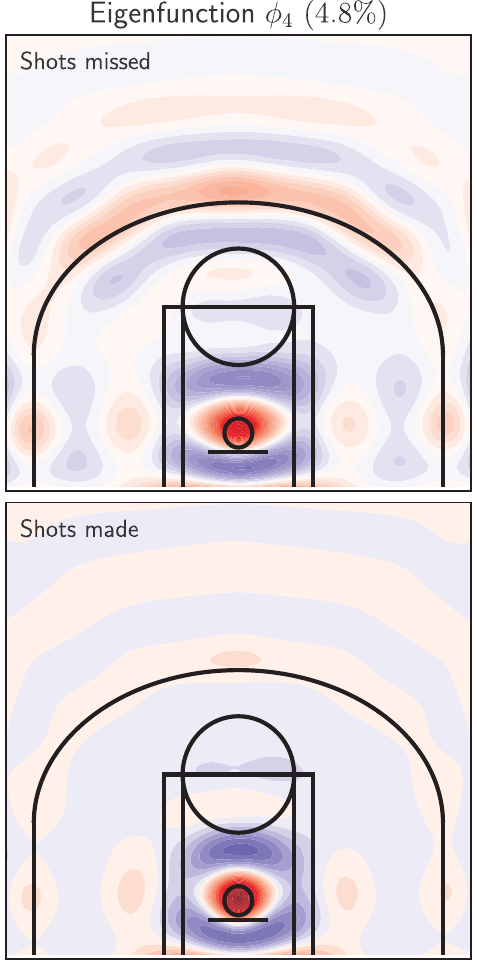}
	\\
	\includegraphics[scale=0.35]{figures/colorbar2.pdf}
	\caption{Estimation of the eigencomponents using a grid of $101 \times 101$.}
	\label{fig:grid-101}
\end{figure}

\begin{figure}
	\centering
	\includegraphics[scale=0.35]{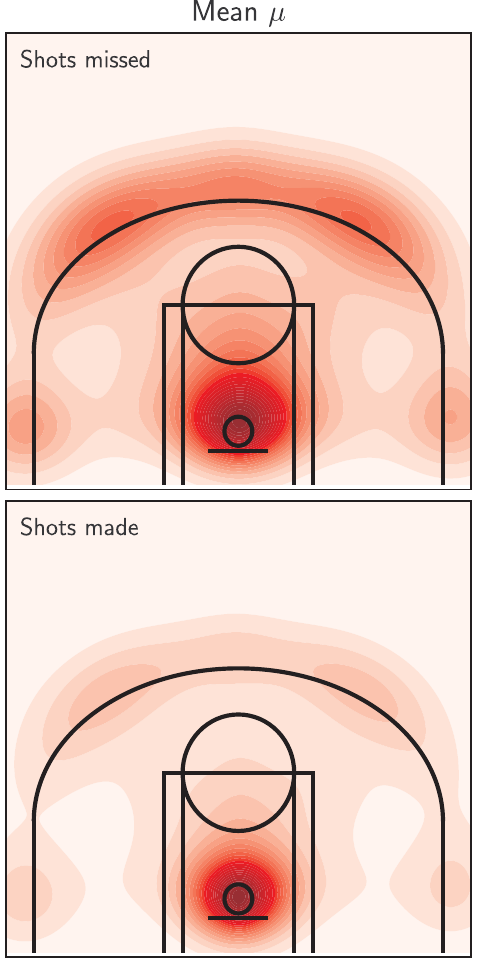}
	\includegraphics[scale=0.35]{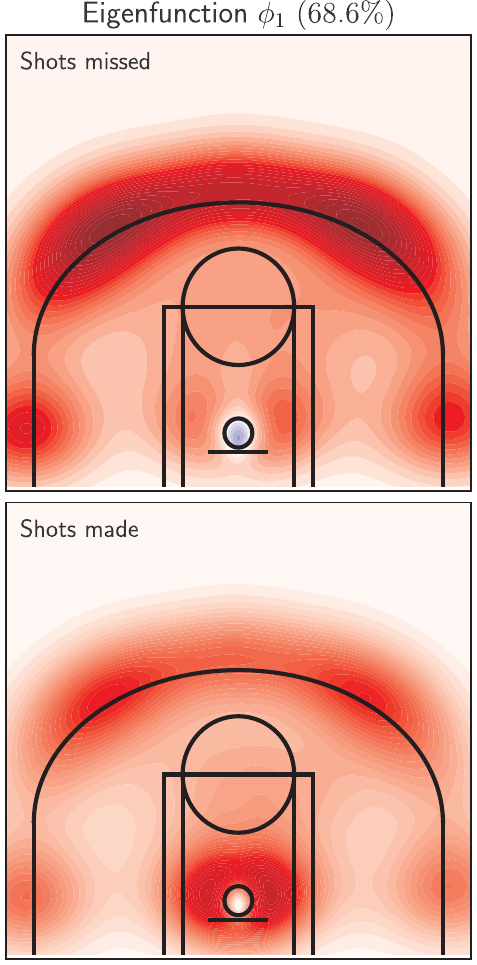}
	\includegraphics[scale=0.35]{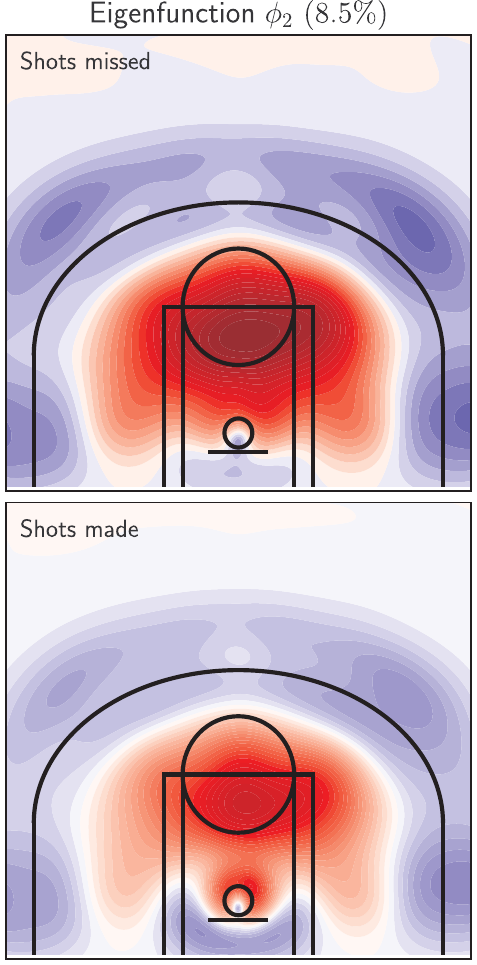}
	\includegraphics[scale=0.35]{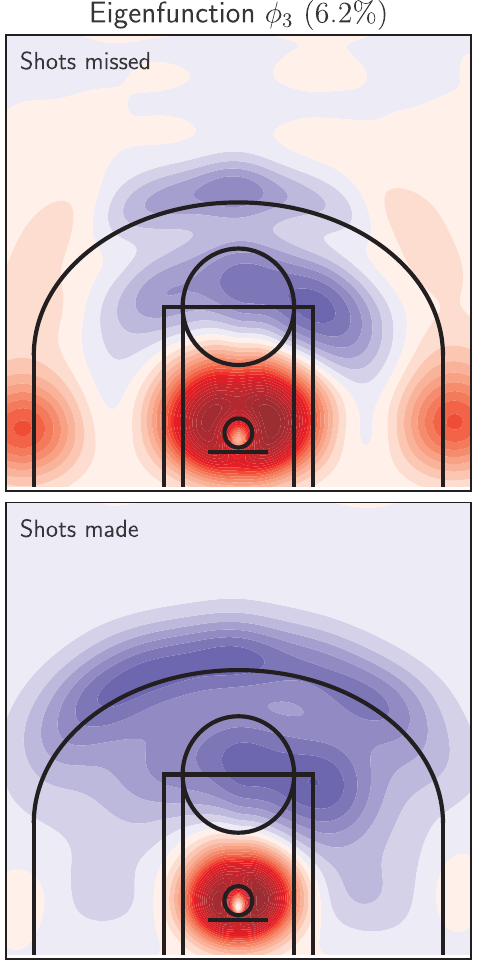}
	\includegraphics[scale=0.35]{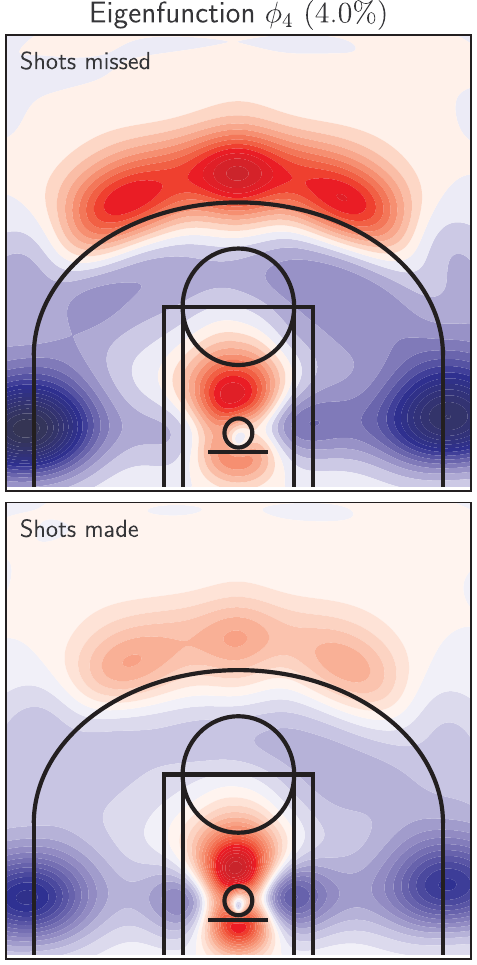}
	\\
	\includegraphics[scale=0.35]{figures/colorbar2.pdf}
	\caption{Estimation of the eigencomponents using a grid of $401 \times 401$.}
	\label{fig:grid-401}
\end{figure}

We conduct a sensitivity check concerning the grid resolution used to evaluate the kernel density estimates and to compute the MFPCA. We repeat the complete estimation procedure using two coarser grids, $51 \times 51$ and $101 \times 101$, and one finer grid, $401 \times 401$. The resulting mean functions and eigenfunctions are shown in Figures~\ref{fig:grid-51}--\ref{fig:grid-401}. Overall, the dominant spatial structures are consistent across the different resolutions. The mean functions and the first eigenfunctions are very similar to those obtained with the $201 \times 201$ grid, indicating that the main conclusions concerning average shooting behavior and the leading mode of variation are not driven by the chosen discretization. The second eigenfunction obtained with the $51 \times 51$ grid is different from the one obtained with the $201 \times 201$ grid. The third and fourth eigenfunctions obtained with the $51 \times 51$ grid are very similar to the second and third eigenfunctions, respectively, obtained with the $201 \times 201$ grid. The results obtained with the $101 \times 101$ grid are very similar to those obtained with the $201 \times 201$ grid, except for the fourth eigenfunction. The results obtained with the $401 \times 401$ grid are nearly indistinguishable from the main analysis, suggesting that the $201 \times 201$ grid provides a sufficiently fine approximation while avoiding unnecessary computational cost. This behavior is expected, as components that explain less variation are more sensitive to numerical approximation errors.


\subsection{Bootstrap results} 
\label{sec:bootstrap_results}

\begin{figure}
	\centering
	\includegraphics[scale=0.35]{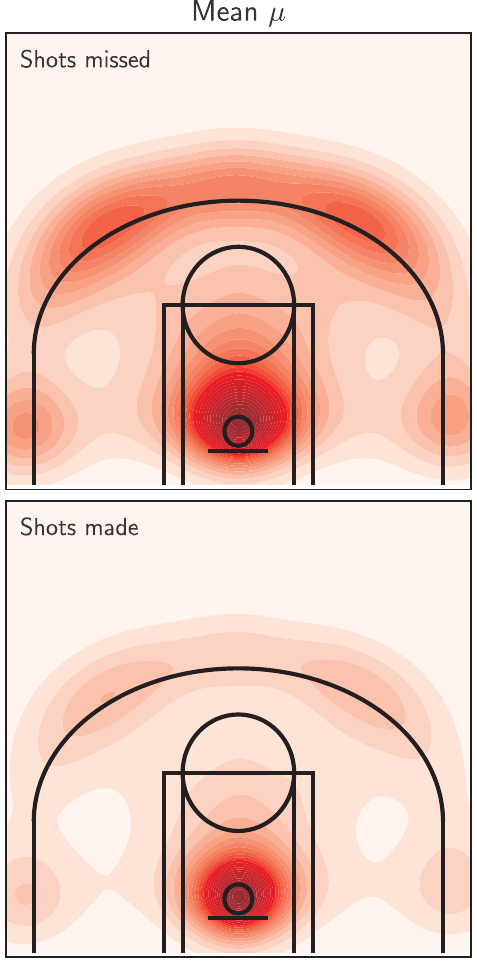}
	\includegraphics[scale=0.35]{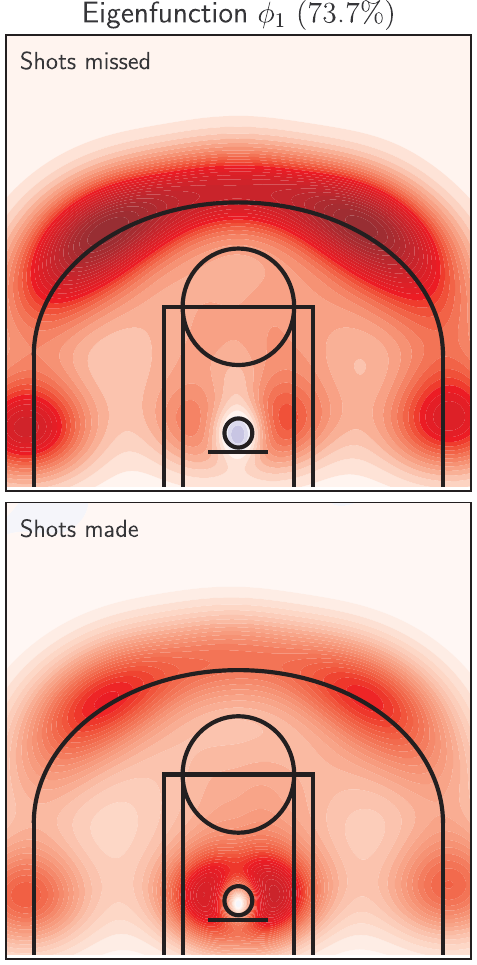}
	\includegraphics[scale=0.35]{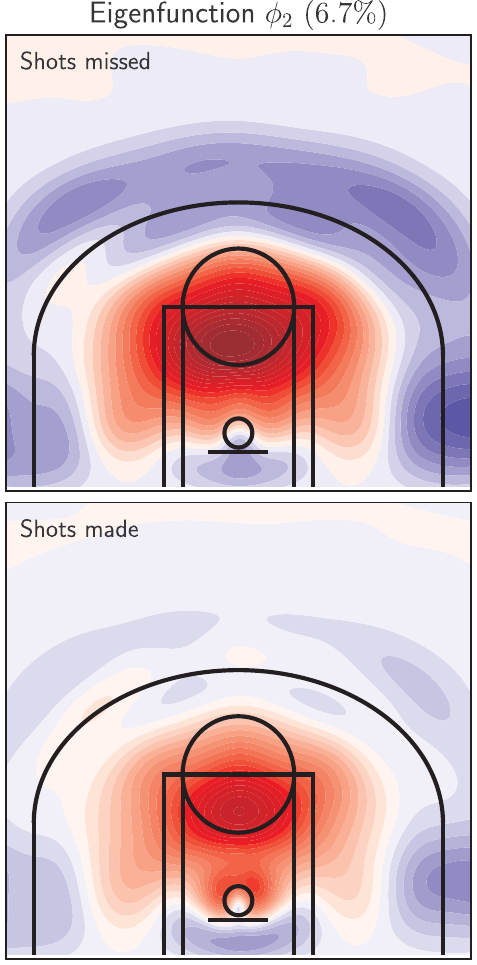}
	\includegraphics[scale=0.35]{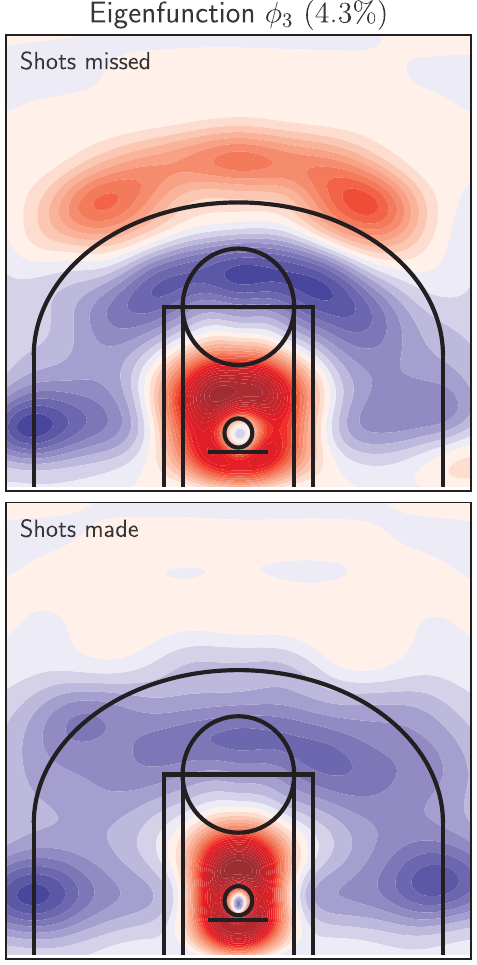}
	\includegraphics[scale=0.35]{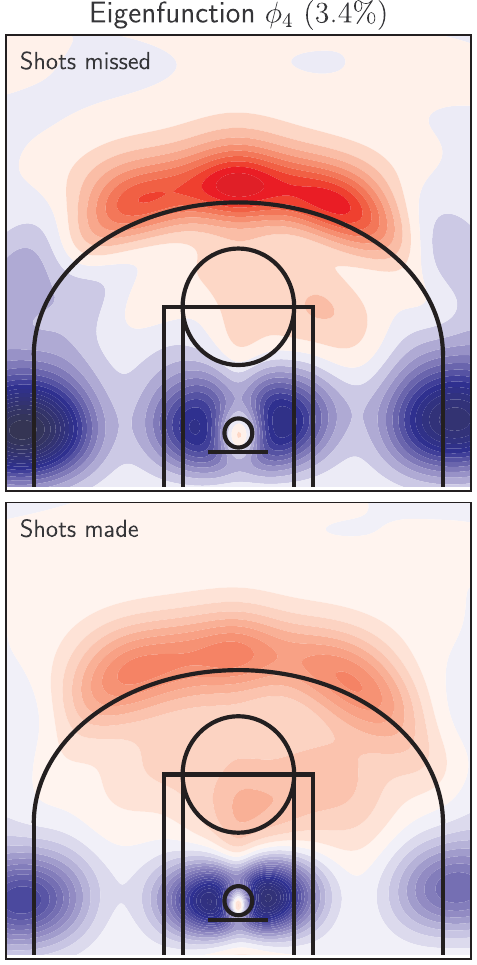}
	\\
	\includegraphics[scale=0.35]{figures/colorbar2.pdf}
	\caption{Bootstrap: First sample results. The normalization is applied to the combined functions across made and missed shots.}
	\label{fig:boostrap_1}
\end{figure}

\begin{figure}
	\centering
	\includegraphics[scale=0.35]{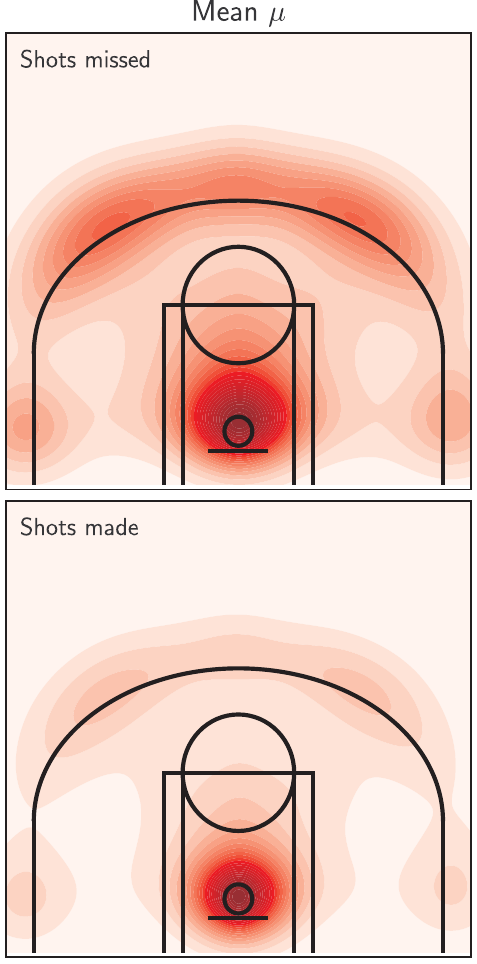}
	\includegraphics[scale=0.35]{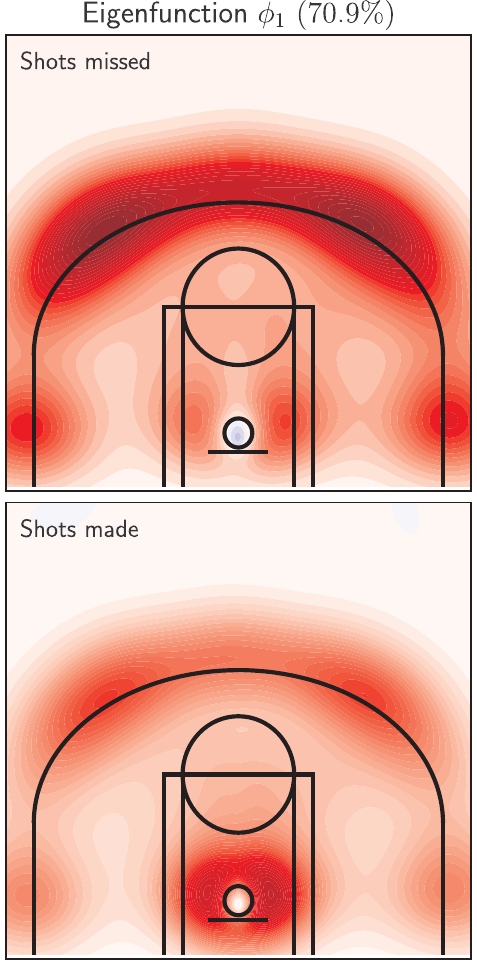}
	\includegraphics[scale=0.35]{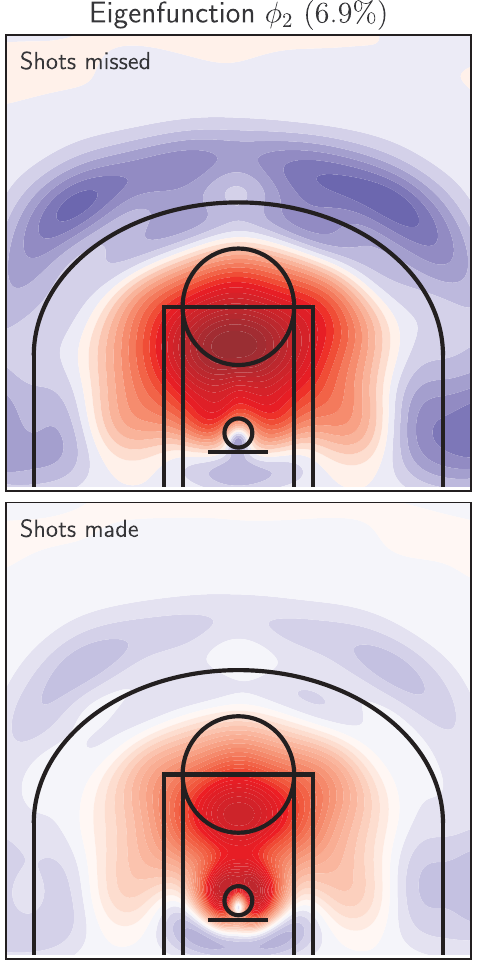}
	\includegraphics[scale=0.35]{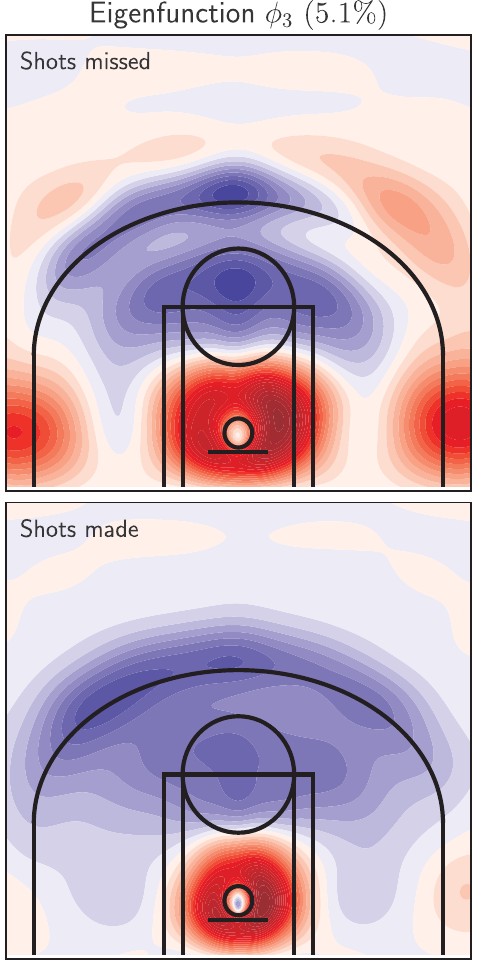}
	\includegraphics[scale=0.35]{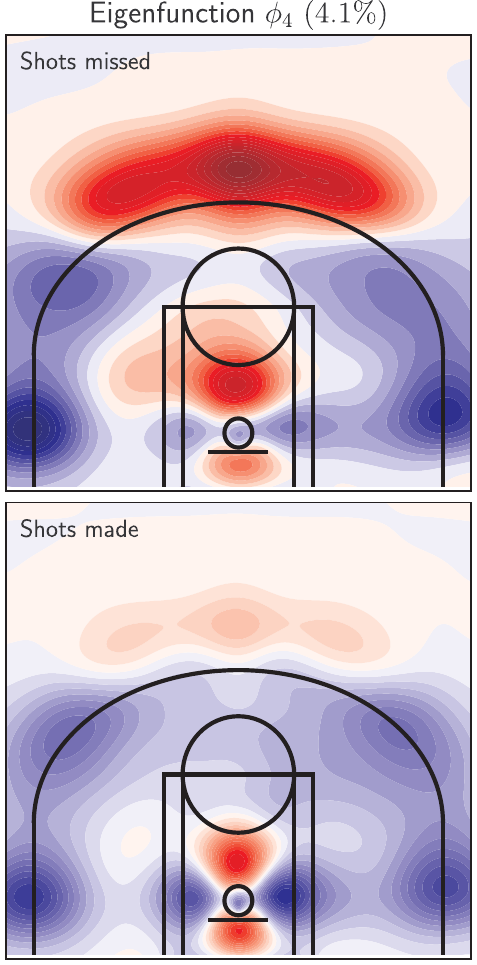}
	\\
	\includegraphics[scale=0.35]{figures/colorbar2.pdf}
	\caption{Bootstrap: Second sample results. The normalization is applied to the combined functions across made and missed shots.}
	\label{fig:boostrap_2}
\end{figure}

\begin{figure}
	\centering
	\includegraphics[scale=0.35]{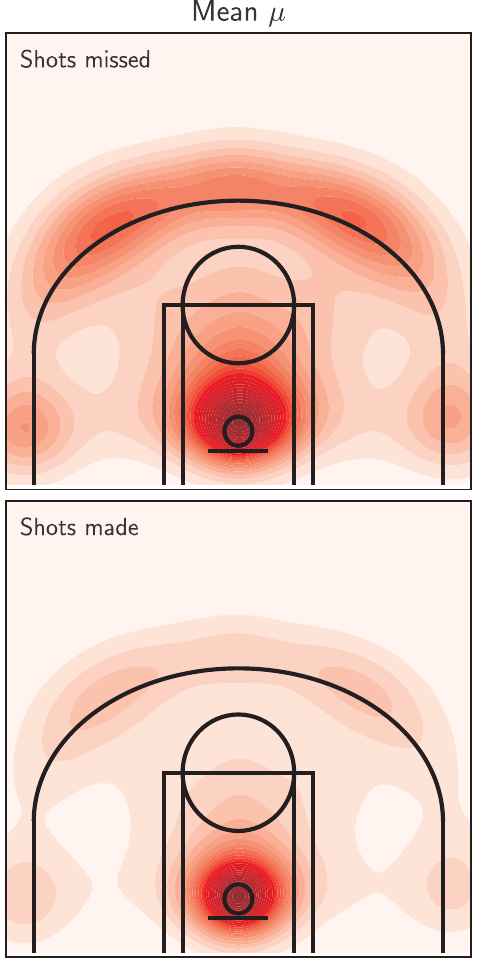}
	\includegraphics[scale=0.35]{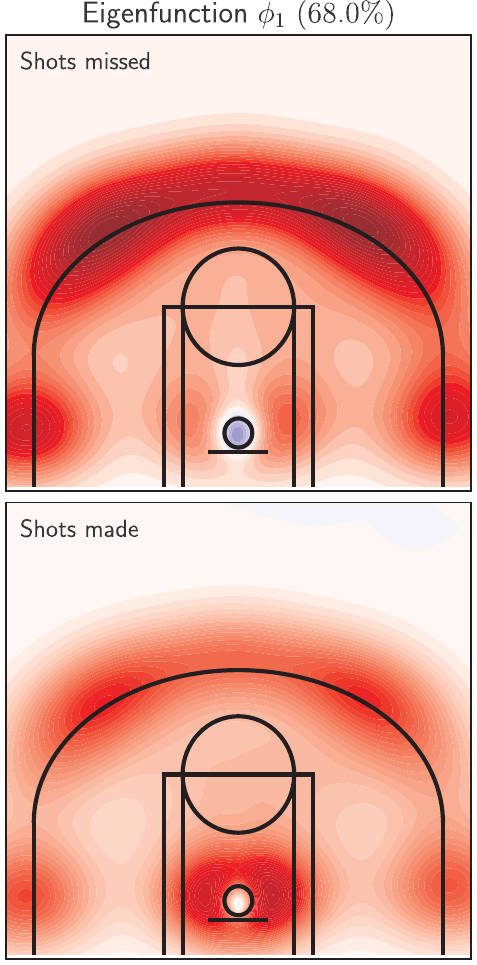}
	\includegraphics[scale=0.35]{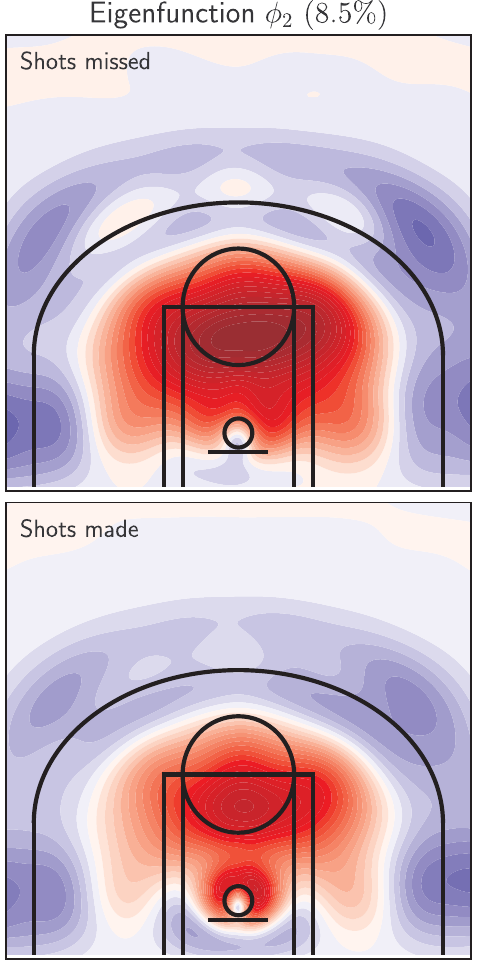}
	\includegraphics[scale=0.35]{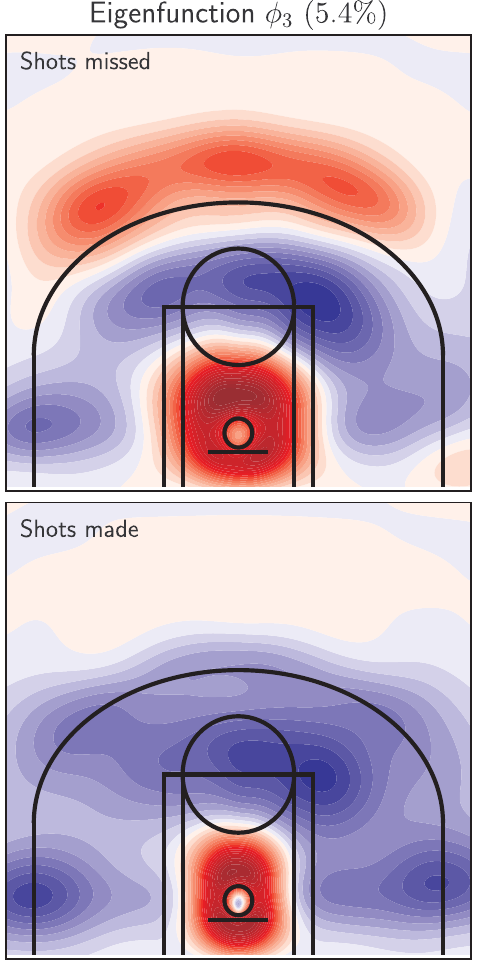}
	\includegraphics[scale=0.35]{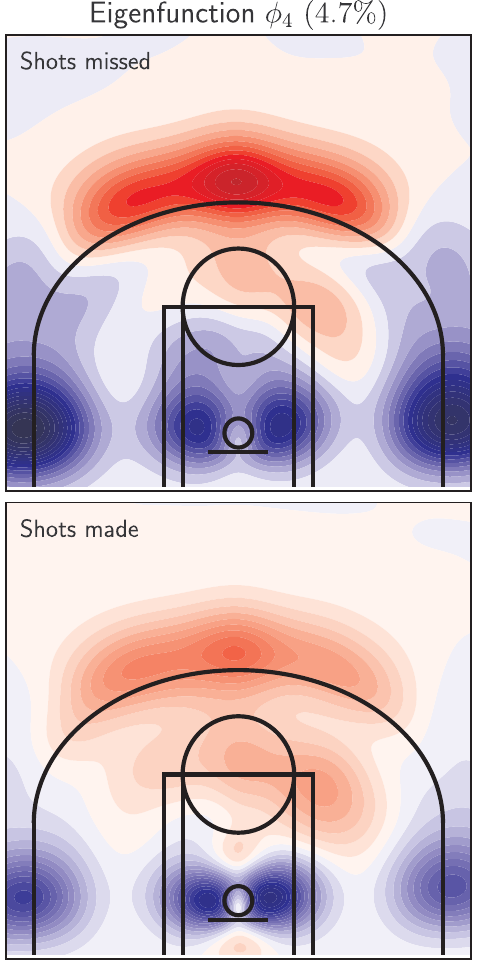}
	\\
	\includegraphics[scale=0.35]{figures/colorbar2.pdf}
	\caption{Bootstrap: Third sample results. The normalization is applied to the combined functions across made and missed shots.}
	\label{fig:boostrap_3}
\end{figure}

\begin{figure}
	\centering
	\includegraphics[scale=0.35]{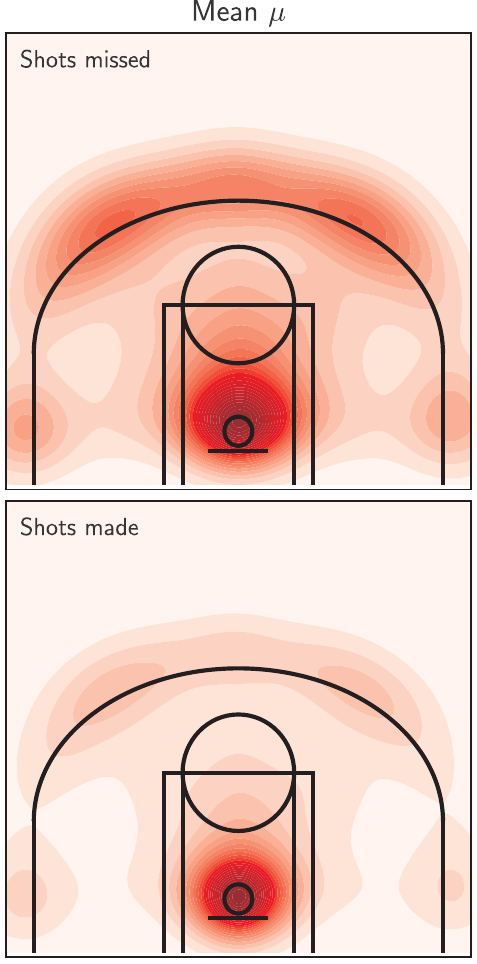}
	\includegraphics[scale=0.35]{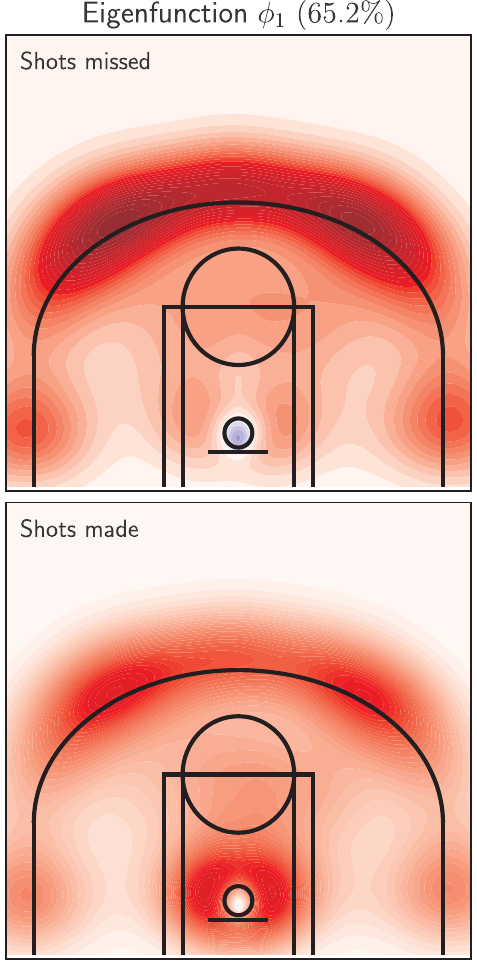}
	\includegraphics[scale=0.35]{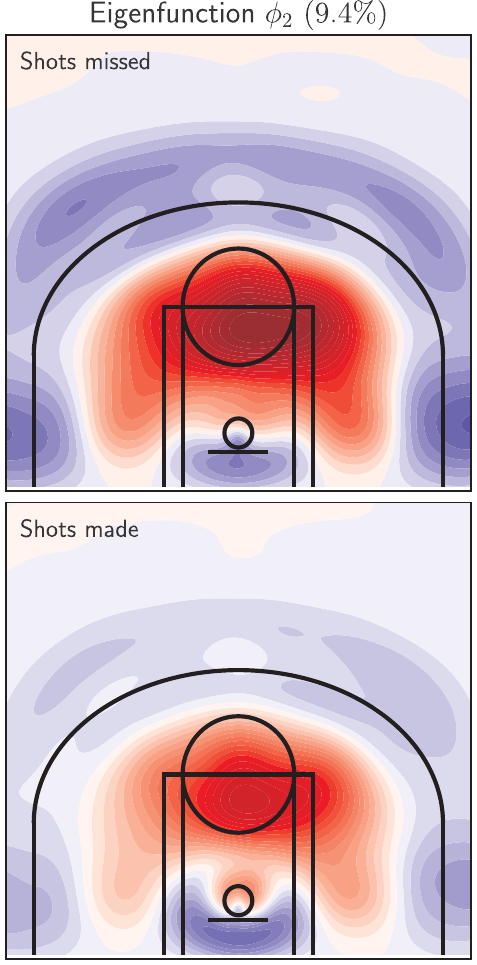}
	\includegraphics[scale=0.35]{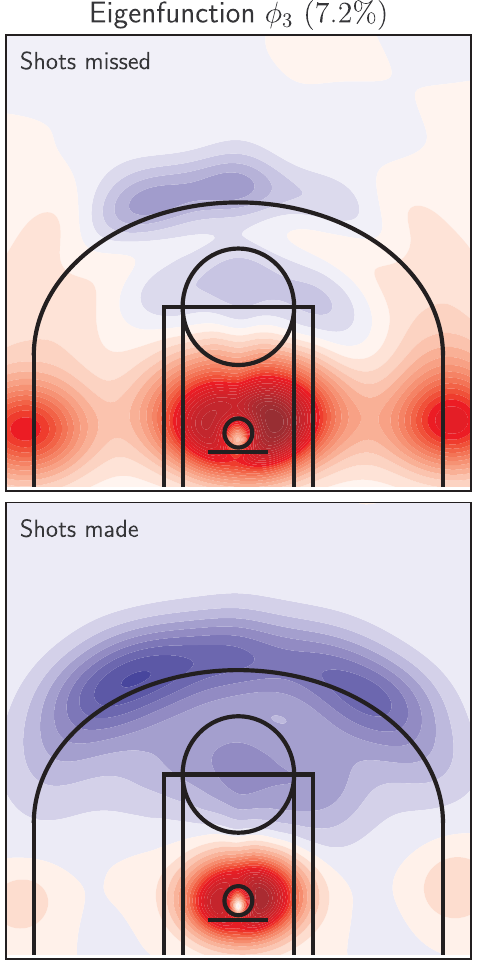}
	\includegraphics[scale=0.35]{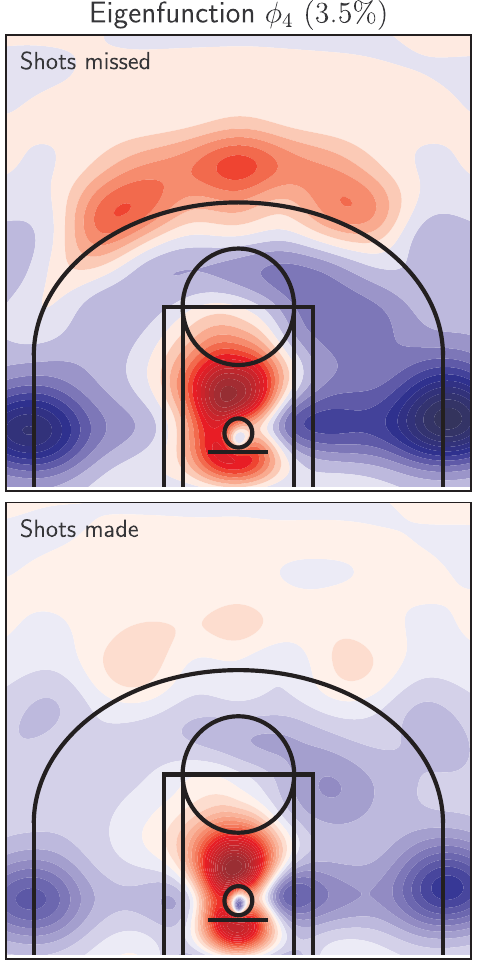}
	\\
	\includegraphics[scale=0.35]{figures/colorbar2.pdf}
	\caption{Bootstrap: Fourth sample results. The normalization is applied to the combined functions across made and missed shots.}
	\label{fig:boostrap_4}
\end{figure}

\begin{figure}
	\centering
	\includegraphics[scale=0.35]{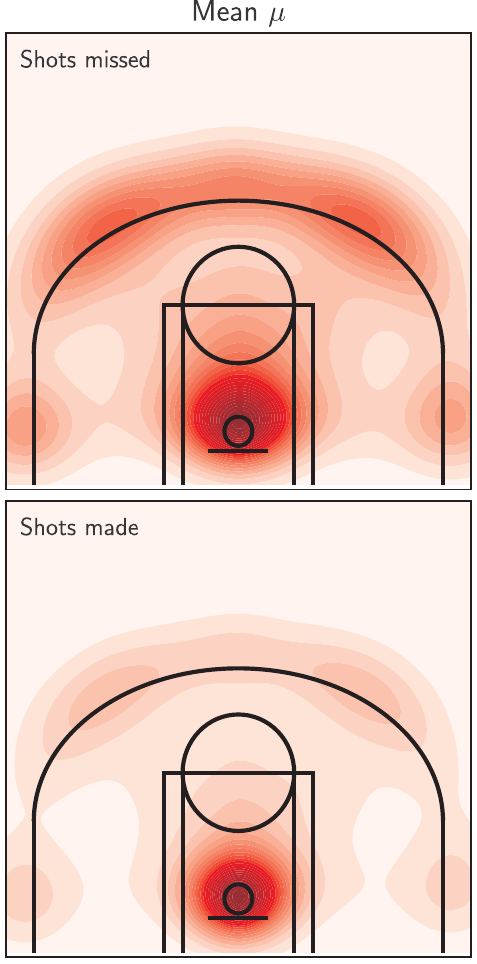}
	\includegraphics[scale=0.35]{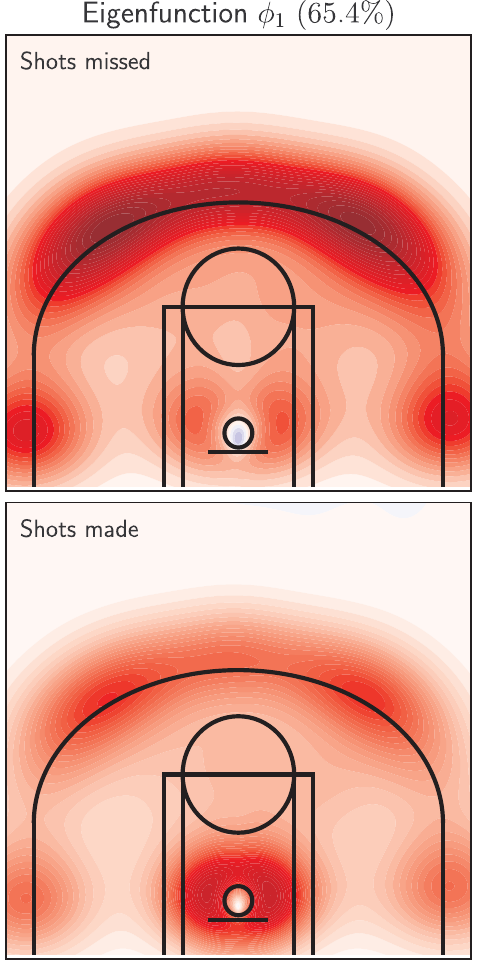}
	\includegraphics[scale=0.35]{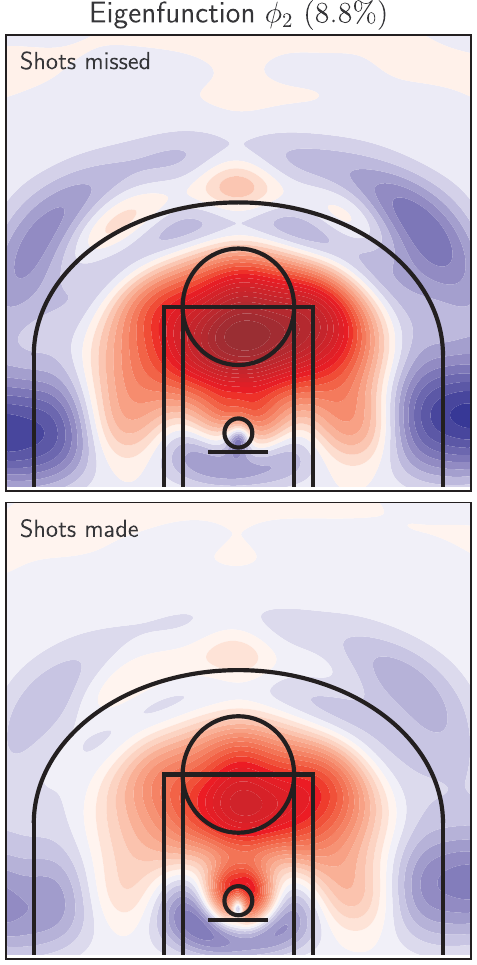}
	\includegraphics[scale=0.35]{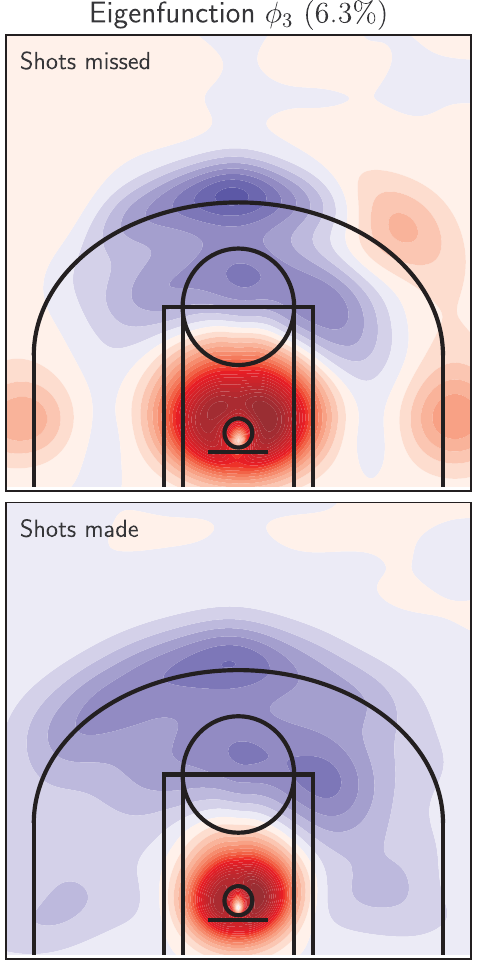}
	\includegraphics[scale=0.35]{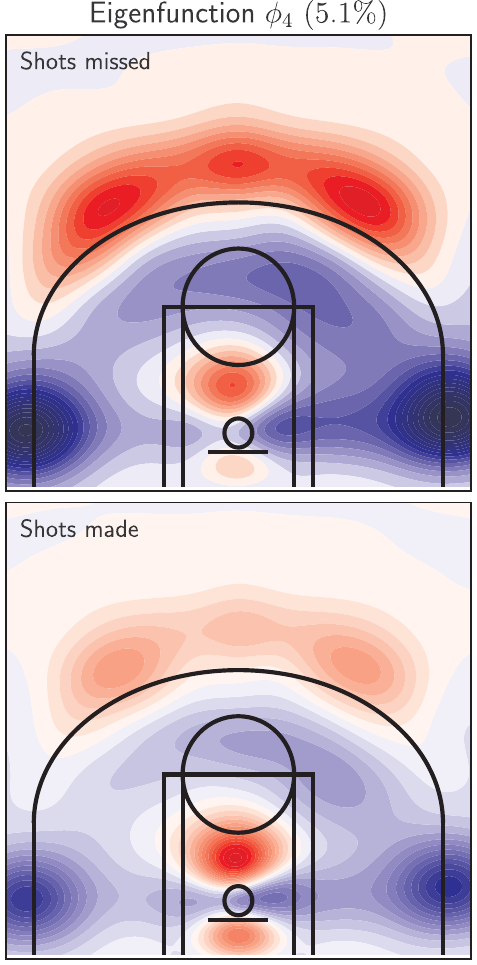}
	\\
	\includegraphics[scale=0.35]{figures/colorbar2.pdf}
	\caption{Bootstrap: Fifth sample results. The normalization is applied to the combined functions across made and missed shots.}
	\label{fig:boostrap_5}
\end{figure}

As a sensitivity check, we performed the following exercise:
	\begin{itemize}
		\item Construct $5$ bootstrap samples.
		\item For each bootstrap sample, compute the MFPCA.
		\item Display the mean and the first $4$ MFPCs from each bootstrap sample (Figures~\ref{fig:boostrap_1} to~\ref{fig:boostrap_5}).
	\end{itemize}
	The shapes of the resulting mean functions and the first two principal components are very consistent.
	There is slight sample-to-sample variability for the third component.
	This result is intuitive. The mean and leading components, explaining the most variability, represent the most common modes of variation in the data and hence are more stable from sample to sample. Overall, this simple sensitivity check indicates that the prominent modes of variation are stable.



\section{Clustering results} 
\label{sec:clustering_results}

\begin{figure}
	\centering
	\begin{subfigure}{0.49\textwidth}
		\centering
		\includegraphics[scale=0.2]{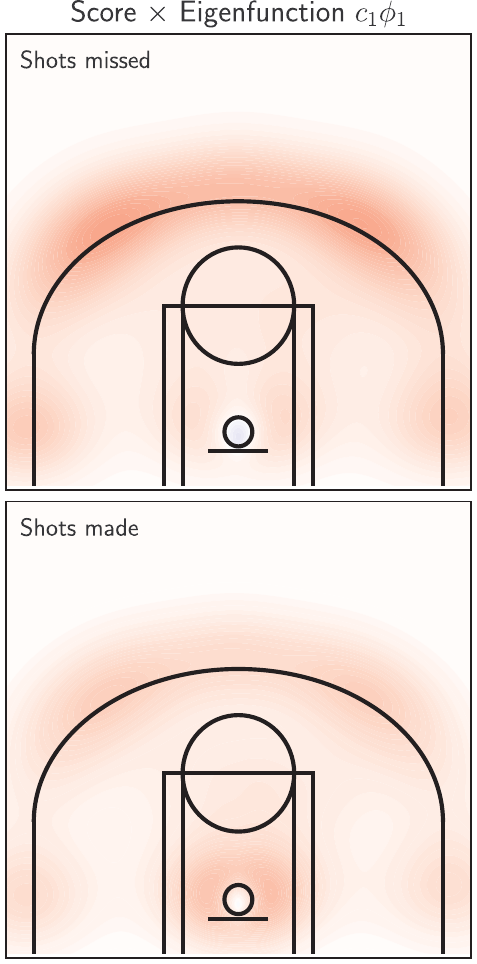}
		\includegraphics[scale=0.2]{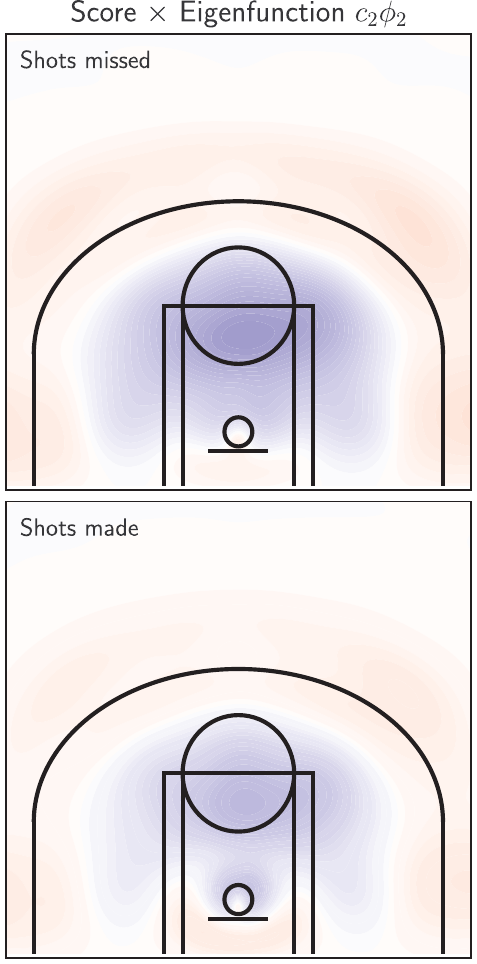}
		\includegraphics[scale=0.2]{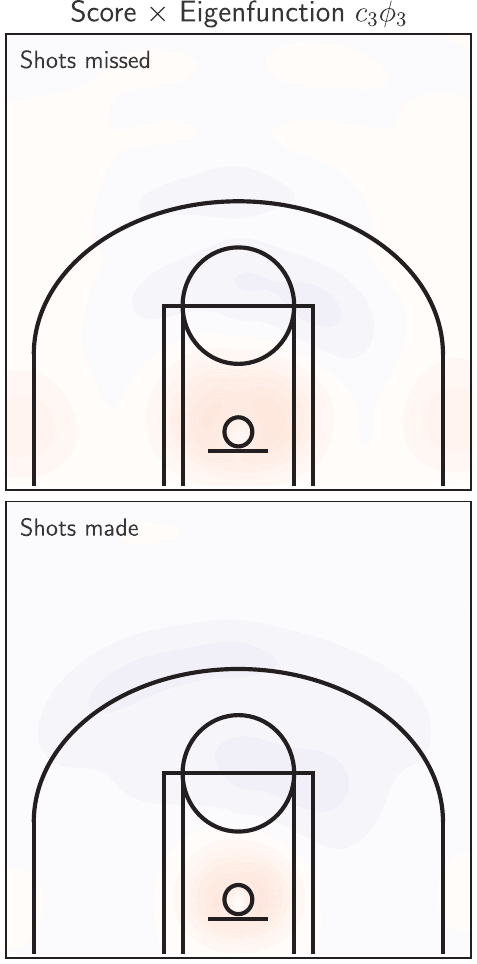}
		\includegraphics[scale=0.2]{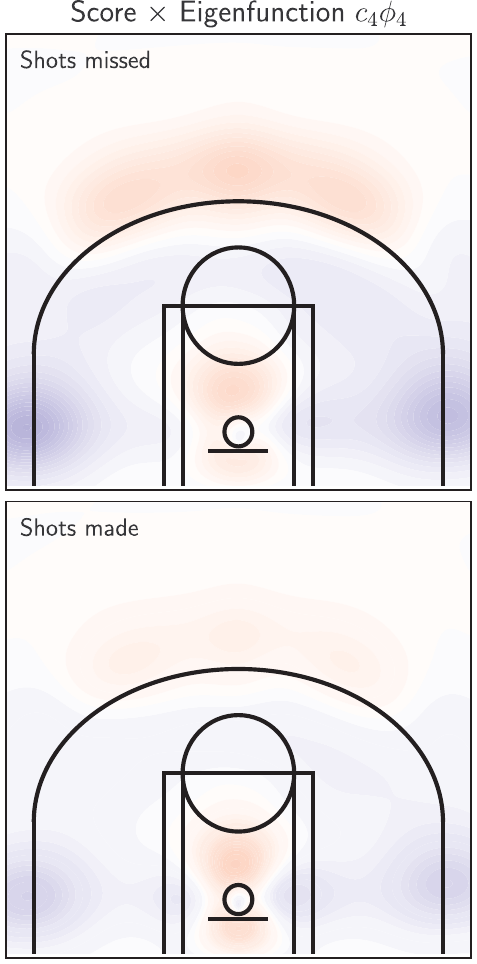}
		\caption{Cluster 1.}
	\end{subfigure}
	\hfill
	\begin{subfigure}{0.49\textwidth}
		\centering
		\includegraphics[scale=0.2]{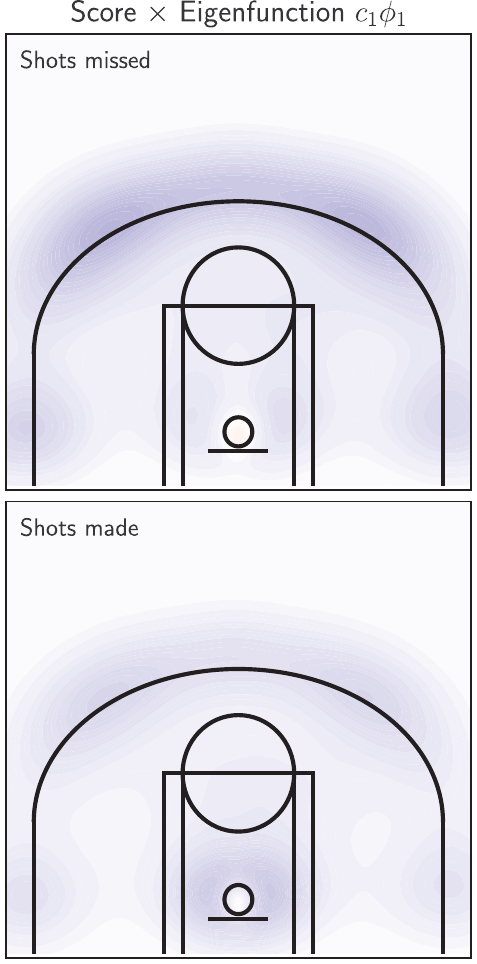}
		\includegraphics[scale=0.2]{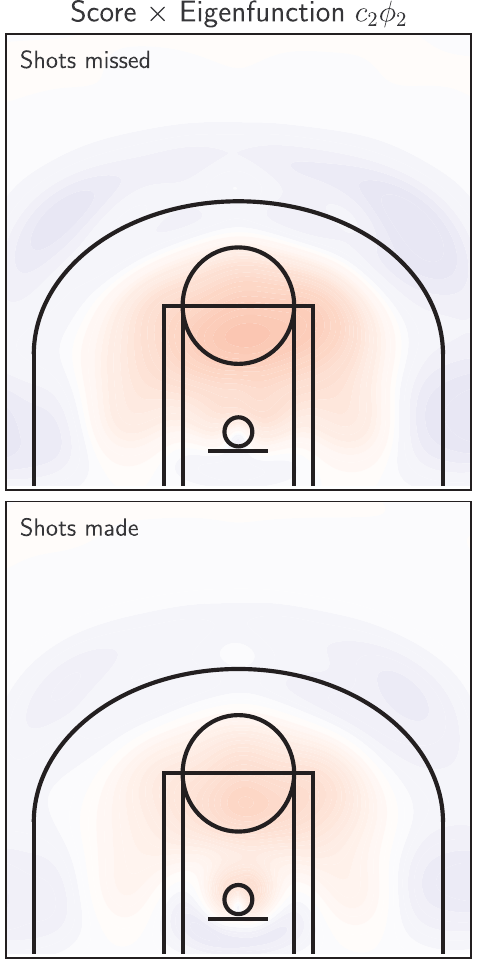}
		\includegraphics[scale=0.2]{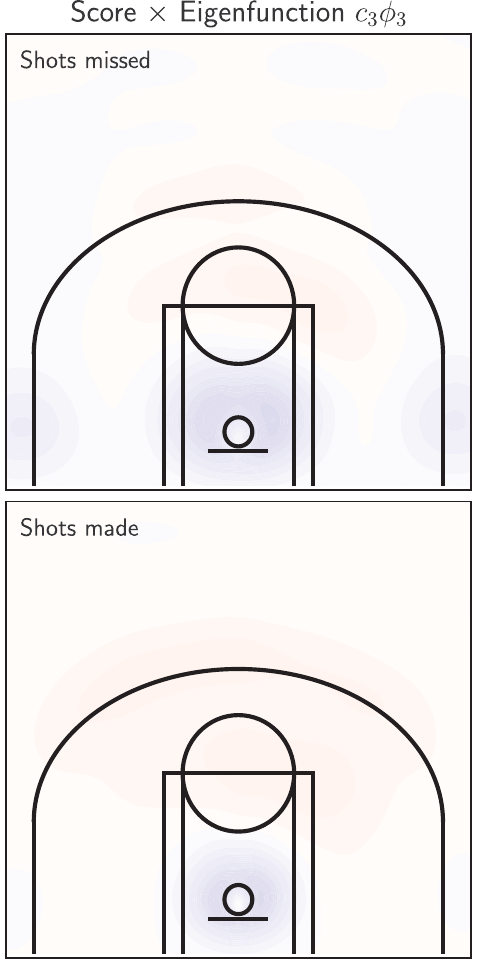}
		\includegraphics[scale=0.2]{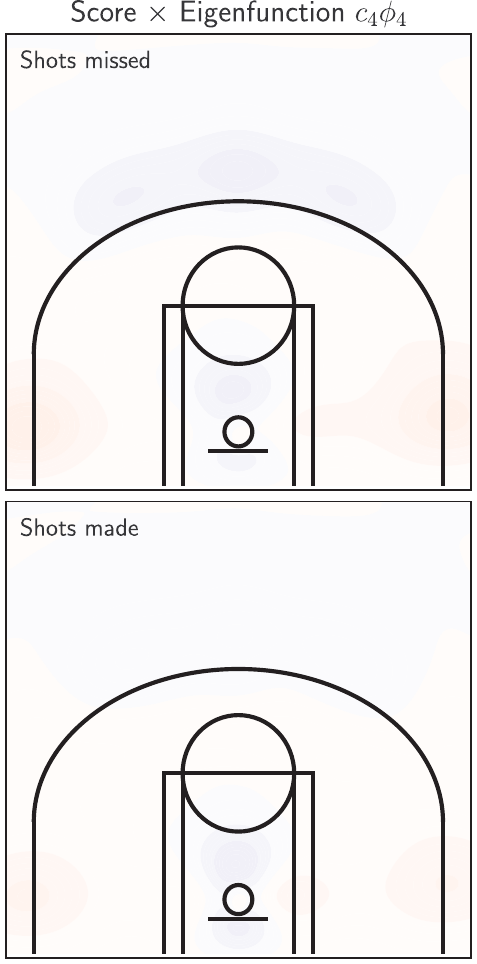}
		\caption{Cluster 2.}
	\end{subfigure}
	\\
	\begin{subfigure}{0.49\textwidth}
		\centering
		\includegraphics[scale=0.2]{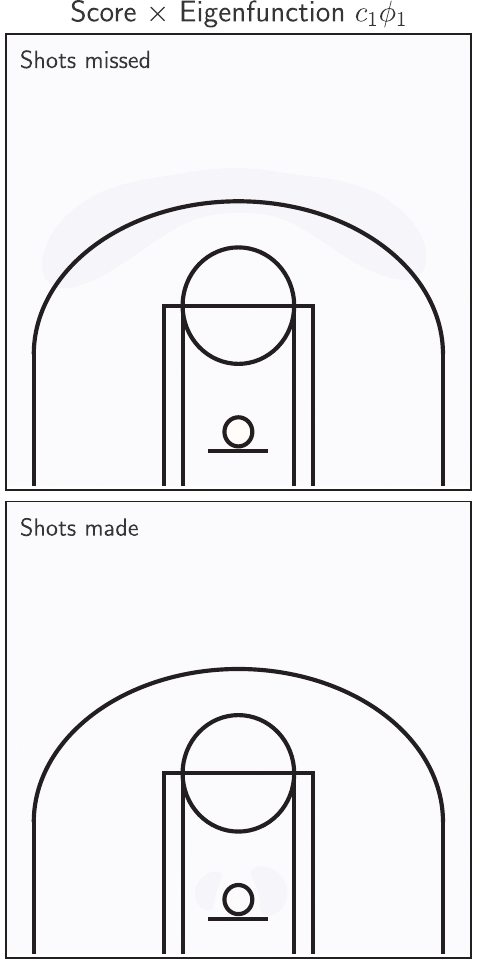}
		\includegraphics[scale=0.2]{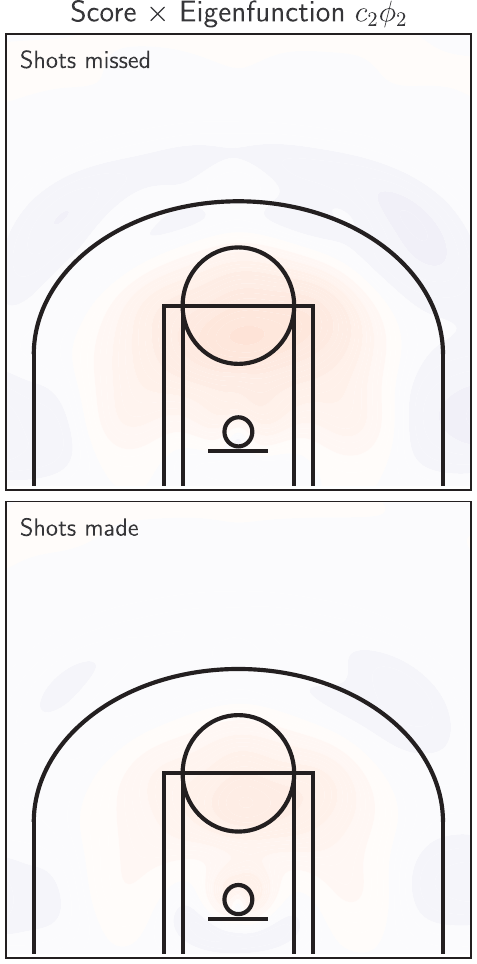}
		\includegraphics[scale=0.2]{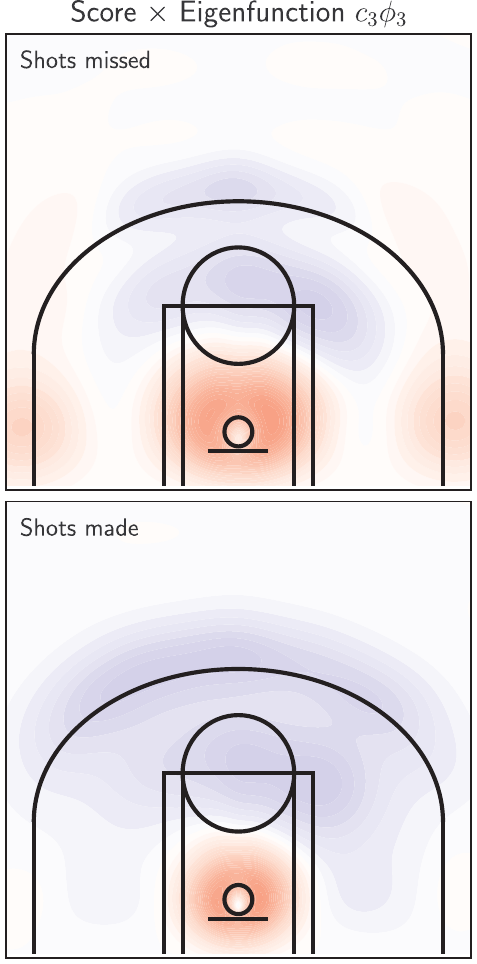}
		\includegraphics[scale=0.2]{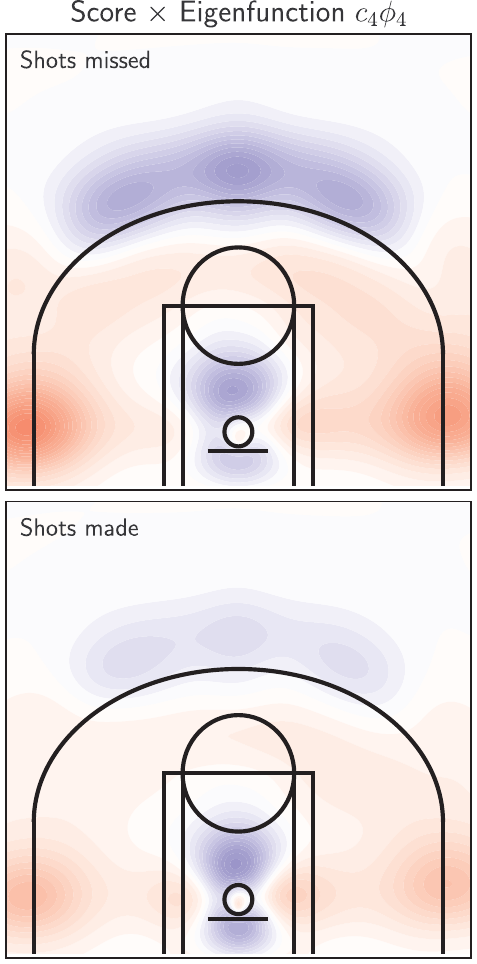}
		\caption{Cluster 3.}
	\end{subfigure}
	\hfill
	\begin{subfigure}{0.49\textwidth}
		\centering
		\includegraphics[scale=0.2]{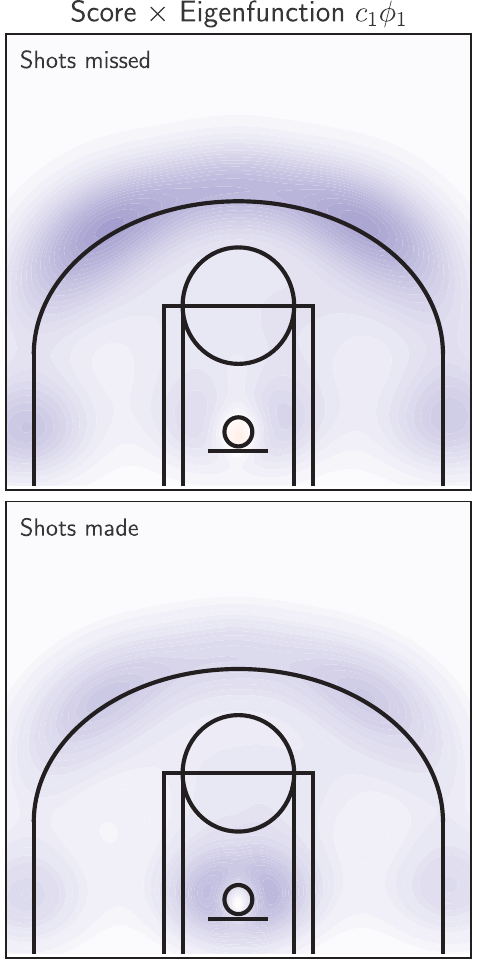}
		\includegraphics[scale=0.2]{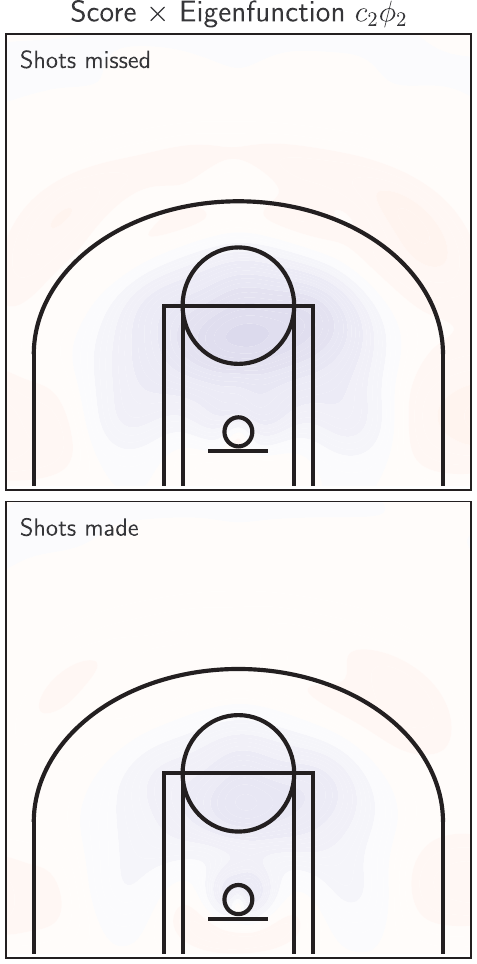}
		\includegraphics[scale=0.2]{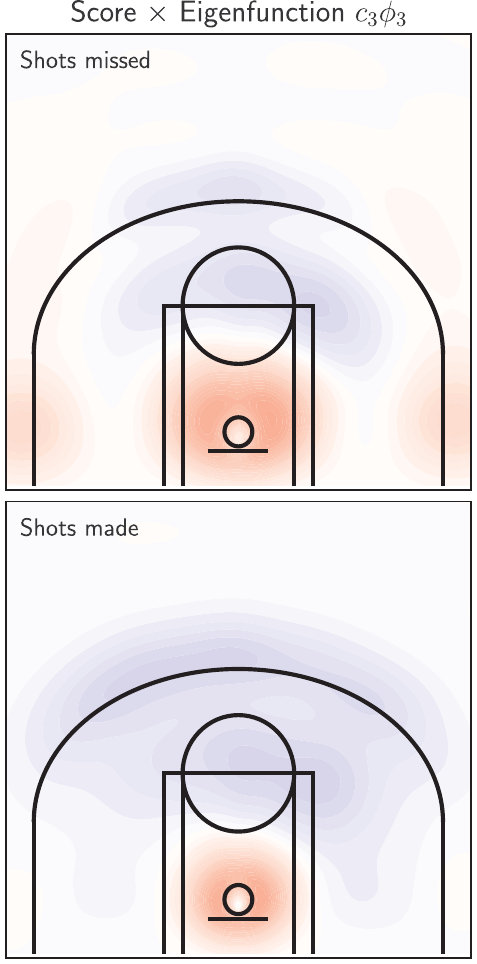}
		\includegraphics[scale=0.2]{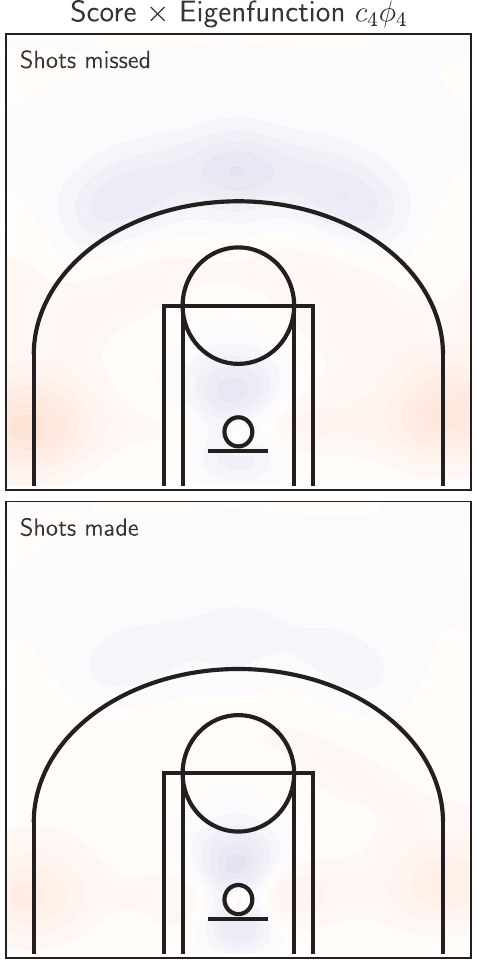}
		\caption{Cluster 4.}
	\end{subfigure}
	\\
	\begin{subfigure}{0.49\textwidth}
		\centering
		\includegraphics[scale=0.2]{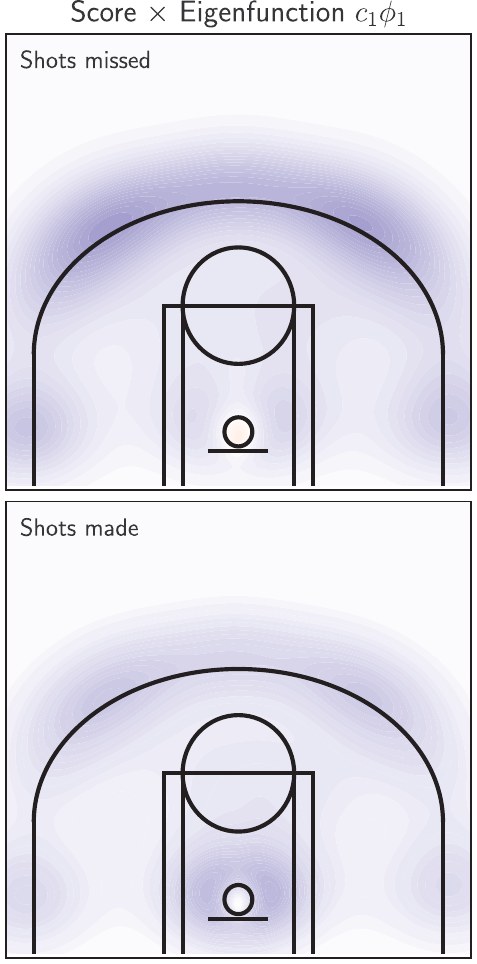}
		\includegraphics[scale=0.2]{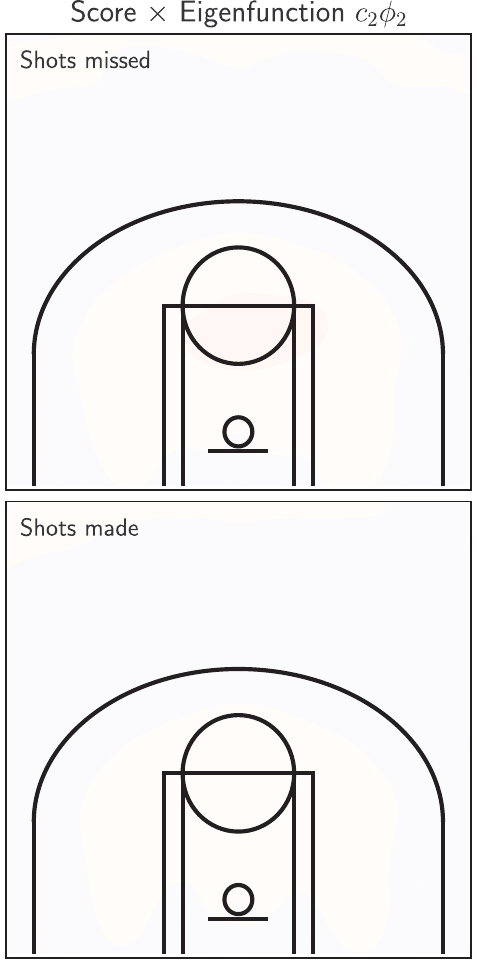}
		\includegraphics[scale=0.2]{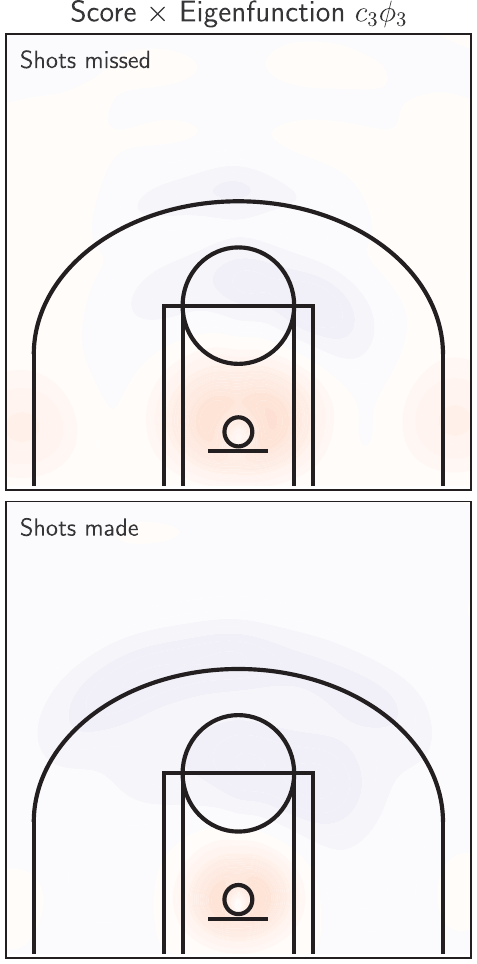}
		\includegraphics[scale=0.2]{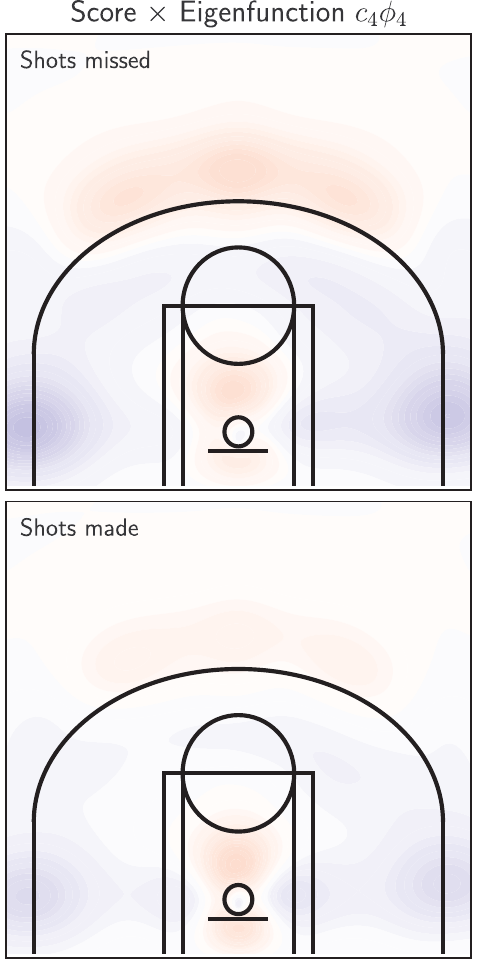}
		\caption{Cluster 5.}
	\end{subfigure}
  \caption{MFPCA decompositions from the clustering using equal weights for each component. The normalization is applied to the combined functions across made and missed shots.}
	\label{fig:cluster_no_weight}
\end{figure}

In this section, we present the complete clustering results. First, we present in which cluster each of the $173$ players belong to when we assign equal weights to all components. Figure \ref{fig:cluster_no_weight} presents the decomposition of each cluster when the clustering is performed when equal weights are given to the components. The clustering results can be interpreted based on shooting tendencies.

\begin{description}
	\item[Cluster 1] Grayson Allen, Malik Beasley, Pat Connaughton, Robert Covington, Jae Crowder, Stephen Curry, Donte DiVincenzo, Evan Fournier, Danilo Gallinari, Eric Gordon, Devonte' Graham, Tim Hardaway Jr., James Harden, Joe Harris, Buddy Hield, Kevin Huerter, Joe Ingles, Damian Lillard, Brook Lopez, Kevin Love, Kyle Lowry, Lauri Markkanen, Doug McDermott, Patty Mills, Royce O'Neale, Jordan Poole, Michael Porter Jr., Duncan Robinson, Landry Shamet, Anfernee Simons, Marcus Smart, Fred VanVleet, Coby White.

	      Number of guards: 18; forward-guard: 5; forward: 7; forward-center: 2; center: 1.

	\item[Cluster 2]  Bam Adebayo, Jarrett Allen, Deandre Ayton, Bradley Beal, Bogdan Bogdanović, Devin Booker, Dillon Brooks, Jalen Brunson, Alec Burks, Clint Capela, Brandon Clarke, Nic Claxton, Mike Conley, Seth Curry, Anthony Davis, DeMar DeRozan, Andre Drummond, Kevin Durant, Joel Embiid, De'Aaron Fox, Markelle Fultz, Daniel Gafford, Paul George, Shai Gilgeous-Alexander, Rudy Gobert, Tobias Harris, Gordon Hayward, Tyler Herro, Richaun Holmes, Brandon Ingram, Kyrie Irving, Reggie Jackson, Tyus Jones, Luke Kennard, Kawhi Leonard, CJ McCollum, T.J. McConnell, JaVale McGee, Khris Middleton, Dejounte Murray, Jamal Murray, Chris Paul, Mason Plumlee, Jakob Pöltl, Immanuel Quickley, Josh Richardson, Mitchell Robinson, Derrick Rose, D'Angelo Russell, Dennis Schröder, Collin Sexton, Pascal Siakam, Ben Simmons, Gary Trent Jr., Russell Westbrook, Andrew Wiggins, Ivica Zubac.

	      Number of guards: 28; forward-guard: 3; forward: 11; forward-center: 7; center: 8.

	\item[Cluster 3] Kyle Anderson, OG Anunoby, Saddiq Bey, Bojan Bogdanović, Mikal Bridges, Reggie Bullock Jr., Jimmy Butler, Kentavious Caldwell-Pope, Dorian Finney-Smith, Jeff Green, Rui Hachimura, Gary Harris, Justin Holiday, De'Andre Hunter, Cameron Johnson, Terance Mann, Monté Morris, Marcus Morris Sr., Georges Niang, Bobby Portis, Taurean Prince, Terry Rozier, Lonnie Walker IV, T.J. Warren.

	      Number of guards: 4; forward-guard: 9; forward: 11; forward-center: 0; center: 0.

	\item[Cluster 4]  Harrison Barnes, RJ Barrett, Chris Boucher, Miles Bridges, Bruce Brown, John Collins, Luguentz Dort, Aaron Gordon, Jerami Grant, Josh Hart, Keldon Johnson, Kyle Kuzma, Larry Nance Jr., Kelly Olynyk, Cedi Osman, Kelly Oubre Jr., Norman Powell, Delon Wright, Thaddeus Young.

	      Number of guards: 4; forward-guard: 4; forward: 8; forward-center: 3; center: 0.

	\item[Cluster 5]  Giannis Antetokounmpo, Marvin Bagley III, Malcolm Brogdon, Jaylen Brown, Thomas Bryant, Wendell Carter Jr., Jordan Clarkson, Spencer Dinwiddie, Luka Dončić, Darius Garland, Draymond Green, Tyrese Haliburton, Jrue Holiday, Al Horford, Jaren Jackson Jr., LeBron James, Nikola Jokić, Zach LaVine, Caris LeVert, De'Anthony Melton, Donovan Mitchell, Malik Monk, Ja Morant, Jusuf Nurkić, Kristaps Porziņģis, Dwight Powell, Julius Randle, Naz Reid, Domantas Sabonis, Dario Šarić, Jayson Tatum, Karl-Anthony Towns, Myles Turner, Jonas Valančiūnas, Nikola Vučević, Moritz Wagner, P.J. Washington, Derrick White, Christian Wood, Trae Young.

	      Number of guards: 14; forward-guard: 3; forward: 6; forward-center: 13; center: 4.

\end{description}

\begin{figure}
	\centering
	\begin{subfigure}{0.49\textwidth}
		\centering
		\includegraphics[scale=0.2]{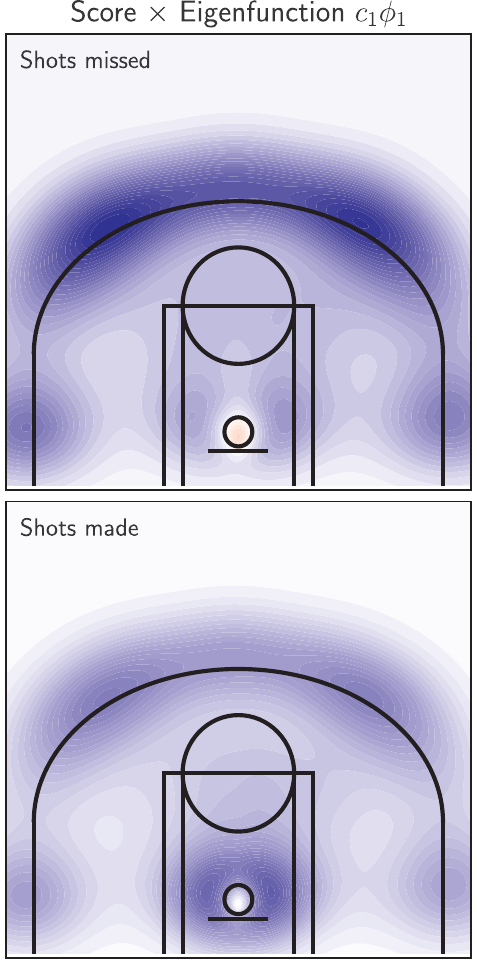}
		\includegraphics[scale=0.2]{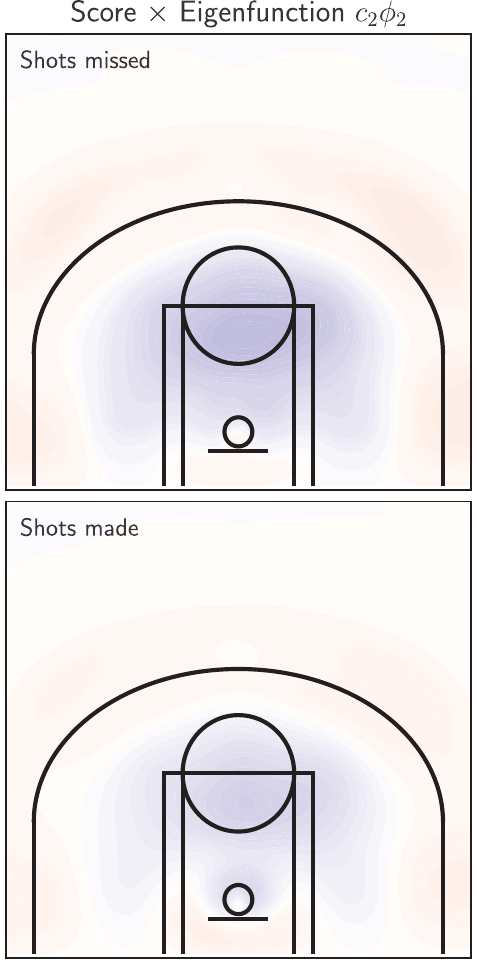}
		\includegraphics[scale=0.2]{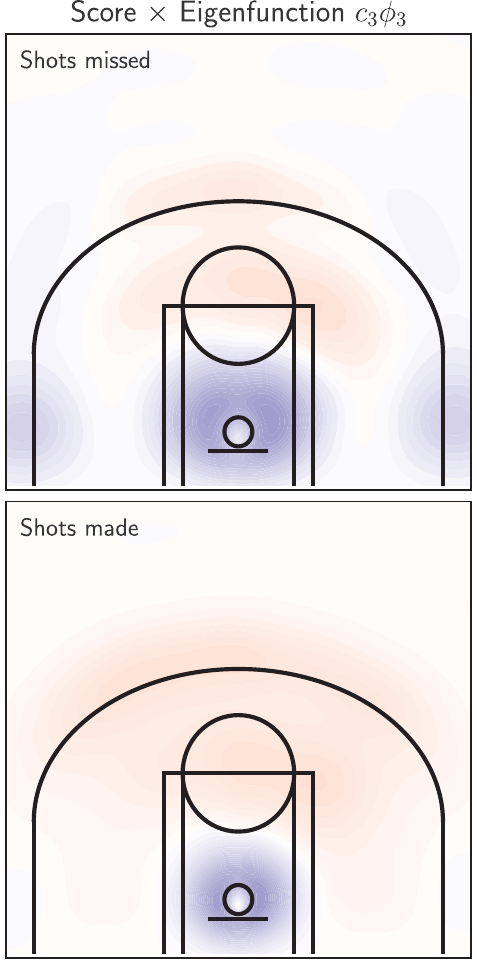}
		\includegraphics[scale=0.2]{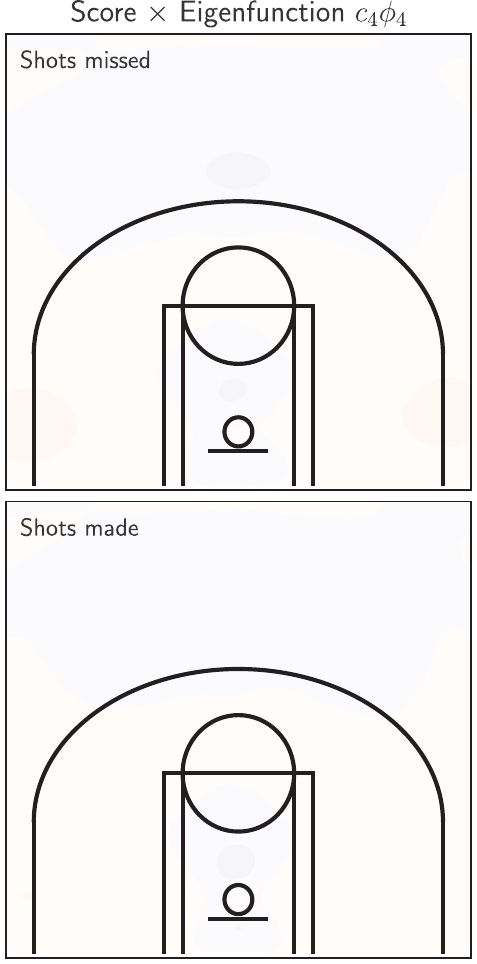}
		\caption{Cluster 1.}
	\end{subfigure}
	\hfill
	\begin{subfigure}{0.49\textwidth}
		\centering
		\includegraphics[scale=0.2]{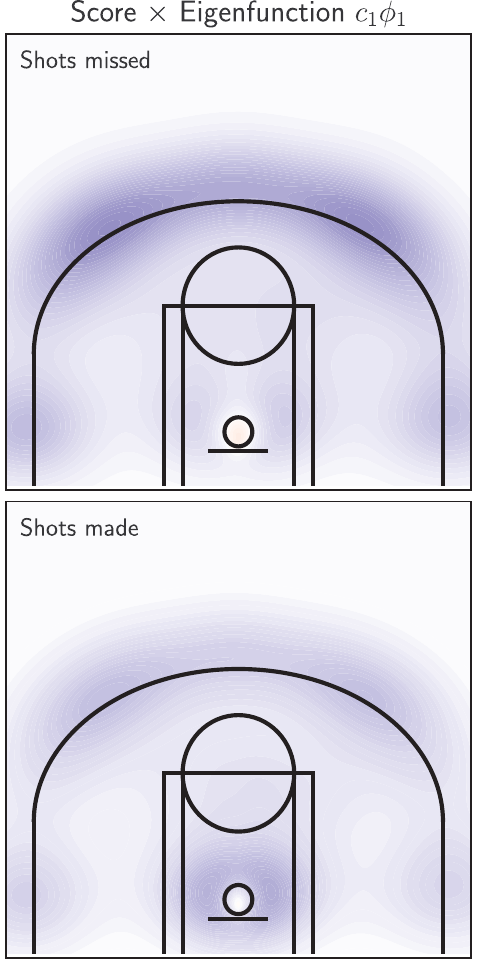}
		\includegraphics[scale=0.2]{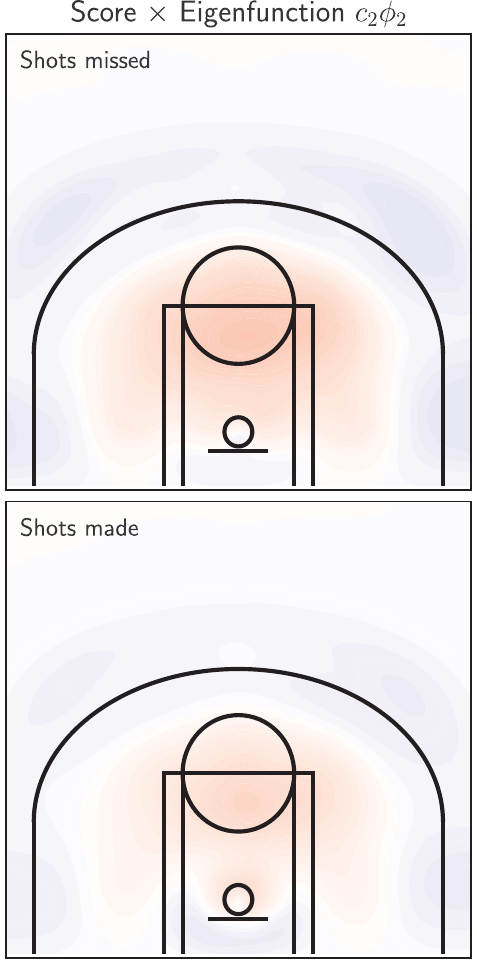}
		\includegraphics[scale=0.2]{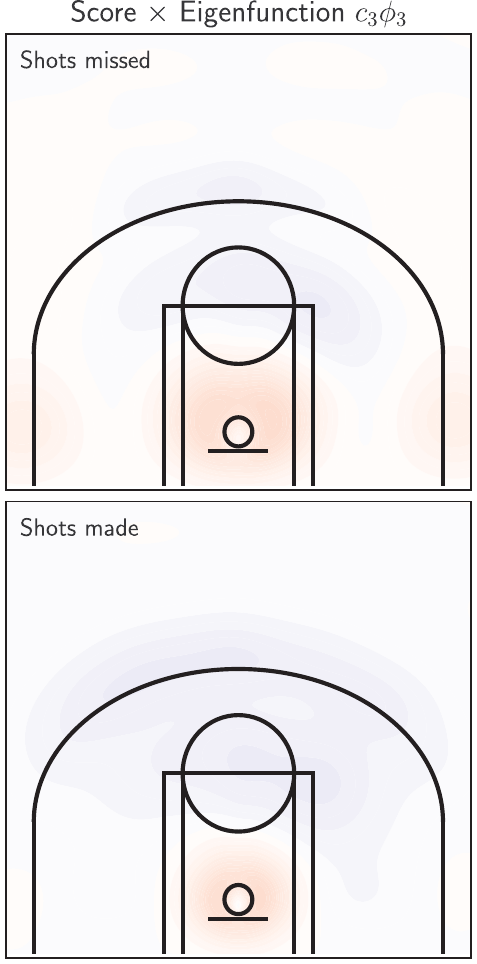}
		\includegraphics[scale=0.2]{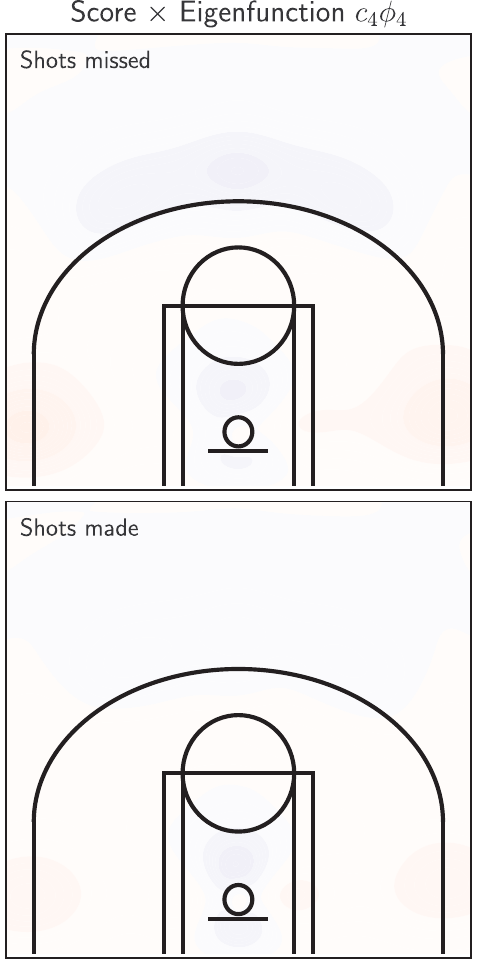}
		\caption{Cluster 2}
	\end{subfigure}
	\\
	\begin{subfigure}{0.49\textwidth}
		\centering
		\includegraphics[scale=0.2]{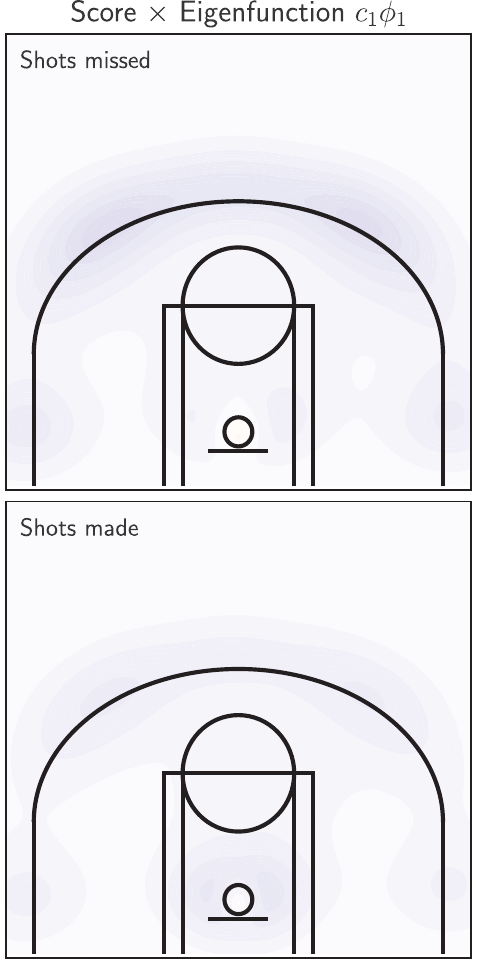}
		\includegraphics[scale=0.2]{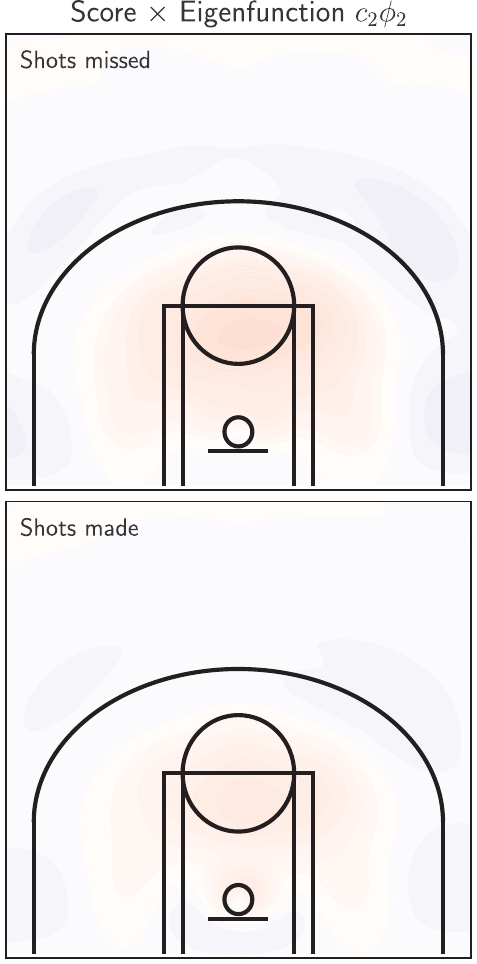}
		\includegraphics[scale=0.2]{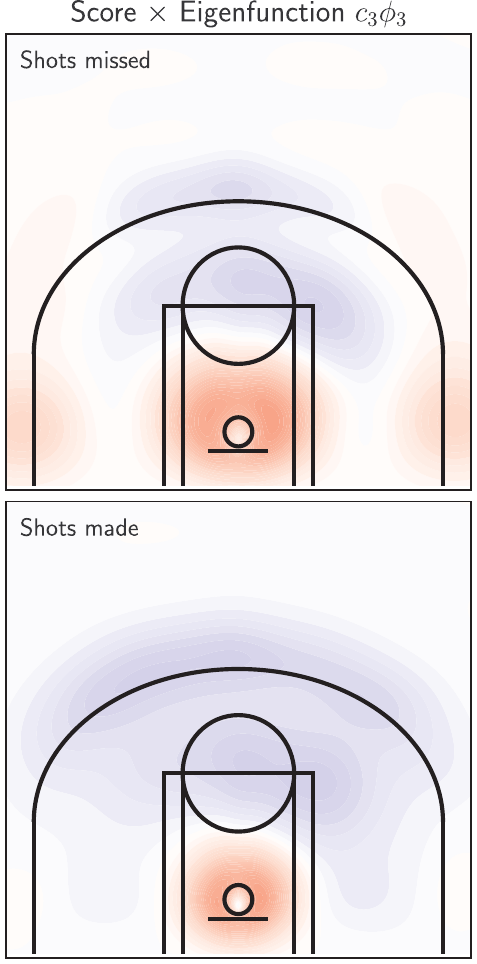}
		\includegraphics[scale=0.2]{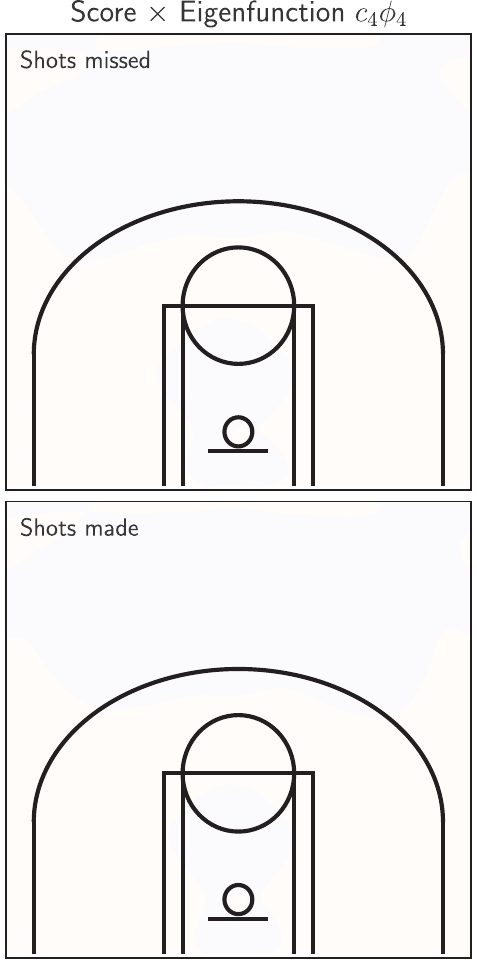}
		\caption{Cluster 3.}
	\end{subfigure}
	\hfill
	\begin{subfigure}{0.49\textwidth}
		\centering
		\includegraphics[scale=0.2]{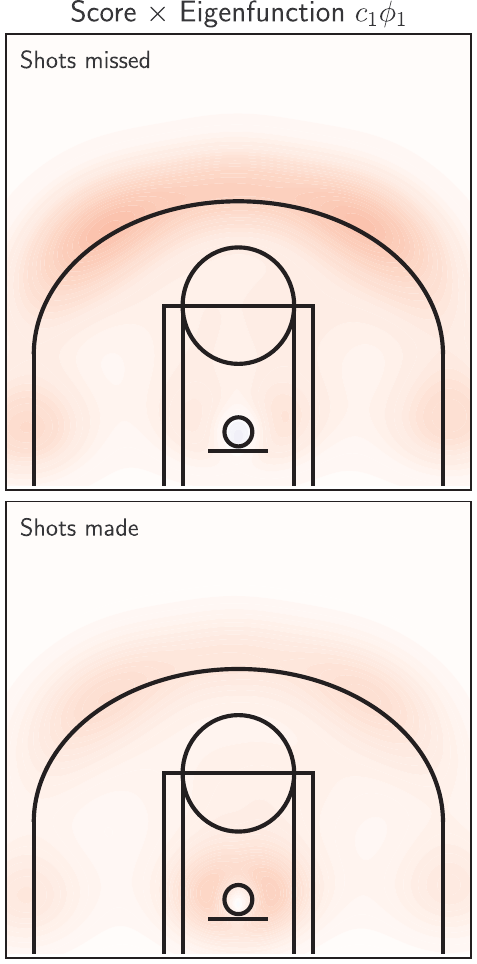}
		\includegraphics[scale=0.2]{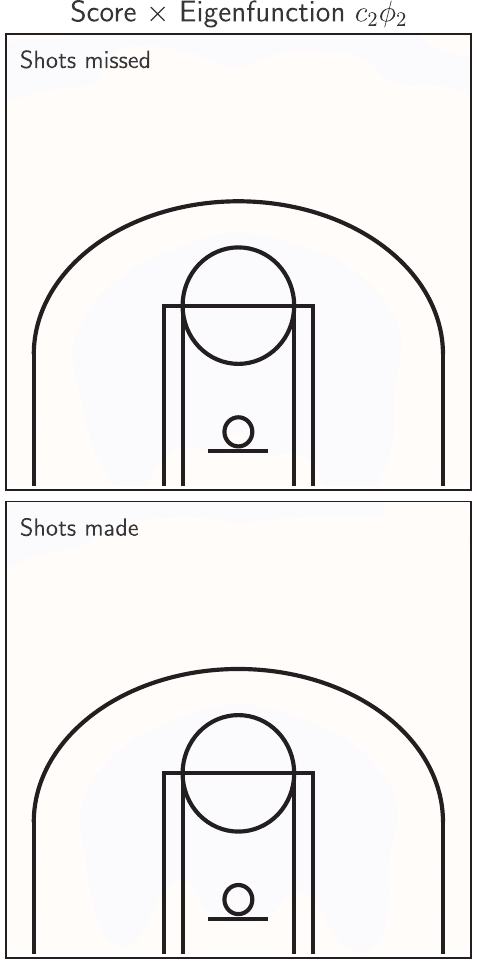}
		\includegraphics[scale=0.2]{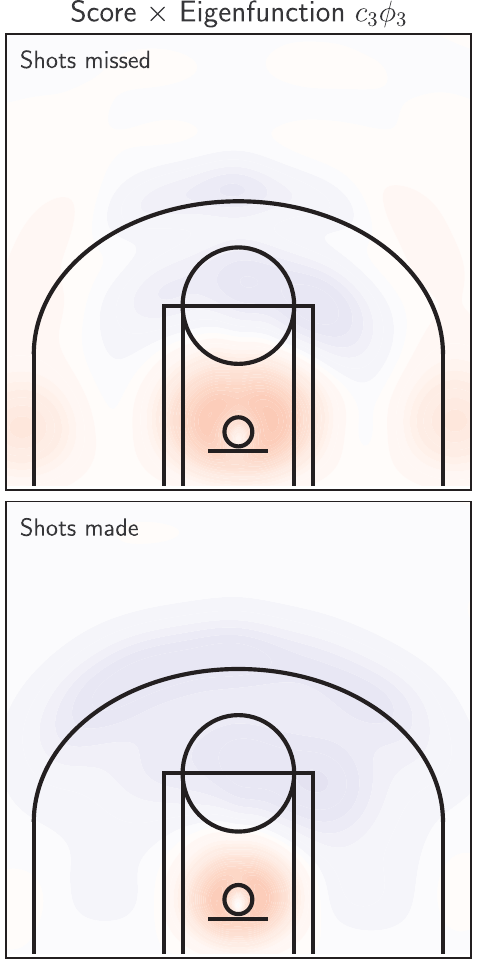}
		\includegraphics[scale=0.2]{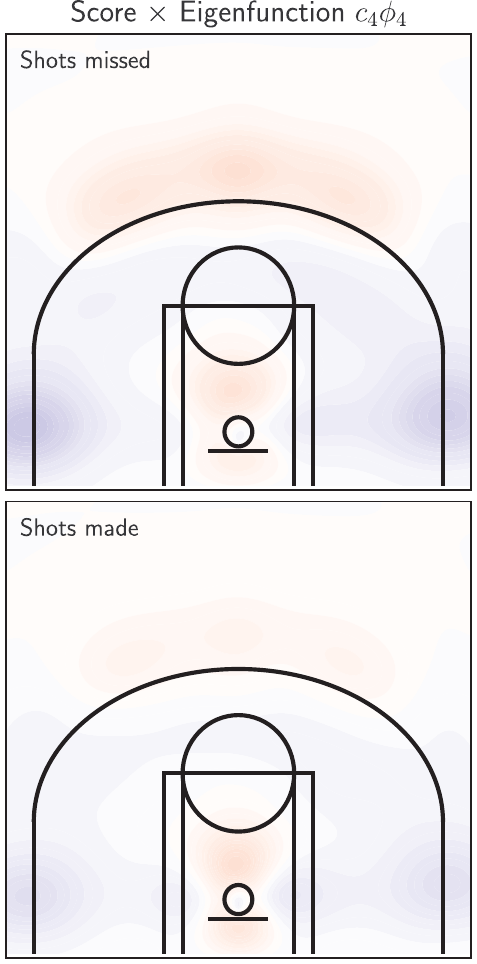}
		\caption{Cluster 4}
	\end{subfigure}
	\\
	\begin{subfigure}{0.49\textwidth}
		\centering
		\includegraphics[scale=0.2]{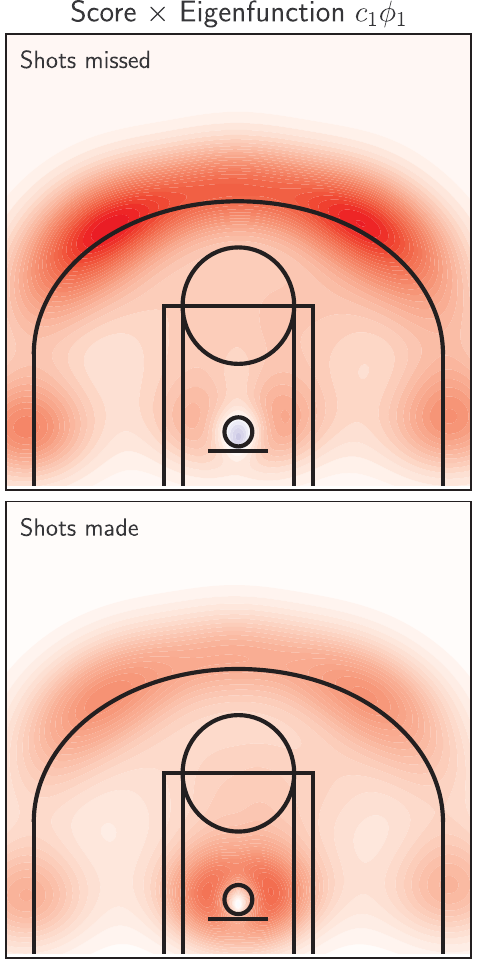}
		\includegraphics[scale=0.2]{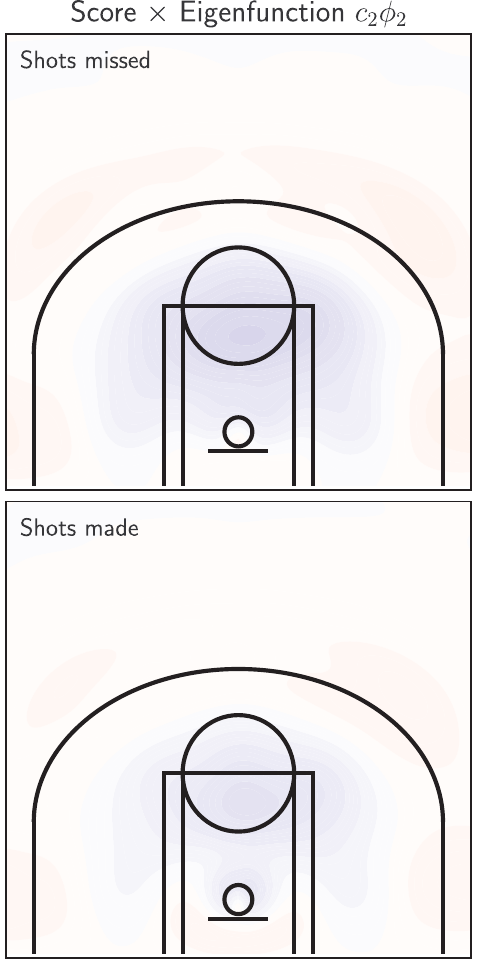}
		\includegraphics[scale=0.2]{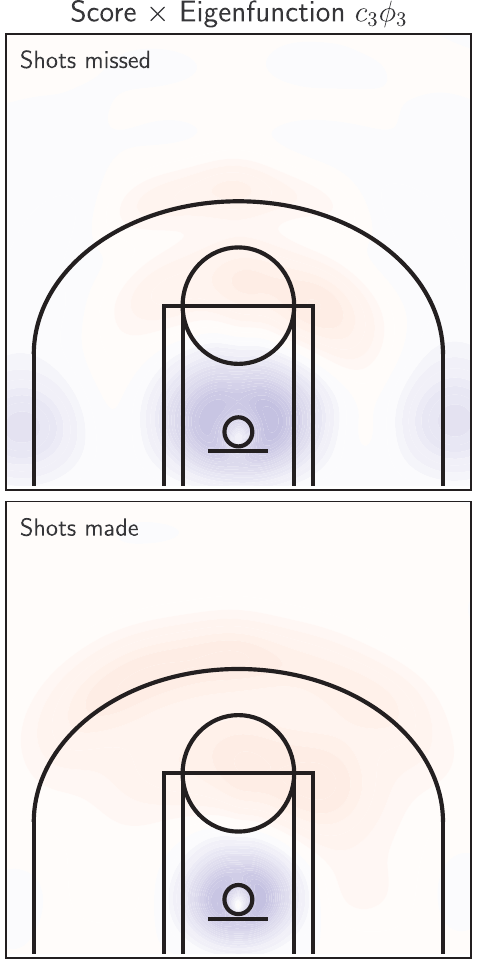}
		\includegraphics[scale=0.2]{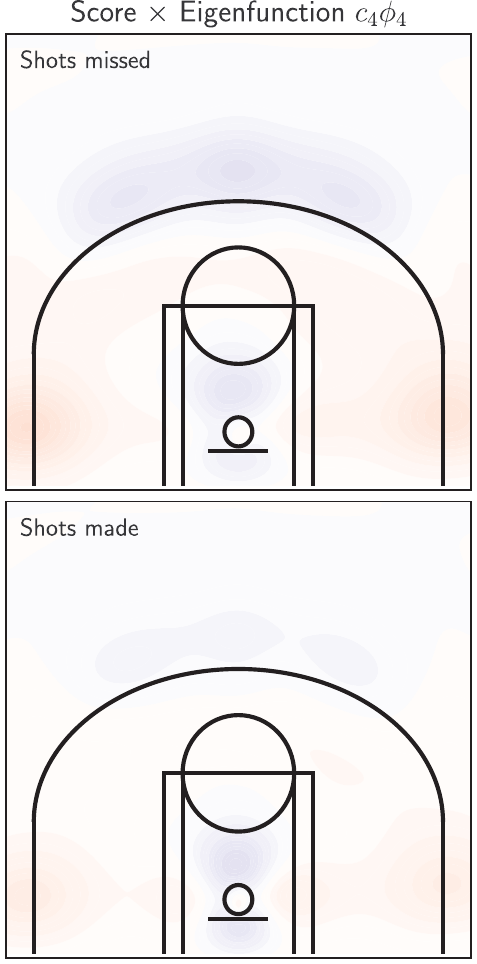}
		\caption{Cluster 5}
	\end{subfigure}
  \caption{MFPCA decompositions from the clustering using variance weights for each component. The normalization is applied to the combined functions across made and missed shots.}
	\label{fig:cluster_weight}
\end{figure}

Second, we present in which cluster each of the $173$ players belong when we assign a weight to each component given the proportion of variance they explain. Figure \ref{fig:cluster_weight} presents the decomposition of each cluster when the clustering is performed when weights reflect the proportion of the variance they explained. Since the first component accounts for approximately $70\%$ of the total variance, the clusters are primarily driven by the first score $c_1$, reflecting the overall shooting frequency of the players.

\begin{description}
	\item[Cluster 1] Bam Adebayo, Jarrett Allen, Giannis Antetokounmpo, Deandre Ayton, Marvin Bagley III, Jimmy Butler, Clint Capela, Brandon Clarke, Nic Claxton, Anthony Davis, Andre Drummond, Markelle Fultz, Daniel Gafford, Rudy Gobert, Richaun Holmes, Nikola Jokić, T.J. McConnell, JaVale McGee, Ja Morant, Jusuf Nurkić, Mason Plumlee, Jakob Pöltl, Dwight Powell, Mitchell Robinson, Domantas Sabonis, Ben Simmons, Jonas Valančiūnas, Russell Westbrook, Thaddeus Young, Ivica Zubac.

	      Number of guards: 4; forward-guard: 1; forward: 6; forward-center: 8; center: 11.

	\item[Cluster 2] Kyle Anderson, OG Anunoby, RJ Barrett, Bradley Beal, Chris Boucher, Miles Bridges, Malcolm Brogdon, Bruce Brown, Jaylen Brown, Thomas Bryant, Wendell Carter Jr., John Collins, DeMar DeRozan, Joel Embiid, De'Aaron Fox, Shai Gilgeous-Alexander, Aaron Gordon, Jerami Grant, Draymond Green, Rui Hachimura, Tobias Harris, Jrue Holiday, Jaren Jackson Jr., LeBron James, Keldon Johnson, Zach LaVine, Caris LeVert, Terance Mann, Dejounte Murray, Larry Nance Jr., Kelly Oubre Jr., Bobby Portis, Julius Randle, Naz Reid, Derrick Rose, Dennis Schröder, Collin Sexton, Pascal Siakam, Karl-Anthony Towns, Nikola Vučević, Moritz Wagner, T.J. Warren, Christian Wood, Delon Wright.

	      Number of guards: 12; forward-guard: 9; forward: 12; forward-center: 10; center: 1.

	\item[Cluster 3] Harrison Barnes, Bojan Bogdanović, Devin Booker, Mikal Bridges, Dillon Brooks, Jalen Brunson, Spencer Dinwiddie, Luguentz Dort, Darius Garland, Jeff Green, James Harden, Gary Harris, Josh Hart, Gordon Hayward, De'Andre Hunter, Brandon Ingram, Kyrie Irving, Kyle Kuzma, Kawhi Leonard, Damian Lillard, Donovan Mitchell, Kristaps Porziņģis, Norman Powell, Terry Rozier, Dario Šarić, Jayson Tatum, Myles Turner, P.J. Washington, Andrew Wiggins, Trae Young.

	      Number of guards: 14; forward-guard: 4; forward: 9; forward-center: 3; center: 0.

	\item[Cluster 4] Saddiq Bey, Alec Burks, Kentavious Caldwell-Pope, Jordan Clarkson, Mike Conley, Donte DiVincenzo, Luka Dončić, Kevin Durant, Dorian Finney-Smith, Paul George, Eric Gordon, Tyrese Haliburton, Joe Harris, Al Horford, Reggie Jackson, Tyus Jones, Brook Lopez, Lauri Markkanen, CJ McCollum, Doug McDermott, De'Anthony Melton, Khris Middleton, Malik Monk, Monté Morris, Jamal Murray, Kelly Olynyk, Cedi Osman, Jordan Poole, Taurean Prince, Josh Richardson, Fred VanVleet, Lonnie Walker IV, Coby White, Derrick White.

	      Number of guards: 19; forward-guard: 4; forward: 7; forward-center: 3; center: 1.

	\item[Cluster 5] Grayson Allen, Malik Beasley, Bogdan Bogdanović, Reggie Bullock Jr., Pat Connaughton, Robert Covington, Jae Crowder, Seth Curry, Stephen Curry, Evan Fournier, Danilo Gallinari, Devonte' Graham, Tim Hardaway Jr., Tyler Herro, Buddy Hield, Justin Holiday, Kevin Huerter, Joe Ingles, Cameron Johnson, Luke Kennard, Kevin Love, Kyle Lowry, Patty Mills, Marcus Morris Sr., Georges Niang, Royce O'Neale, Chris Paul, Michael Porter Jr., Immanuel Quickley, Duncan Robinson, D'Angelo Russell, Landry Shamet, Anfernee Simons, Marcus Smart, Gary Trent Jr..

	      Number of guards: 19; forward-guard: 6; forward: 9; forward-center: 1; center: 0.

\end{description}

\end{appendix}

\section*{Acknowledgment}

S. Golovkine did this work while he was at the University of Limerick and was partially supported by Science Foundation Ireland under Grant No. 19/FFP/7002 and co-funded under the European Regional Development Fund.

\bibliographystyle{apalike}
\bibliography{./biblio.bib}

@inproceedings{millerFactorizedPointProcess2014,
  title = {Factorized Point Process Intensities: A Spatial Analysis of Professional},
  shorttitle = {Factorized Point Process Intensities},
  booktitle = {Proceedings of the 31st {{International Conference}} on {{International Conference}} on {{Machine Learning}} - {{Volume}} 32},
  author = {Miller, Andrew and Bornn, Luke and Adams, Ryan and Goldsberry, Kirk},
  year = {2014},
  month = jun,
  series = {{{ICML}}'14},
  pages = {I-235--I-243},
  publisher = {JMLR.org},
  address = {Beijing, China},
  urldate = {2024-06-18}
}

@article{jiaoBayesianMarkedSpatial2021,
  title = {A {{Bayesian}} Marked Spatial Point Processes Model for Basketball Shot Chart},
  author = {Jiao, Jieying and Hu, Guanyu and Yan, Jun},
  year = {2021},
  month = jun,
  journal = {Journal of Quantitative Analysis in Sports},
  volume = {17},
  number = {2},
  pages = {77--90},
  publisher = {De Gruyter},
  issn = {1559-0410},
  doi = {10.1515/jqas-2019-0106},
  urldate = {2024-06-18}
}

@article{franksCharacterizingSpatialStructure2015,
  title = {Characterizing the {{Spatial Structure}} of {{Defensive Skill}} in {{Professional Basketball}}},
  author = {Franks, Alexander and Miller, Andrew and Bornn, Luke and Goldsberry, Kirk},
  year = {2015},
  journal = {The Annals of Applied Statistics},
  volume = {9},
  number = {1},
  eprint = {24522412},
  eprinttype = {jstor},
  pages = {94--121},
  publisher = {Institute of Mathematical Statistics},
  issn = {1932-6157},
  urldate = {2024-06-18}
}

@article{maddoxBayesianEstimationIngame2022,
  title = {Bayesian Estimation of In-Game Home Team Win Probability for College Basketball},
  author = {Maddox, Jason T. and Sides, Ryan and Harvill, Jane L.},
  year = {2022},
  month = sep,
  journal = {Journal of Quantitative Analysis in Sports},
  volume = {18},
  number = {3},
  pages = {201--213},
  publisher = {De Gruyter},
  issn = {1559-0410},
  doi = {10.1515/jqas-2021-0086},
}

@article{ternerModelingPlayerTeam2021,
  title = {Modeling {{Player}} and {{Team Performance}} in {{Basketball}}},
  author = {Terner, Zachary and Franks, Alexander},
  year = {2021},
  month = mar,
  journal = {Annual Review of Statistics and Its Application},
  volume = {8},
  number = {Volume 8, 2021},
  pages = {1--23},
  publisher = {Annual Reviews},
  issn = {2326-8298, 2326-831X},
  doi = {10.1146/annurev-statistics-040720-015536},
}

@article{cervoneMultiresolutionStochasticProcess2016,
  title = {A {{Multiresolution Stochastic Process Model}} for {{Predicting Basketball Possession Outcomes}}},
  author = {Cervone, Daniel and D'Amour, Alex and Bornn, Luke and Goldsberry, Kirk},
  year = {2016},
  month = apr,
  journal = {Journal of the American Statistical Association},
  volume = {111},
  number = {514},
  pages = {585--599},
  publisher = {Taylor \& Francis},
  issn = {0162-1459},
  doi = {10.1080/01621459.2016.1141685},
}

@article{fuHoopInSightAnalyzingComparing2024,
  title = {{{HoopInSight}}: {{Analyzing}} and {{Comparing Basketball Shooting Performance Through Visualization}}},
  shorttitle = {{{HoopInSight}}},
  author = {Fu, Yu and Stasko, John},
  year = {2024},
  month = jan,
  journal = {IEEE Transactions on Visualization and Computer Graphics},
  volume = {30},
  number = {1},
  pages = {858--868},
  issn = {1941-0506},
  doi = {10.1109/TVCG.2023.3326910},
}

@article{liuChangesBasketballShooting1999,
  title = {Changes in {{Basketball Shooting Patterns}} as a {{Function}} of {{Distance}}},
  author = {Liu, Suyen and Burton, Allen W.},
  year = {1999},
  month = dec,
  journal = {Perceptual and Motor Skills},
  volume = {89},
  number = {3},
  pages = {831--845},
  publisher = {SAGE Publications Inc},
  issn = {0031-5125},
  doi = {10.2466/pms.1999.89.3.831},
}

@article{csataljayEffectsDefensivePressure2013,
  title = {Effects of Defensive Pressure on Basketball Shooting Performance},
  author = {Csataljay, Gabor and James, Nic and Hughes, Mike and Dancs, Henriette},
  year = {2013},
  month = dec,
  journal = {International Journal of Performance Analysis in Sport},
  volume = {13},
  number = {3},
  pages = {594--601},
  publisher = {Routledge},
  issn = {2474-8668},
  doi = {10.1080/24748668.2013.11868673},
  urldate = {2024-07-01},
}

@article{okazakiReviewBasketballJump2015,
  title = {A Review on the Basketball Jump Shot},
  author = {Okazaki, Victor H.A. and Rodacki, Andr{\'e} L.F. and Satern, Miriam N.},
  year = {2015},
  month = apr,
  journal = {Sports Biomechanics},
  volume = {14},
  number = {2},
  pages = {190--205},
  publisher = {Routledge},
  issn = {1476-3141},
  doi = {10.1080/14763141.2015.1052541},
}

@article{zuccolottoSpatialPerformanceAnalysis2023,
  title = {Spatial Performance Analysis in Basketball with {{CART}}, Random Forest and Extremely Randomized Trees},
  author = {Zuccolotto, Paola and Sandri, Marco and Manisera, Marica},
  year = {2023},
  journal = {Annals of Operations Research},
  volume = {325},
  number = {1},
  pages = {495--519},
  issn = {0254-5330},
  doi = {10.1007/s10479-022-04784-3},
}

@article{reichSpatialAnalysisBasketball2006,
  title = {A {{Spatial Analysis}} of {{Basketball Shot Chart Data}}},
  author = {Reich, Brian J. and Hodges, James S. and Carlin, Bradley P. and Reich, Adam M.},
  year = {2006},
  month = feb,
  journal = {The American Statistician},
  publisher = {Taylor \& Francis},
  doi = {10.1198/000313006X90305},
  urldate = {2024-07-02},
  copyright = {{\copyright} American Statistical Association},
  langid = {english},
}

@article{wong-toiJointAnalysisField2022,
  title = {A {{Joint Analysis}} for {{Field Goal Attempts}} and {{Percentages}} of {{Professional Basketball Players}}: {{Bayesian Nonparametric Resource}}},
  shorttitle = {A {{Joint Analysis}} for {{Field Goal Attempts}} and {{Percentages}} of {{Professional Basketball Players}}},
  author = {{Wong-Toi}, Eliot and Yang, Hou-Cheng and Shen, Weining and Hu, Guanyu},
  year = {2022},
  month = aug,
  journal = {Journal of Data Science},
  volume = {21},
  number = {1},
  pages = {68--86},
  publisher = {School of Statistics, Renmin University of China},
  issn = {1680-743X, 1683-8602},
  doi = {10.6339/22-JDS1062},
  urldate = {2024-07-02},
}

@article{yinAnalysisProfessionalBasketball2023,
  title = {Analysis of {{Professional Basketball Field Goal Attempts}} via a {{Bayesian Matrix Clustering Approach}}},
  author = {Yin, Fan and Hu, Guanyu and Shen, Weining},
  year = {2023},
  month = jan,
  journal = {Journal of Computational and Graphical Statistics},
  volume = {32},
  number = {1},
  pages = {49--60},
  publisher = {Taylor \& Francis},
  issn = {1061-8600},
  doi = {10.1080/10618600.2022.2085727},
  urldate = {2024-07-02},
}

@article{huZeroinflatedPoissonModel2023,
  title = {Zero-Inflated {{Poisson}} Model with Clustered Regression Coefficients: {{Application}} to Heterogeneity Learning of Field Goal Attempts of Professional Basketball Players},
  shorttitle = {Zero-Inflated {{Poisson}} Model with Clustered Regression Coefficients},
  author = {Hu, Guanyu and Yang, Hou-Cheng and Xue, Yishu and Dey, Dipak K.},
  year = {2023},
  journal = {Canadian Journal of Statistics},
  volume = {51},
  number = {1},
  pages = {157--172},
  issn = {1708-945X},
  doi = {10.1002/cjs.11684},
  urldate = {2024-10-03},
}

@misc{golovkineUseGramMatrix2024,
  title = {On the Use of the {{Gram}} Matrix for Multivariate Functional Principal Components Analysis},
  author = {Golovkine, Steven and Gunning, Edward and Simpkin, Andrew J. and Bargary, Norma},
  year = {2024},
  month = jun,
  number = {arXiv:2306.12949},
  eprint = {2306.12949},
  primaryclass = {stat},
  publisher = {arXiv},
  doi = {10.48550/arXiv.2306.12949},
  urldate = {2024-07-18},
}

@book{silvermanDensityEstimationStatistics1986,
  title = {Density {{Estimation}} for {{Statistics}} and {{Data Analysis}}},
  author = {Silverman, Bernard W.},
  year = {1986},
  month = apr,
  publisher = {CRC Press},
  isbn = {978-0-412-24620-3},

}

@inproceedings{yinBayesianNonparametricLearning2022,
  title = {Bayesian {{Nonparametric Learning}} for {{Point Processes}} with {{Spatial Homogeneity}}: {{A Spatial Analysis}} of {{NBA Shot Locations}}},
  shorttitle = {Bayesian {{Nonparametric Learning}} for {{Point Processes}} with {{Spatial Homogeneity}}},
  booktitle = {Proceedings of the 39th {{International Conference}} on {{Machine Learning}}},
  author = {Yin, Fan and Jiao, Jieying and Yan, Jun and Hu, Guanyu},
  year = {2022},
  month = jun,
  pages = {25523--25551},
  publisher = {PMLR},
  issn = {2640-3498},
  urldate = {2024-10-24},
}

@article{huBayesianGroupLearning2021,
  title = {Bayesian Group Learning for Shot Selection of Professional Basketball Players},
  author = {Hu, Guanyu and Yang, Hou-Cheng and Xue, Yishu},
  year = {2021},
  journal = {Stat},
  volume = {10},
  number = {1},
  pages = {e324},
  issn = {2049-1573},
  doi = {10.1002/sta4.324},
  urldate = {2024-10-24},
}

@article{millerRelationshipBasketballShooting1996,
  title = {The Relationship between Basketball Shooting Kinematics, Distance and Playing Position},
  author = {Miller, Stuart and Bartlett, Roger},
  year = {1996},
  month = jun,
  journal = {Journal of Sports Sciences},
  publisher = {Taylor \& Francis Group},
  doi = {10.1080/02640419608727708},
  urldate = {2024-07-01},
  copyright = {Copyright Taylor and Francis Group, LLC},
  langid = {english}
}

@article{dixonModellingAssociationFootball1997,
  title = {Modelling {{Association Football Scores}} and {{Inefficiencies}} in the {{Football Betting Market}}},
  author = {Dixon, Mark J. and Coles, Stuart G.},
  year = {1997},
  month = jun,
  journal = {Journal of the Royal Statistical Society Series C: Applied Statistics},
  volume = {46},
  number = {2},
  pages = {265--280},
  issn = {0035-9254},
  doi = {10.1111/1467-9876.00065}
}

@article{babootaPredictiveAnalysisModelling2019,
  title = {Predictive Analysis and Modelling Football Results Using Machine Learning Approach for {{English Premier League}}},
  author = {Baboota, Rahul and Kaur, Harleen},
  year = {2019},
  month = apr,
  journal = {International Journal of Forecasting},
  volume = {35},
  number = {2},
  pages = {741--755},
  issn = {0169-2070},
  doi = {10.1016/j.ijforecast.2018.01.003}
}

@article{bunkerApplicationMachineLearning2022,
  title = {The {{Application}} of {{Machine Learning Techniques}} for {{Predicting Match Results}} in {{Team Sport}}: {{A Review}}},
  shorttitle = {The {{Application}} of {{Machine Learning Techniques}} for {{Predicting Match Results}} in {{Team Sport}}},
  author = {Bunker, Rory and Susnjak, Teo},
  year = {2022},
  month = apr,
  journal = {Journal of Artificial Intelligence Research},
  volume = {73},
  pages = {1285--1322},
  issn = {1076-9757},
  doi = {10.1613/jair.1.13509}
}

@article{horvatUseMachineLearning2020,
  title = {The Use of Machine Learning in Sport Outcome Prediction: {{A}} Review},
  shorttitle = {The Use of Machine Learning in Sport Outcome Prediction},
  author = {Horvat, Tomislav and Job, Josip},
  year = {2020},
  journal = {WIREs Data Mining and Knowledge Discovery},
  volume = {10},
  number = {5},
  pages = {e1380},
  issn = {1942-4795},
  doi = {10.1002/widm.1380}
}

@article{pappalardoPlayeRankDatadrivenPerformance2019,
  title = {{{PlayeRank}}: {{Data-driven Performance Evaluation}} and {{Player Ranking}} in {{Soccer}} via a {{Machine Learning Approach}}},
  shorttitle = {{{PlayeRank}}},
  author = {Pappalardo, Luca and Cintia, Paolo and Ferragina, Paolo and Massucco, Emanuele and Pedreschi, Dino and Giannotti, Fosca},
  year = {2019},
  month = sep,
  journal = {ACM Trans. Intell. Syst. Technol.},
  volume = {10},
  number = {5},
  pages = {59:1--59:27},
  issn = {2157-6904},
  doi = {10.1145/3343172}
}

@inproceedings{decroosActionsSpeakLouder2019,
  title = {Actions {{Speak Louder}} than {{Goals}}: {{Valuing Player Actions}} in {{Soccer}}},
  shorttitle = {Actions {{Speak Louder}} than {{Goals}}},
  booktitle = {Proceedings of the 25th {{ACM SIGKDD International Conference}} on {{Knowledge Discovery}} \& {{Data Mining}}},
  author = {Decroos, Tom and Bransen, Lotte and Van Haaren, Jan and Davis, Jesse},
  year = {2019},
  month = jul,
  series = {{{KDD}} '19},
  pages = {1851--1861},
  publisher = {Association for Computing Machinery},
  address = {New York, NY, USA},
  doi = {10.1145/3292500.3330758}
}

@article{schmidSimulatingDefensiveTrajectories2021,
  title = {Simulating {{Defensive Trajectories}} in {{American Football}} for {{Predicting League Average Defensive Movements}}},
  author = {Schmid, Marc and Blauberger, Patrick and Lames, Martin},
  year = {2021},
  month = jul,
  journal = {Frontiers in Sports and Active Living},
  volume = {3},
  publisher = {Frontiers},
  issn = {2624-9367},
  doi = {10.3389/fspor.2021.669845}
}

@book{ramsayFunctionalDataAnalysis2005,
  title = {Functional {{Data Analysis}}},
  author = {Ramsay, J. O. and Silverman, B. W.},
  year = {2005},
  series = {Springer {{Series}} in {{Statistics}}},
  publisher = {Springer},
  address = {New York, NY},
  doi = {10.1007/b98888},
}

@book{kokoszkaIntroductionFunctionalData2017,
  title = {Introduction to {{Functional Data Analysis}}},
  author = {Kokoszka, Piotr and Reimherr, Matthew},
  year = {2017},
  month = sep,
  publisher = {{Chapman and Hall/CRC}},
  address = {New York},
}

@book{kaufmanFindingGroupsData2009,
  title = {Finding {{Groups}} in {{Data}}: {{An Introduction}} to {{Cluster Analysis}}},
  shorttitle = {Finding {{Groups}} in {{Data}}},
  author = {Kaufman, Leonard and Rousseeuw, Peter J.},
  year = {2009},
  month = sep,
  publisher = {John Wiley \& Sons}
}

@article{happMultivariateFunctionalPrincipal2018,
  title = {Multivariate {{Functional Principal Component Analysis}} for {{Data Observed}} on {{Different}} ({{Dimensional}}) {{Domains}}},
  author = {Happ, Clara and Greven, Sonja},
  year = {2018},
  month = apr,
  journal = {Journal of the American Statistical Association},
  volume = {113},
  number = {522},
  pages = {649--659},
  publisher = {Taylor \& Francis},
  issn = {0162-1459},
  doi = {10.1080/01621459.2016.1273115},
}

@article{kovalchikPlayerTrackingData2023,
  title = {Player {{Tracking Data}} in {{Sports}}},
  author = {Kovalchik, Stephanie A.},
  year = {2023},
  month = nov,
  journaltitle = {Annual Review of Statistics and Its Application},
  shortjournal = {Annu. Rev. Stat. Appl.},
  volume = {10},
  number = {1},
  pages = {677--697},
  issn = {2326-8298, 2326-831X},
  doi = {10.1146/annurev-statistics-033021-110117},
}

@article{hubertComparingPartitions1985,
  title = {Comparing Partitions},
  author = {Hubert, Lawrence and Arabie, Phipps},
  year = 1985,
  month = dec,
  journal = {Journal of Classification},
  volume = {2},
  number = {1},
  pages = {193--218},
  issn = {1432-1343},
  doi = {10.1007/BF01908075},
  urldate = {2025-12-23},
}

@article{rousseeuwSilhouettesGraphicalAid1987,
  title = {Silhouettes: {{A}} Graphical Aid to the Interpretation and Validation of Cluster Analysis},
  shorttitle = {Silhouettes},
  author = {Rousseeuw, Peter J.},
  year = 1987,
  month = nov,
  journal = {Journal of Computational and Applied Mathematics},
  volume = {20},
  pages = {53--65},
  issn = {0377-0427},
  doi = {10.1016/0377-0427(87)90125-7},
  urldate = {2025-12-24},
}

@article{petersenFunctionalDataAnalysis2016,
  title = {Functional Data Analysis for Density Functions by Transformation to a {{Hilbert}} Space},
  author = {Petersen, Alexander and M{\"u}ller, Hans-Georg},
  year = 2016,
  month = feb,
  journal = {The Annals of Statistics},
  volume = {44},
  number = {1},
  pages = {183--218},
  publisher = {Institute of Mathematical Statistics},
  issn = {0090-5364, 2168-8966},
  doi = {10.1214/15-AOS1363},
}

@misc{HawkeyeInnovation2025,
  author = {{Hawkeye Innovation, LLC}},
  howpublished = {\url{https://www.hawkeyeinnovation.com/}},
  year         = {2025},
  organization = {Hawkeye Innovation, LLC},
  address      = {Agawam, Massachusetts, United States},
}

\end{document}